\documentclass[a4paper,11pt]{article}
\pdfoutput=1
\usepackage{jheppub}
\makeatletter
\gdef\@fpheader{}
\makeatother

\usepackage{bm,microtype,booktabs}
\newcommand{\etas}{\eta_s}
\newcommand{\lacc}{\lambda^*}
\newcommand{\bgtemp}{T^{(0)}}
\newcommand{\tempresponse}{T^{(1)}}
\newcommand{\fieldratio}{\ensuremath{B_0^2/(3a_1T_0^4)}}
\newcommand{\flowresponse}{\Omega^{(1)}}
\newcommand{\figW}{\textwidth}
\newcommand{\ReportFigureOne}{\includegraphics[width=0.82\figW]{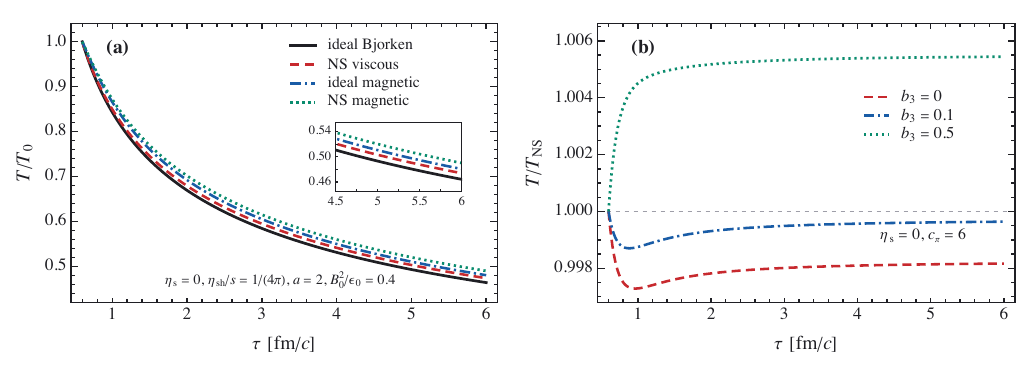}}
\newcommand{\ReportFigureTen}{\includegraphics[width=0.55\figW]{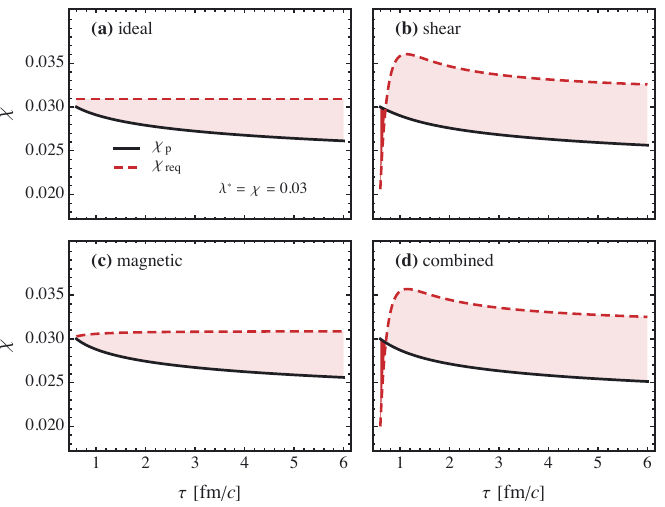}}
\newcommand{\ReportFigureTwo}{\includegraphics[width=0.82\figW]{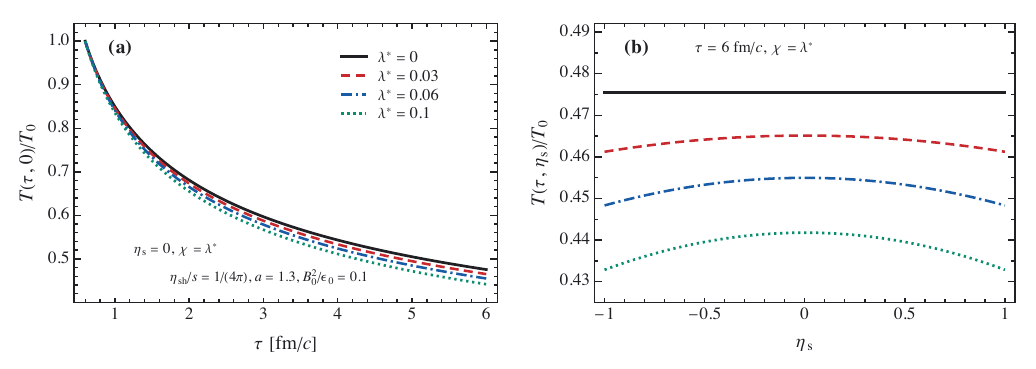}}
\newcommand{\ReportFigureThree}{\includegraphics[width=0.82\figW]{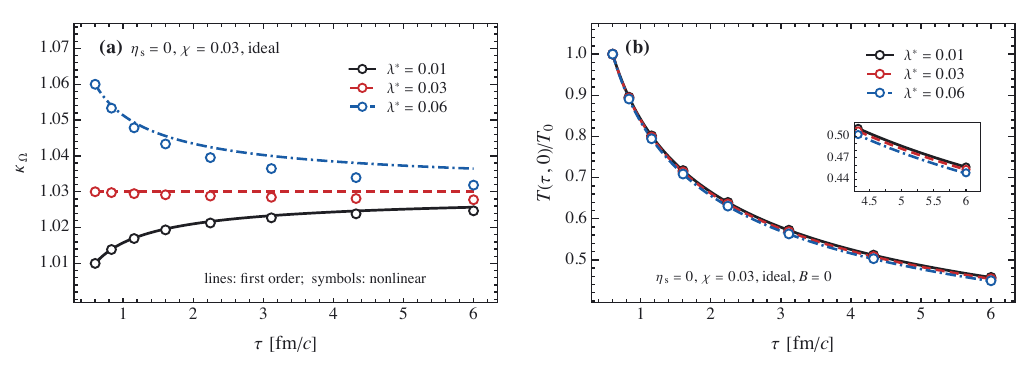}}
\newcommand{\ReportFigureFour}{\includegraphics[width=0.82\figW]{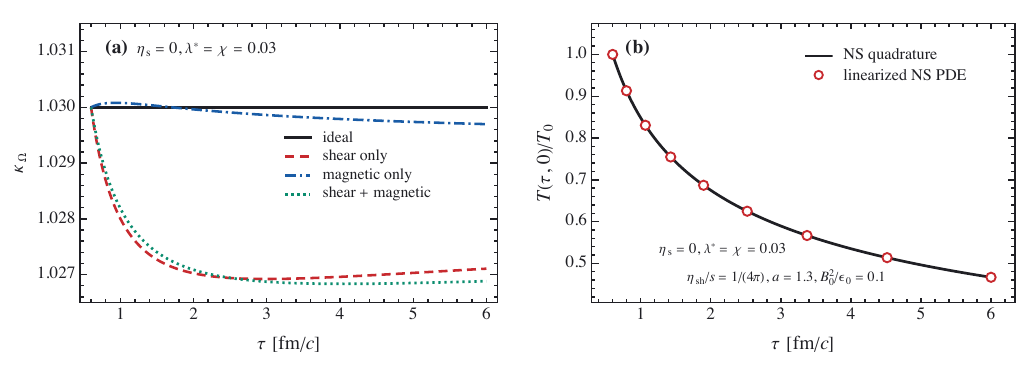}}
\newcommand{\ReportFigureFive}{\includegraphics[width=0.82\figW]{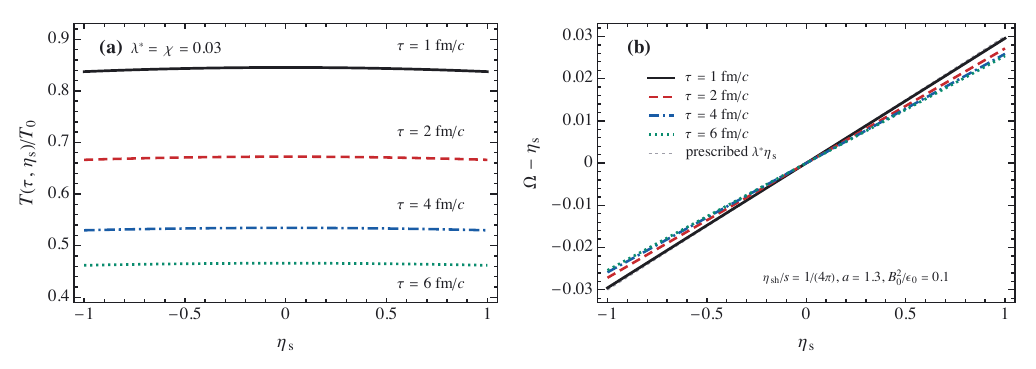}}
\newcommand{\ReportFigureEleven}{\includegraphics[width=\figW]{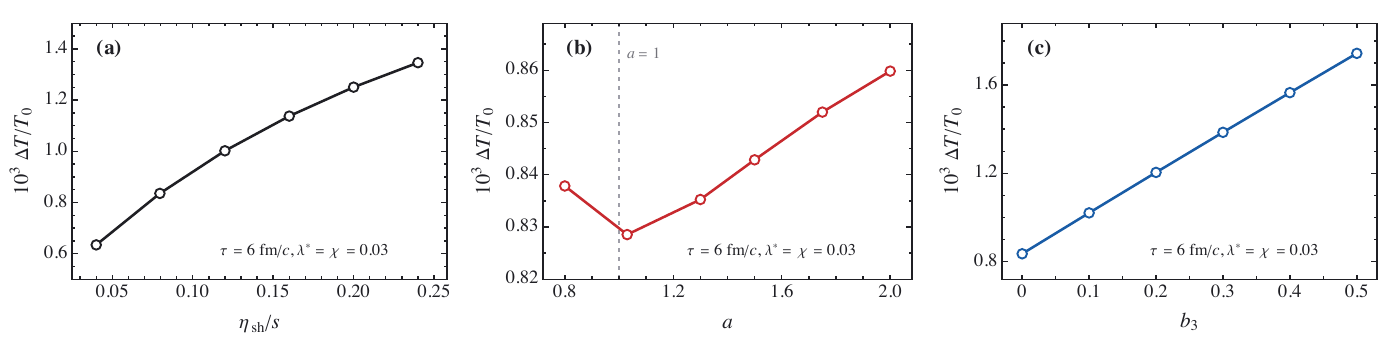}}
\newcommand{\ReportFigureTwelve}{\includegraphics[width=0.82\figW]{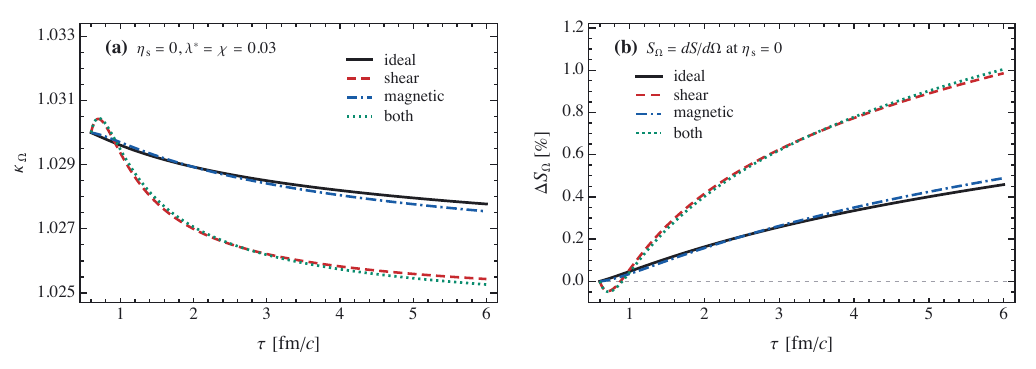}}
\newcommand{\ReportFigureSix}{\includegraphics[width=0.82\figW]{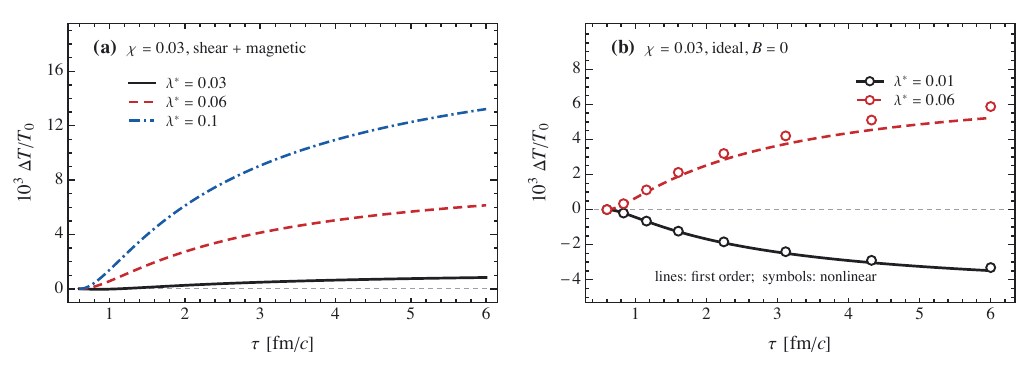}}
\newcommand{\ReportFigureSeven}{\includegraphics[width=0.82\figW]{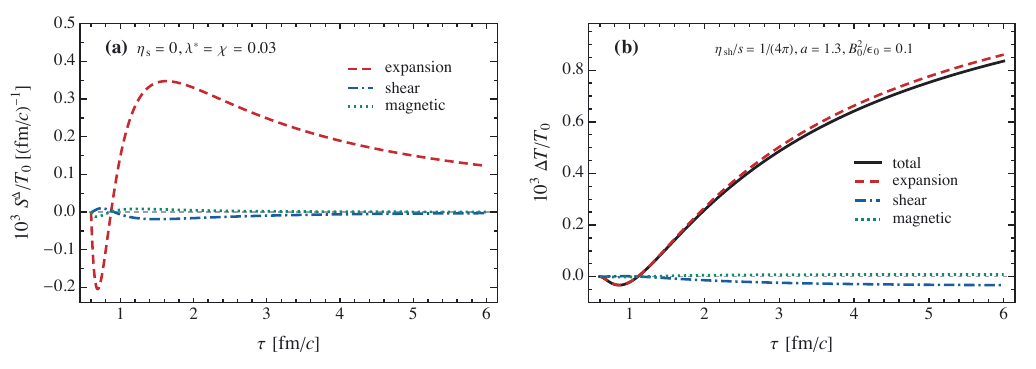}}
\newcommand{\ReportFigureEight}{\includegraphics[width=0.82\figW]{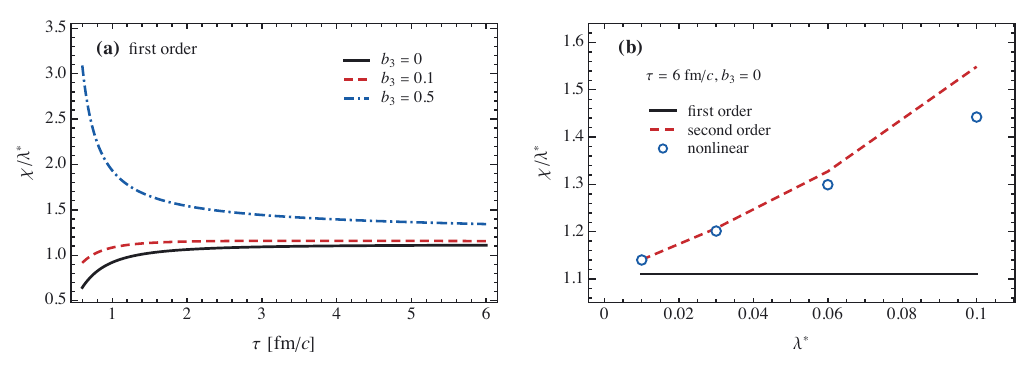}}
\newcommand{\ReportFigureNine}{\includegraphics[width=0.82\figW]{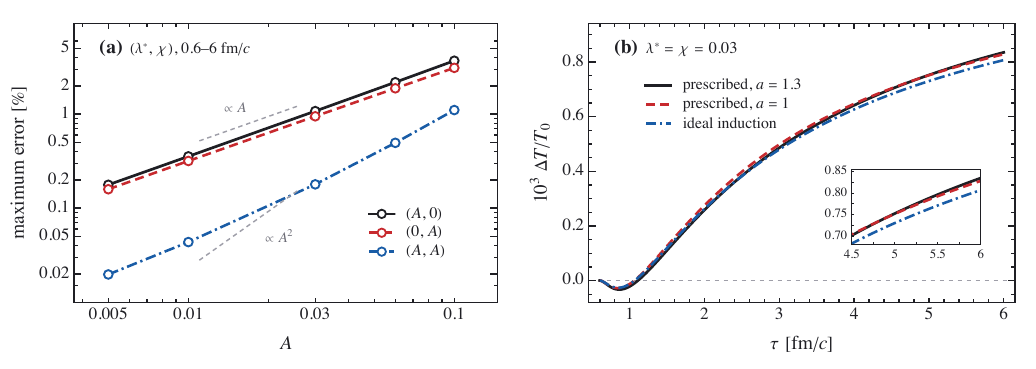}}

\title{1+1 dimensional relativistic viscous magnetohydrodynamics with self-consistent longitudinal acceleration}
\author[1,*]{Shuoyan Liu\note[*]{Corresponding author.}}
\author[2,3]{and Ze-Fang Jiang}
\affiliation[1]{Department of Physics and Siyuan Laboratory, Jinan University,
  Guangzhou 510632, China}
\affiliation[2]{Department of Physics and Electronic-Information Engineering,
  Hubei Engineering University, Xiaogan, Hubei, 432000, China}
\affiliation[3]{Institute of Particle Physics and Key Laboratory of Quark and
  Lepton Physics (MOE), Central China Normal University, Wuhan,
  Hubei, 430079, China}
\emailAdd{lsy0304@stu2026.jnu.edu.cn}
\emailAdd{jiangzf@mails.ccnu.edu.cn}
\abstract{
We study 1+1 dimensional relativistic viscous magnetohydrodynamics (MHD) with self-consistent longitudinal acceleration in a transverse magnetic field. Energy-momentum conservation and shear relaxation equations for fluid temperature, fluid rapidity $\Omega(\tau,\eta_s)$, and shear stress under the conformal equation of state (EoS) $\varepsilon=3p$ are solved. Extending boost-invariant ($\Omega=\eta_s$) and prescribed ($\Omega=(1+\lambda^*)\eta_s$) flows, first-order (Navier--Stokes) perturbative analytical solutions are first derived, and second-order (Israel--Stewart-type) nonlinear numerical solutions are compared with prescribed flow under identical initial conditions. Longitudinal pressure gradient from initial temperature curvature $\chi$ changes rapidity slope and expansion work, giving a different cooling rate of fluid temperature. Under power-law magnetic field $B(\tau)\propto\tau^{-a}$ at reference parameters, our numerical results show that central temperature difference between the two flows comes mainly from shear viscosity rather than magnetic stress, and equal-temperature ratio $\chi/\lambda^*$ depends on proper time $\tau$ and initial shear stress. At $\chi=\lambda^*=0.03$, prescribed flow carries a non-zero momentum residual, while maximum relative difference of central fluid temperature during $\tau=0.6$--$6$~fm/$c$ is found to be only $0.18\%$. Due to inverse rapidity slope $(\partial_{\eta_s}\Omega)^{-1}$, entropy per unit fluid rapidity $dS/d\Omega$ still shows non-zero difference at equal central temperatures. In the ideal MHD (flux-freezing) limit, direct magnetic work in fluid energy equation cancels out, with magnetic pressure and magnetic inertia retained in longitudinal momentum equation.
}

\begin{document}
\maketitle
\flushbottom

\section{Introduction}
\label{sec:introduction}
In relativistic heavy-ion collisions, hot quark--gluon plasma (QGP) expanding under strong longitudinal gradients is produced. Strong magnetic fields of $10^{18}$--$10^{19}$~G are also generated in non-central collisions and affect both stress tensor and spacetime evolution of QGP~\cite{Kharzeev2008,DengHuang2012}. Thermal photon production in QGP depends on fluid temperature evolution, with dynamical uncertainty of cooling history directly affecting electromagnetic probe observables~\cite{Jing2026}.

In boost-invariant Bjorken expansion~\cite{Bjorken1983}, fluid rapidity equals spacetime rapidity ($\Omega=\etas$) and a conformal ideal fluid cools as $T\propto\tau^{-1/3}$. When a transverse magnetic field is frozen into the fluid in ideal magnetohydrodynamics (MHD), it decays as $B\propto\tau^{-1}$ and magnetic energy drops out of the fluid energy equation, so the ideal Bjorken cooling law holds without modification~\cite{Roy2015}. This cooling behavior is modified once finite magnetization is taken into account~\cite{Pu2016}, or when shear viscosity is combined with a prescribed power-law magnetic field $B\propto\tau^{-a}$~\cite{Jiang2025}. In Ref.~\cite{Jiang2025}, first-order (Navier--Stokes) and second-order (Israel--Stewart) shear stress equations were solved in boost-invariant flow, with initial shear stress affecting fluid temperature evolution in the second-order case.

To include longitudinal acceleration, longitudinal boost invariance ($\Omega=\etas$) has to be relaxed. In ideal relativistic hydrodynamics, exact accelerating solutions were obtained in the Cs\"org\H{o}--Nagy--Csan\'ad (CNC)~\cite{Csorgo2008,Nagy2008} and Cs\"org\H{o}--Kasza--Csan\'ad--Jiang (CKCJ)~\cite{CKCJ2018} families. When a transverse magnetic field or shear viscosity is added, weakly accelerating flows with a fixed rapidity profile $\Omega=(1+\lacc)\etas$ at all proper times have been studied perturbatively~\cite{She2019,Jiang2020}. Magnetic accelerating solutions were derived in Ref.~\cite{She2019}, and viscous perturbative solutions were derived and compared with shear-relaxation models in Ref.~\cite{Jiang2020}.

In heavy-ion collisions, viscous hydrodynamic codes describe QGP evolution up to final-state hadrons. For $(2+1)$-dimensional boost-invariant expansion, VISH2+1~\cite{SongHeinz2008} computes transverse flow, and iEBE-VISHNU~\cite{ShenVISHNU} couples event-by-event hydrodynamics to hadronic transport. Full $(3+1)$-dimensional evolution is implemented in MUSIC~\cite{MusicIdeal,MusicViscous}, vHLLE~\cite{KarpenkoVHLLE}, and ECHO-QGP~\cite{DelZannaECHO}, with three-dimensional viscous calculations used for hadron spectra, femtoscopic radii~\cite{BozekViscous}, and longitudinal flow decorrelation from rapidity-dependent initial conditions and velocities in CLVisc~\cite{PangCLVisc}. From fits to hadron data, shear viscosity and initial conditions are constrained together~\cite{SongEta,BernhardBayes,JetscapeBayes}. Rapidity-dependent dissipative hydrodynamics is also used to test different shear-stress closures~\cite{FlorkowskiRapidity}. In our 1+1 dimensional longitudinal setup, transverse flow is omitted so that pressure-driven changes of the rapidity slope $\partial_{\etas}\Omega$ and their effect on expansion work and cooling can be isolated.

In this work, fluid rapidity $\Omega(\tau,\etas)$ is not fixed in advance but solved together with temperature and shear stress, using energy conservation, longitudinal momentum conservation and shear relaxation equation in prescribed transverse magnetic stress. How to get flow rapidity depending on both $\tau$ and $\etas$ was left as open problem in Ref.~\cite{Jiang2020}, since general longitudinal acceleration need numerical solution of coupled equations. For small initial gradients, coupled nonlinear equations are solved numerically, and analytical solutions of perturbative coefficients in Navier--Stokes limit are also given. Longitudinal pressure gradient changes fluid velocity, so expansion rate $\theta=\nabla_\mu u^\mu$ and expansion work $(\varepsilon+p)\theta$ are changed, which leads to different cooling of fluid element. We study this effect with shear relaxation and magnetic stress.

In Ref.~\cite{Jiang2018Fits}, acceleration parameters were extracted from charged-particle pseudorapidity distributions, and they correspond to a prescribed velocity profile. In our calculation, initial temperature profile and initial velocity profile are given separately, so the effect of thermal curvature can be distinguished from that of initial expansion. Near midrapidity, initial temperature curvature gives the longitudinal pressure gradient, and it also changes the rapidity slope at late times. For longitudinal expansion, exact solutions of second-order conformal hydrodynamics were obtained in Ref.~\cite{Hatta2014}. With general rapidity dependence, exact ideal-fluid and kinetic solutions were found in Refs.~\cite{Shi2022,Chen2024}, and anisotropic and viscous hydrodynamics were studied in Refs.~\cite{Chen2025a,Chen2025b}. The relation between pressure gradient and flow-rapidity evolution was also discussed in perturbative accelerating hydrodynamics~\cite{Kurgyis2017}. In resistive MHD with finite conductivity and ideal matter stress tensor, longitudinal acceleration comes from electromagnetic force~\cite{Moghaddam2020,Kord2022}, and such a resistive longitudinal solution was reproduced numerically in Milne coordinates in Ref.~\cite{Nakamura2023}. In this work, prescribed flow and self-consistent flow are compared with the same initial temperature, velocity, shear stress and magnetic stress profiles.
Section~\ref{sec:formalism} sets out the conservation equations, field and shear prescriptions, and initial data.
Section~\ref{sec:baselines} introduces shear and magnetic stress in Bjorken flow.
Section~\ref{sec:prescribedFlow} tests the momentum balance of a prescribed accelerating profile.
We then evolve the rapidity in Sec.~\ref{sec:selfConsistent} and compare the resulting cooling, expansion work, and entropy distributions in Sec.~\ref{sec:comparison}.
Appendix~\ref{sec:numericalMethod} gives the numerical methods and convergence tests.
\section{Longitudinal flow with a prescribed magnetic stress}
\label{sec:formalism}

We consider a conformal fluid undergoing longitudinal expansion in the presence of shear viscosity and a prescribed transverse magnetic stress.
Energy and momentum conservation determine the temperature and rapidity. A relaxation equation governs the longitudinal shear stress.

\subsection{Longitudinal kinematics and thermodynamics}
We use natural units, $\hbar=c=k_B=1$, and Milne coordinates
$x^\mu=(\tau,x,y,\etas)$, where
\begin{equation}
 \tau=\sqrt{t^2-z^2},\qquad
 \etas=\frac12\ln\frac{t+z}{t-z}.
\end{equation}
The metric is $g_{\mu\nu}=\operatorname{diag}(1,-1,-1,-\tau^2)$.
All fluid variables depend on $\tau$ and $\etas$; transverse flow and
transverse gradients are neglected. The laboratory longitudinal
velocity is $v_z=\tanh\Omega$. The four-velocity and longitudinal
unit vector are
\begin{align}
 u^\mu&=\left(\cosh(\Omega-\etas),0,0,
                   \frac{\sinh(\Omega-\etas)}{\tau}\right),\\
 n^\mu&=\left(\sinh(\Omega-\etas),0,0,
                   \frac{\cosh(\Omega-\etas)}{\tau}\right).
\end{align}
They obey $u^\mu u_\mu=1$, $n^\mu n_\mu=-1$ and $u^\mu n_\mu=0$.
The projector $\Delta^{\mu\nu}=g^{\mu\nu}-u^\mu u^\nu$
selects directions orthogonal to the fluid four-velocity,
$\Delta^{\mu\nu}u_\nu=0$; it defines the local rest space used
in the shear tensor below.
The operator $D$ differentiates along a fluid element, while
$\theta=\nabla_\mu u^\mu$ measures its fractional local-rest-frame
volume growth per unit material proper time. They are
\begin{align}
 D&=\cosh(\Omega-\etas)\partial_\tau
       +\frac{\sinh(\Omega-\etas)}{\tau}\partial_{\etas},\\
 \theta&=\sinh(\Omega-\etas)\partial_\tau\Omega
       +\frac{\cosh(\Omega-\etas)}{\tau}\partial_{\etas}\Omega.
 \label{eq:kinematics}
\end{align}
In particular, $n_\nu Du^\nu=-D\Omega$.

At zero chemical potential we take
\begin{equation}
 \varepsilon=3p=3a_1T^4,\quad s=4a_1T^3,\quad
 \eta_{\rm sh}=4a_1(\eta_{\rm sh}/s) T^3,
 \label{eq:eos}
\end{equation}
where $\varepsilon$, $p$, and $s$ denote the matter energy density, equilibrium pressure, and entropy density, respectively, excluding magnetic contributions. The shear-viscosity-to-entropy-density ratio $\eta_{\rm sh}/s$ is taken to be constant, and the subscript in $\eta_{\rm sh}$ distinguishes the shear viscosity from the spacetime rapidity $\etas$. The coefficient $a_1=g_*\pi^2/90$ is determined by the effective number of massless degrees of freedom, $g_*=16+21N_f/2$ for gluons and $N_f$ quark flavors, giving $a_1\simeq5.21$ for $N_f=3$. The conformal equation of state fixes the speed of sound at $c_s^2=1/3$ and leaves the temperature as the sole independent thermodynamic variable, allowing analytical boost-invariant baseline solutions and direct comparison with Refs.~\cite{She2019,Jiang2020,Jiang2025}. We use this approximation to describe matter above the QCD crossover. The HotQCD chiral analysis gives $T_c\simeq154$~MeV~\cite{HotQCDCrossover}. The lattice equation of state retains the temperature dependence of the trace anomaly $(\varepsilon-3p)/T^4$~\cite{HotQCD2014}.
\subsection{Magnetic and shear stresses}
Following Refs.~\cite{She2019,Jiang2025}, we prescribe the proper-time dependence of the comoving field strength $B$, with $B_0=B(\tau_0)$, as
\begin{equation}
 B^2(\tau)=B_0^2\left(\frac{\tau_0}{\tau}\right)^{2a}.
 \label{eq:field}
\end{equation}
We specify the field strength by the initial ratio
$B_0^2/\varepsilon_0=\fieldratio$, where
$\varepsilon_0=\varepsilon(\tau_0,0)=3a_1T_0^4$; the initial
magnetic-to-matter energy ratio is therefore $B_0^2/(2\varepsilon_0)$.
The exponent $a$ fixes the rate of decay.
To connect with the field convention of Ref.~\cite{Jiang2025}, we define
\begin{equation}
 \widetilde T=T/T_0,\qquad \sigma=B_0^2/T_0^4,\qquad
 \frac{B_0^2}{\varepsilon_0}=\frac{\sigma}{3a_1}.
\end{equation}
The parameters of that reference, $\epsilon_1=(a-1)\sigma/(12a_1)$ and
$\epsilon_2=4(\eta_{\rm sh}/s)/(9T_0)$ express the same magnetic and viscous sources; $\epsilon_2$ has dimensions of time in natural units. 
The scalar $\Phi\equiv-n_\mu n_\nu\pi^{\mu\nu}$ measures the shear correction to the longitudinal matter pressure, $p-\Phi$.
Transverse isotropy, tracelessness, and orthogonality to $u^\mu$ give
\begin{equation}
 \pi^{\mu\nu}=\frac{\Phi}{2}
              (-\Delta^{\mu\nu}-3n^\mu n^\nu).
 \label{eq:PhiTensor}
\end{equation}
It gives $\pi^{\mu\nu}\sigma_{\mu\nu}=\Phi\theta$ and
$n_\nu\nabla_\mu\pi^{\mu\nu}=n^\mu\partial_\mu\Phi+\Phi D\Omega$.
\subsection{Energy and longitudinal momentum conservation}
Both the cooling and the pressure-driven flow are governed by the conservation law $\nabla_\mu T^{\mu\nu}=0$, where the effective total energy-momentum tensor includes the matter contribution and the prescribed magnetic stress:
\begin{align}
 T^{\mu\nu}={}&(\varepsilon+p+B^2)u^\mu u^\nu
 -(p+B^2/2)g^{\mu\nu} -B^\mu B^\nu+\pi^{\mu\nu},\qquad
 B^2=-B^\mu B_\mu\ge0.
 \label{eq:tensor}
\end{align}
The magnetic four-vector $B^\mu$ represents the magnetic field measured in the fluid rest frame and satisfies $u_\mu B^\mu=0$. The field is transverse to the beam axis, so $n_\mu B^\mu=0$. In Heaviside--Lorentz units, the comoving magnetic energy density is $B^2/2$, while the field strength in heavy-ion phenomenology is conventionally quoted as $eB$, with $e^2=4\pi\alpha$. The matter energy density $\varepsilon$ excludes the magnetic contribution $B^2/2$. Bulk viscosity vanishes in the conformal limit, and we neglect the magnetization of the medium.
Projection along $u_\nu$ gives the comoving energy balance,
\begin{equation}
\begin{aligned}
 u_\nu\nabla_\mu T^{\mu\nu}=0
 &\ \Rightarrow\ D\varepsilon+(\varepsilon+p)\theta =\Phi\theta-D(B^2/2)-B^2\theta, \\
 \varepsilon=3a_1T^4
 &\ \Rightarrow\ 12a_1T^3DT =-4a_1T^4\theta+\Phi\theta-D(B^2/2)-B^2\theta.
\end{aligned}
\end{equation}
The corresponding momentum projection is
\begin{equation}
 n_\nu\nabla_\mu T^{\mu\nu}=0
 \ \Rightarrow\ (4a_1T^4+B^2-\Phi)D\Omega +n^\mu\partial_\mu(a_1T^4+B^2/2-\Phi)=0.
\end{equation}
Dividing the energy projection by $12a_1T^3$ and inserting the prescribed history yields the temperature and momentum equations
\begin{equation}
 DT+\frac{T\theta}{3}-\frac{\Phi\theta}{12a_1T^3} +\frac{B^2}{12a_1T^3}
   \left[\theta-\frac{a\cosh(\Omega-\etas)}{\tau}\right]=0.
 \label{eq:relaxE}
\end{equation}
\begin{equation}
 n^\mu\partial_\mu(a_1T^4+B^2/2) +(4a_1T^4+B^2-\Phi)D\Omega
 -n^\mu\partial_\mu\Phi=0.
 \label{eq:relaxM}
\end{equation}
\subsection{Background work and electromagnetic closure}

The matter balance associated with Eq.~\eqref{eq:tensor} is
\begin{align}
 \nabla_\mu T_{\rm matter}^{\mu\nu}
 =-\nabla_\mu\left[B^2u^\mu u^\nu-\frac{B^2}{2}g^{\mu\nu}
                                      -B^\mu B^\nu\right],
 \label{eq:force}
\end{align}
where $T_{\rm matter}^{\mu\nu}$ contains the fluid and shear stresses.
Equation~\eqref{eq:force} fixes the force and work exerted by the
prescribed magnetic stress from its divergence.

The field generated by the incoming nuclei depends on the collision geometry and varies rapidly in space and time~\cite{SkokovField,DengHuang2012}. Its subsequent evolution depends on the electromagnetic response of the matter~\cite{TuchinField,MclerranSkokov}, for which lattice estimates of the electrical conductivity provide a microscopic input~\cite{AartsConductivity}. Spatial gradients in a prescribed field can also change the flow, as demonstrated by transverse-flow solutions in Bjorken expansion~\cite{PuYangField}. In the longitudinal geometry considered here, the prescribed power law provides a controlled way to compare the effects of different decay rates on the energy balance.

A dynamical electromagnetic closure relates the fields and currents to the evolving plasma~\cite{HernandezKovtun,Nakamura2023}. The ideal induction limit of that closure supplies a reference for the power-law histories.

For a physical field with vanishing comoving electric field, the
homogeneous Maxwell equation instead requires
\begin{equation}
 \nabla_\mu(B^\mu u^\nu-B^\nu u^\mu)=0.
 \label{eq:maxwell}
\end{equation}
In this transverse geometry Eq.~\eqref{eq:maxwell} gives $D\ln B=-\theta$.
Ideal induction therefore removes the direct magnetic source $-B^2(D\ln B+\theta)$ from the matter energy equation.
Magnetic pressure and inertia still enter Eq.~\eqref{eq:relaxM}. For the prescribed power law, Eq.~\eqref{eq:force} specifies the exchange with the matter.
An electromagnetic model of a different decay history would also have to specify the associated electric field and currents.
\subsection{Shear relaxation and stress memory}
\label{sec:earlyRelaxation}
The shear stress can start below its Navier--Stokes (NS) value and build up as the fluid expands. We evolve it with the conformal relaxation law
\begin{equation}
\begin{aligned}
 \tau_\pi D\Phi+\Phi
  &=\frac43\eta_{\rm sh}\theta-\frac43\tau_\pi\theta\Phi,\\
 \tau_\pi&=\frac{c_\pi(\eta_{\rm sh}/s)}{T}.
\end{aligned}
 \label{eq:relaxPhi}
\end{equation}
The relaxation-time convention of Ref.~\cite{Jiang2025} follows from
\begin{equation}
 \tau_\pi=2\eta_{\rm sh}\beta_2,\qquad \beta_2=\frac{3}{4p}
 \quad\Longrightarrow\quad \tau_\pi=\frac{6(\eta_{\rm sh}/s)}{T},
\end{equation}
using $s=4p/T$, which gives $c_\pi=6$. This value corresponds to the original Israel--Stewart choice of moments for a massless Boltzmann gas~\cite{IsraelStewart,Romatschke2010}. The 14-moment truncation of Ref.~\cite{Denicol2012} gives $c_\pi=5$ for an ultrarelativistic classical gas with a constant cross section (Table~2 of that reference). Its higher-moment truncations give different values. Here $c_\pi$ specifies the shear relaxation time in the scalar model.
Transient relativistic hydrodynamics evolves the shear stress as a dynamical variable~\cite{IsraelStewart,BetzMoments}. Conformal second-order hydrodynamics allows additional transport terms~\cite{BaierBRSSS}. Kinetic derivations determine their coefficients from the collision kernel and the closure~\cite{Denicol2012,JaiswalRTA}. Our scalar equation retains shear relaxation and conformal expansion damping in the longitudinal sector.

We initialize the shear stress as $\Phi(\tau_0,\etas)=b_3p(\tau_0,\etas)$. The reference calculation uses $b_3=0$, allowing the viscous stress to build up from zero. At $\etas=0$, $\Phi$ coincides with the Cartesian combination $\pi^{00}-\pi^{33}$ used in Ref.~\cite{Jiang2025}. Equation~\eqref{eq:relaxPhi} defines the scalar, transversely isotropic shear model employed in the nonlinear calculation.

\subsection{Instantaneous-shear limit and projected equations}
Setting $\tau_\pi D\Phi$ and the term proportional to $\tau_\pi\theta\Phi$ to zero gives $\Phi=4\eta_{\rm sh}\theta/3$. We use this instantaneous limit for the analytical comparison.

The corresponding tensor relation is
\begin{equation}
 \pi^{\mu\nu}=2\eta_{\rm sh}\sigma^{\mu\nu}.
\end{equation}
Here $\pi^{\mu\nu}$ is the shear correction to the stress tensor,
with the rest-space projector defined above.
The symmetric, traceless velocity-gradient tensor is
\begin{align}
 \sigma^{\mu\nu}={}&\Delta^{\mu\alpha}\Delta^{\nu\beta}
 \left[\frac{\nabla_\alpha u_\beta+\nabla_\beta u_\alpha}{2}
              -\frac{\Delta_{\alpha\beta}}{3}\theta\right].
\end{align}
For the longitudinal geometry,
\begin{align}
 \pi^{\mu\nu}&=\frac{2\eta_{\rm sh}\theta}{3}
                     (-\Delta^{\mu\nu}-3n^\mu n^\nu),\\
 \pi^{\mu\nu}\sigma_{\mu\nu}&=\frac43\eta_{\rm sh}\theta^2.
\end{align}
Using the equation of state, the two independent projections of
$\nabla_\mu T^{\mu\nu}=0$ can be written in $T$ and $\Omega$:
\begin{equation}
 DT+\frac{T\theta}{3}-\frac{4(\eta_{\rm sh}/s)}{9}\theta^2 +\frac{D(B^2/2)+B^2\theta}{12a_1T^3}=0.
 \label{eq:energy}
\end{equation}
\begin{equation}
\begin{aligned}
 &n^\mu\partial_\mu(a_1T^4+B^2/2) +\left(4a_1T^4+B^2-\frac{16}{3}a_1(\eta_{\rm sh}/s) T^3\theta\right)D\Omega \\
 &\quad-\frac{16}{3}a_1(\eta_{\rm sh}/s) n^\mu\partial_\mu(T^3\theta)=0.
\end{aligned}
 \label{eq:momentum}
\end{equation}
The coefficient of $D\Omega$ in Eq.~\eqref{eq:momentum} is the effective longitudinal inertia, $\varepsilon+p+B^2-\Phi$. The last term contains $n^\mu\partial_\mu(T^3\theta)$, the longitudinal gradient of the NS stress. This gradient vanishes in Bjorken flow.
\subsection{Initial profiles, central symmetry, and domain of application}
The initial profiles are
\begin{equation}
\begin{aligned}
 T(\tau_0,\etas)&=T_0(1-\chi\etas^2/3),\\
 \Omega(\tau_0,\etas)&=(1+\lacc)\etas.
\end{aligned}
\label{eq:initial}
\end{equation}
The two parameters in Eq.~\eqref{eq:initial} are independent initial data.
The thermal curvature
$\chi=-3[\partial_{\etas}^2T]_{\tau_0,0}/(2T_0)$ sets the initial pressure
gradient: for $\chi>0$ the temperature is maximal at $\etas=0$, and the
pressure gradient pushes the fluid away from the center.  Following
Refs.~\cite{She2019,Jiang2020}, $\lacc$ is the initial longitudinal
acceleration parameter; it measures the excess of the initial rapidity
slope $\partial_{\etas}\Omega|_{\tau_0}=1+\lacc$ over the Bjorken value.
Three-dimensional initial-state models assign energy and longitudinal momentum through participant geometry, stopping, and deposition over a range of rapidities and proper times~\cite{ShenSchenkeInitial,ShenSchenkeDynamic,ShenAlzhrani}. A tilted source can generate directed flow~\cite{BozekTilt}, and event-by-event simulations with initial parton velocities show sensitivity to both the density and velocity profiles~\cite{PangInitialFlow}. Equation~\eqref{eq:initial} isolates two features of these profiles near midrapidity: the curvature that sets the initial pressure force and the velocity slope that sets the initial expansion.

Unless stated otherwise, the parameters are
\begin{equation}
\begin{aligned}
 T_0&=0.65\ {\rm GeV},&\tau_0&=0.6\ {\rm fm}/c,\\
 \eta_{\rm sh}/s&=1/(4\pi),&c_\pi&=6,\\
 a&=1.3,&B_0^2/\varepsilon_0&=0.1,
\end{aligned}
\label{eq:referenceParameters}
\end{equation}
with $\Phi_0=b_3p_0=0$. The values $\eta_{\rm sh}/s=0.04,0.16$ provide viscosity comparisons. The reference value is the holographic value $1/(4\pi)$~\cite{Kovtun2005}. We use $\lambda^*=0.03$ and $0.06$ as representative small accelerations and extend the range to $0.1$.
At the reflection center $\etas=0$ we define the central rapidity slope
$\kappa_\Omega\equiv[\partial_{\etas}\Omega]_0$.  Reflection symmetry gives
$\Omega=0$ there at all times, so that $D=\partial_\tau$, $D\Omega=0$, and
$\theta=\kappa_\Omega/\tau$.  The central fluid element is therefore not
accelerated; the slope $\kappa_\Omega$ instead measures its expansion rate
in units of the Bjorken value $1/\tau$.  The evolution of $\kappa_\Omega$
follows from the $\etas$-derivative of the momentum equation at the
center, so momentum balance controls how much expansion work the central
element performs.
At the center the energy projection becomes
\begin{equation}
 \tau\frac{d\varepsilon}{d\tau}
 =-(\varepsilon+p-\Phi)\kappa_\Omega
   +B^2(a-\kappa_\Omega).
 \label{eq:powerExchangeCenter}
\end{equation}
The shear correction $\Phi$ follows Eq.~\eqref{eq:relaxPhi}.  At the
center, the energy projection contains the adiabatic expansion-cooling
term $-\kappa_\Omega T/(3\tau)$ and two nonadiabatic sources, from shear
heating and magnetic work:
\begin{equation}
\begin{aligned}
 S_\eta&=\frac{\Phi\kappa_\Omega}{12a_1\tau T^3},\\
 S_B&=\frac{B^2(a-\kappa_\Omega)}{12a_1\tau T^3},\\
 \frac{dT}{d\tau}&=-\frac{\kappa_\Omega T}{3\tau}+S_\eta+S_B.
\end{aligned}
\label{eq:absoluteSources}
\end{equation}
At the center, the NS temperature equation is
\begin{equation}
 \frac{dT}{d\tau}=-\frac{\kappa_\Omega T}{3\tau}
 +\frac{4(\eta_{\rm sh}/s)\kappa_\Omega^2}{9\tau^2}
 +\frac{B^2(a-\kappa_\Omega)}{12a_1\tau T^3}.
 \label{eq:centerBalance}
\end{equation}
Unless stated otherwise, the finite-gradient calculation evolves the
scalar shear-relaxation system from $\Phi(\tau_0,\etas)=0$, and the NS limit
provides the instantaneous-shear comparison.  The model assumes a conformal
equation of state, purely longitudinal flow, vanishing magnetization and
bulk viscosity, a constant $\eta_{\rm sh}/s$, and a transversely isotropic
shear stress.  In a magnetized plasma the shear viscosity splits into
several anisotropic coefficients~\cite{Denicol2018MHD}; the longitudinal
shear considered here couples to a combination of them, which we represent
by the single coefficient $\eta_{\rm sh}$.  We restrict the analysis to $|\etas|\le1$,
$0.6\le\tau\le6$~fm/$c$, and initial gradients $\lacc,\chi\lesssim0.1$; the
accuracy of the gradient expansion in this window is tested in
Sec.~\ref{sec:accuracy}. For the nuclear radii $R=6.38$~fm (Au) and $6.62$~fm (Pb)~\cite{Loizides2015}, the edge-to-center acoustic travel time is approximately $R/c_s\simeq11$~fm/$c$. We compare two longitudinal evolutions within the same transversely uniform model. Their central temperatures remain of order $0.3$~GeV or higher in this window.

Temperatures are plotted as $T/T_0$ and temperature differences as
$\Delta T/T_0$; dimensional differences are quoted in MeV.  Proper times in
fm/$c$ are converted to natural units with $\hbar c=0.1973$~GeV\,fm, i.e.,
$1~{\rm fm}=5.0677~{\rm GeV}^{-1}$.

\section{Boost-invariant limits: from Bjorken to viscous magnetic flow}
\label{sec:baselines}

We begin with ideal Bjorken flow and incorporate shear and magnetic
stress in turn. Setting $\lambda^{*}=\chi=0$ gives $\Omega=\eta_s$,
$D=\partial_\tau$, and $\theta=1/\tau$. With $T$, $\Phi$, and $B^2$
independent of $\eta_s$, every term in the longitudinal momentum
equation vanishes, and the Bjorken velocity profile remains fixed.
The energy equation is therefore the only nontrivial conservation
equation, supplemented by the shear-relaxation equation when $\Phi$
is evolved dynamically. These boost-invariant limits isolate the
thermal effects of shear and magnetic stress at fixed Bjorken flow
and provide direct comparisons with the known solutions of
Refs.~\cite{Bjorken1983,Muronga2004,Roy2015,Jiang2025}.

\subsection{Ideal Bjorken flow}
With vanishing shear and magnetic stress, energy conservation gives
\begin{equation}
 \frac{dT}{d\tau}+\frac{T}{3\tau}=0,
 \qquad T(\tau)=T_0\left(\frac{\tau_0}{\tau}\right)^{1/3}.
 \label{eq:baselineBjorken}
\end{equation}
For a comoving fluid element of volume $V$, $D\ln V=\theta=1/\tau$ gives
$V\propto\tau$, so Eq.~\eqref{eq:baselineBjorken} is equivalent to
$TV^{1/3}=\text{const}$.  With $s\propto T^3$, this expresses conservation
of the entropy $sV$ of each element: the expansion is adiabatic, and the
temperature falls because the element does work $p\,dV$ on its
surroundings as it stretches along the beam axis~\cite{Bjorken1983}.  This
solution is the reference against which the shear and magnetic sources
below are measured.

\subsection{Viscous Bjorken flow}
Shear changes the longitudinal matter pressure from $p$ to $p-\Phi$.
The uniform equations become
\begin{equation}
\begin{aligned}
 \frac{dT}{d\tau}+\frac{T}{3\tau}
   -\frac{\Phi}{12a_1\tau T^3}&=0,\\
 \tau_\pi\frac{d\Phi}{d\tau}+\Phi
   &=\frac{4\eta_{\rm sh}}{3\tau}
     -\frac{4\tau_\pi\Phi}{3\tau}.
\end{aligned}
\label{eq:baselineViscous}
\end{equation}
A positive $\Phi$ reduces the longitudinal matter pressure to $p-\Phi$.
An expanding fluid element therefore does less work, $(p-\Phi)\,dV$,
and cools more slowly. In the matter energy balance, this reduction in
expansion work appears as the shear-heating term $\Phi\theta$. If
$\Phi(\tau_0)=0$, the stress builds up on the relaxation timescale
$\tau_\pi$. For $\Phi(\tau_0)=b_3p_0$, the shear stress modifies the
work from the outset. For the same initial temperature and expansion rate,
$b_3>0$ slows the initial cooling, while $b_3<0$ speeds it up.
At finite $\tau_\pi$, the shear stress depends on both its initial
value and the subsequent expansion history. Below, we examine how
this memory affects cooling when the longitudinal flow also evolves
dynamically.

The instantaneous NS value $\Phi=4\eta_{\rm sh}/(3\tau)$ gives
\begin{equation}
\begin{aligned}
 \frac{dT}{d\tau}+\frac{T}{3\tau}&=\frac{4(\eta_{\rm sh}/s)}{9\tau^2},\\
 T_{\rm NS}(\tau)&=\left(\frac{\tau_0}{\tau}\right)^{1/3}
 \left[T_0+\frac{2(\eta_{\rm sh}/s)}{3\tau_0}
 \left(1-\left(\frac{\tau_0}{\tau}\right)^{2/3}\right)\right].
\end{aligned}
\label{eq:baselineNS}
\end{equation}
Here $\tau_0$ is expressed in natural units, so that
$(\eta_{\rm sh}/s)/\tau_0$ has units of temperature. The second term
in brackets gives the viscous correction after the ideal Bjorken
cooling factor has been factored out. This term grows from zero
and approaches $2(\eta_{\rm sh}/s)/(3\tau_0)$ at late times.
For the reference parameters, the fractional temperature excess
relative to ideal Bjorken flow approaches
$2(\eta_{\rm sh}/s)/(3T_0\tau_0)\simeq0.027$ and is about $2\%$
at $\tau=6$~fm/$c$. Equation~\eqref{eq:baselineNS} gives the
Navier--Stokes solution for Bjorken flow obtained by
Muronga~\cite{Muronga2004}. The weakly accelerating viscous solution
of Ref.~\cite{Jiang2020} recovers this result in the boost-invariant
limit. The ratio $(\eta_{\rm sh}/s)/(T\tau)$ measures the viscous gradient scale in Bjorken flow. The instantaneous Navier--Stokes approximation also requires relaxation to be fast on the expansion timescale. For the
reference parameters, this ratio reaches its maximum value of
$0.040$ at $\tau_0$. In Eq.~\eqref{eq:baselineViscous}, the
relaxation time $\tau_\pi$ sets the timescale on which the
longitudinal shear stress responds to the expansion.

The reduction of longitudinal pressure is also present in boost-invariant calculations with transverse expansion~\cite{SongHeinz2008}. The Bjorken limit isolates its effect on longitudinal cooling.

Bjorken symmetry is also used to compare transient and anisotropic hydrodynamics with exact kinetic evolution~\cite{FlorkowskiExact,FlorkowskiAniso}. Studies of the Bjorken hydrodynamic attractor organize the constitutive expansion by the ratio of microscopic to expansion scales~\cite{HellerSpalinski}. The orders in the initial gradients used below instead count powers of the departure from the boost-invariant initial profiles at fixed shear closure. With transverse expansion included, Gubser symmetry supplies a second class of hydrodynamic and kinetic benchmarks~\cite{GubserSymmetry,DenicolGubser}.

\subsection{Viscous magnetic Bjorken flow and shear relaxation}
Adding the prescribed transverse field to the same relaxation law gives
\begin{equation}
 \frac{dT}{d\tau}+\frac{T}{3\tau}
 -\frac{\Phi+(a-1)B^2}{12a_1\tau T^3}=0.
 \label{eq:baselineMagnetic}
\end{equation}
Setting $\Phi=0$ first identifies the magnetic work in the limit studied by Roy \emph{et al.}~\cite{Roy2015}. With $x=\tau/\tau_0$, the energy equation and its solution are
\begin{equation}
\begin{aligned}
 \frac{d\varepsilon}{d\tau}+\frac{4\varepsilon}{3\tau}
   &=\frac{(a-1)B_0^2}{\tau}x^{-2a},\\
 \widetilde T^4
   &=x^{-4/3}+\frac{(a-1)\sigma}{3a_1}
      \frac{x^{-2a}-x^{-4/3}}{4/3-2a}.
\end{aligned}
\label{eq:idealMagneticBaseline}
\end{equation}
At $a=2/3$ the fraction is replaced by its limit, $x^{-4/3}\ln x$.
Equation~\eqref{eq:idealMagneticBaseline} is the power-law magnetic
Bjorken solution of Ref.~\cite{Roy2015}, also used in Ref.~\cite{Jiang2025}.
The magnetic source is proportional to $a-1$: for $a>1$ the field decays
faster than flux conservation and heats the matter, for $a<1$ it enhances
the cooling, and at $a=1$ the ideal-induction result of Ref.~\cite{Roy2015}
is recovered.  For the reference value $(a-1)B_0^2/\varepsilon_0=0.03$, the magnetic field slows the initial cooling. In the ideal fluid, the initial temperature increases when this ratio exceeds $4/3$.

With NS stress, Eq.~\eqref{eq:baselineMagnetic} reduces to the
magnetic--viscous cooling equation of Ref.~\cite{Jiang2025}; with
Eq.~\eqref{eq:relaxPhi} it gives the relaxation curve
(Sec.~\ref{sec:earlyRelaxation}).  These comparisons fix the field
normalization and relaxation convention before longitudinal gradients are
introduced.

\begin{figure}[tbp]
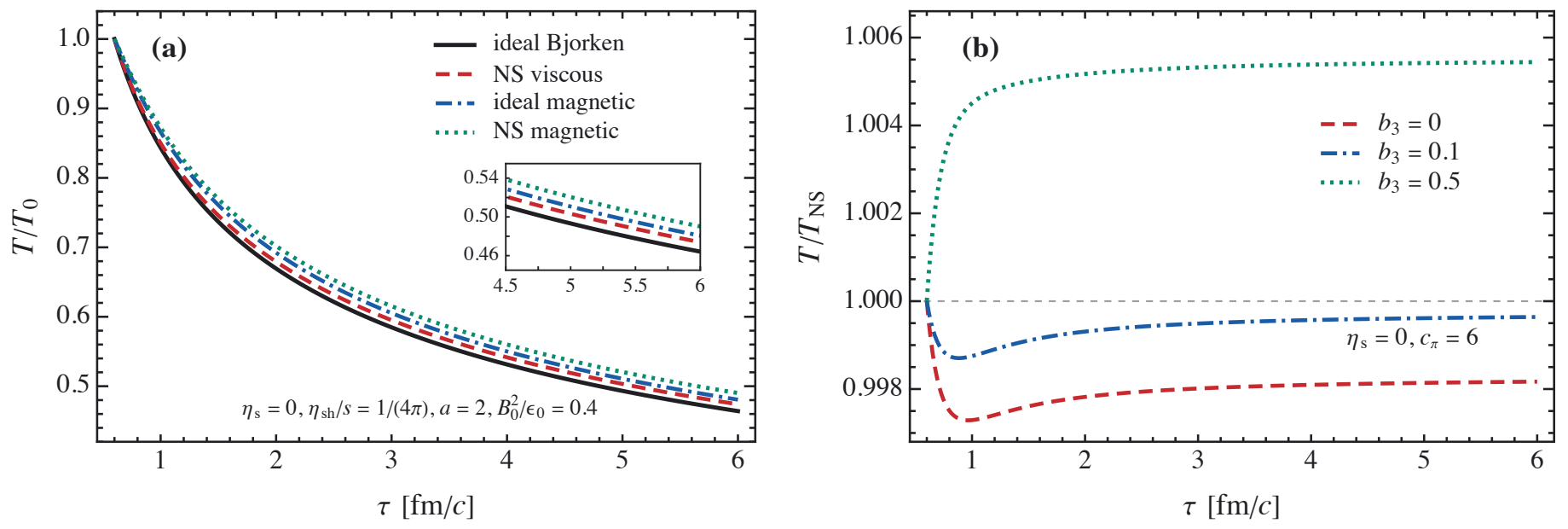

\centering\ReportFigureOne
\caption{Boost-invariant temperature limits. Left: normalized temperatures for ideal Bjorken, NS viscous, ideal magnetic, and NS magnetic flow. Right: relaxation temperatures divided by the NS magnetic temperature for $b_3=0,0.1,0.5$. We use $a=2$, $B_0^2/\varepsilon_0=0.4$, and $\eta_{\rm sh}/s=1/(4\pi)$.}
\label{fig:baselineLadder}
\end{figure}

Both calculations use $c_\pi=6$. In Bjorken flow, Ref.~\cite{Jiang2025} retains the relaxation term $-\tfrac12\tau_\pi\Phi(1/\tau-5\,d\ln T/d\tau)$. It reduces to the term in Eq.~\eqref{eq:relaxPhi} for ideal Bjorken cooling, $d\ln T/d\tau=-1/(3\tau)$. Shear heating and magnetic work can change this cooling rate.
For $\lacc=\chi=0$, the first-order NS--MHD solution of
Ref.~\cite{Jiang2025} provides a perturbative check.  Its zero-field
background $T_{\rm NS}^{(0)}$ is Eq.~\eqref{eq:baselineNS}, and expanding
the magnetic source to first order in $B_0^2$ gives the quadrature form of
their Eq.~(40),
\begin{align}
T_{\rm NS}(\tau)
&=T_{\rm NS}^{(0)}(\tau)+\delta T_B(\tau)+O(\sigma^2),\\
\delta T_B(\tau)
&=\tau^{-1/3}\int_{\tau_0}^{\tau}
\frac{(a-1)B_0^2\,t^{1/3}}{12a_1t[T_{\rm NS}^{(0)}(t)]^3}
\left(\frac{\tau_0}{t}\right)^{2a}dt .
\label{eq:Jiang2025NS}
\end{align}
This expression follows by integrating the linearized NS temperature
equation with integrating factor $\tau^{1/3}$. It tends to the viscous
Bjorken solution when $B_0\to0$, and to ideal Bjorken cooling when both
$B_0$ and $\eta_{\rm sh}/s$ vanish. At the reference field,
$\sigma=1.56$ and $a=1.3$ give $\epsilon_1=0.0075$, within the small-field
regime of the perturbative result. Figure~\ref{fig:baselineLadder} shows a stronger-field comparison at $a=2$ and $B_0^2/\varepsilon_0=0.4$; its $\epsilon_1$ is $0.10$. The NS and relaxation curves share the same initial temperature and magnetic stress. Differences at early time
arise because the relaxation stress requires a finite time to approach
its instantaneous NS value.

\begin{table}[tbp]
\caption{Comparison of the first-order magnetic perturbative solution with shear relaxation and the unexpanded NS equation. The temperatures are normalized by $T_0$ and multiplied by $10^3$. The relative differences are $(T_{\rm pert}-T_{\rm relax})/T_{\rm relax}$ and $(T_{\rm pert}-T_{\rm NS})/T_{\rm NS}$; the last column is given in parts per million. We use $B_0^2/\varepsilon_0=0.10$ for $a=1.30$ and $0.40$ for $a=2.00$.}
\label{tab:jiangCPCcomparison}
\centering\small
\begin{tabular}{cccccc}
\toprule
$a$ & $\tau$ [fm/$c$] & $10^3T_{\rm pert}/T_0$ & $10^3T_{\rm relax}/T_0$ & $\Delta_{\rm relax}$ [\%] & $\Delta_{\rm NS}$ [ppm] \\
\midrule
1.30 & 1.00 & 852.32 & 849.89 & 0.29 & 11.4 \\
1.30 & 2.00 & 682.39 & 680.80 & 0.23 & 29.8 \\
1.30 & 4.00 & 544.36 & 543.24 & 0.21 & 39.6 \\
1.30 & 6.00 & 476.47 & 475.53 & 0.20 & 42.6 \\
2.00 & 1.00 & 873.27 & 870.00 & 0.38 & 1036.3 \\
2.00 & 2.00 & 703.09 & 700.40 & 0.38 & 1653.5 \\
2.00 & 4.00 & 561.06 & 559.01 & 0.37 & 1753.8 \\
2.00 & 6.00 & 491.03 & 489.27 & 0.36 & 1764.1 \\
\bottomrule
\end{tabular}
\end{table}

For the parameters in Table~\ref{tab:jiangCPCcomparison}, the perturbative solution differs from the full NS evolution by at most $0.0043\%$ in the reference field. In the stronger field, the difference is at most $0.18\%$. The comparison with relaxation additionally reflects the evolution of shear stress. The same initial temperature and magnetic history are used in each comparison.

Figure~\ref{fig:baselineLadder} compares these temperature histories. Positive $b_3$ raises the early ratio above unity: initial shear reduces the longitudinal pressure before the NS value is reached.

\section{Longitudinal acceleration with a prescribed flow}
\label{sec:prescribedFlow}
Throughout Sec.~\ref{sec:baselines}, boost invariance fixes the expansion rate at $1/\tau$ and the longitudinal momentum equation is satisfied identically.
We now take the accelerating profile $\Omega_p=(1+\lacc)\etas$ of Refs.~\cite{She2019,Jiang2020}. We first calculate its cooling history and then test whether the resulting pressure gradient can sustain the prescribed motion.

\subsection{Ideal accelerating flow}
For the prescribed flow in the ideal, zero-field limit, the expansion
rate at midrapidity is $\theta=(1+\lambda^{*})/\tau$, and the
temperature equation gives $\widetilde T_p=x^{-(1+\lambda^{*})/3}$.
This solution agrees with Eq.~(12) of Ref.~\cite{She2019} for
$\sigma_0=0$ and $c_s^2=1/3$ and recovers the Bjorken cooling law as
$\lambda^{*}\to0$. A larger $\lambda^{*}$ increases the longitudinal
expansion rate and thus accelerates cooling. For $\lambda^{*}=0.03$,
the temperature at midrapidity is about $2\%$ below the Bjorken value
at $\tau=6$~fm/$c$.

In the conformal ideal limit, expansion of the Euler equation through first order in $\lacc$ gives
\begin{equation}
\begin{aligned}
 \partial_{\etas}\ln T_p&=-\frac{2\lacc}{3}\etas,\\
 \frac{T_p(\tau,\etas)}{T_p(\tau,0)}
    &=\exp(-\lacc\etas^2/3)+O((\lacc)^2).
\end{aligned}
\label{eq:prescribedProfile}
\end{equation}
This is Eq.~(23) of Ref.~\cite{Jiang2020} for $\Pi_d=0$ and $\kappa=3$, to
first order in $\lacc$, and it agrees with Eq.~\eqref{eq:initial} when
$\chi=\lacc$.  At this order the ideal pressure gradient exactly sustains a
constant rapidity slope, so the prescribed flow solves both conservation
equations. At initialization the polynomial profile in Eq.~\eqref{eq:initial} and this Gaussian first differ at order $(\lacc)^2\etas^4$ along $\chi=\lacc$. Higher-order terms, shear, and magnetic stress change the
momentum balance, and the prescribed slope then no longer satisfies the
momentum equation.  We evaluate the momentum residual for independently specified $\chi$ and $\lacc$ in Sec.~\ref{sec:prescribedResidual}.

Maintaining the prescribed velocity in the presence of a momentum residual requires a longitudinal constraint force. Its projection along $u^\mu$ vanishes, so the comoving energy equation still determines $T_p$.

\subsection{Viscous accelerating flow}

At zero magnetic field, substituting the prescribed expansion rate into the NS energy equation gives
\begin{align*}
 \Phi_p&=\frac{4\eta_{\rm sh}}{3}\theta_p,\qquad
 \theta_p=\frac{1+\lacc}{\tau}\\
 &\Rightarrow\quad \frac{dT_p}{d\tau}
 +\frac{(1+\lacc)T_p}{3\tau}
 =\frac{4(\eta_{\rm sh}/s)(1+\lacc)^2}{9\tau^2}.
\end{align*}
For $m\ne1$, multiplication by $\tau^m$ and integration from $\tau_0$ give
\begin{align*}
 T_p={}&\left(\frac{\tau_0}{\tau}\right)^m
 \left[T_0+\frac{\mathcal C}{(1-m)\tau_0}
 \left(1-\left(\frac{\tau_0}{\tau}\right)^{1-m}\right)\right],\\
 m={}&\frac{1+\lacc}{3},\qquad
 \mathcal C=\frac{4(\eta_{\rm sh}/s)(1+\lacc)^2}{9}.
\end{align*}
The solution assumes a constant $\eta_{\rm sh}/s$. For weak acceleration, $m=(1+\lacc)/3$ lies near $1/3$.
The solution reduces to the viscous Bjorken result,
Eq.~\eqref{eq:baselineNS}, as $\lambda^{*}\to0$ and to the ideal
accelerating solution above as $\eta_{\rm sh}/s\to0$. For $\kappa=3$
and $\Pi_d=4(\eta_{\rm sh}/s)/3$, the viscous source agrees with that
in Eq.~(21) of Ref.~\cite{Jiang2020} through first order in
$\lambda^{*}$. Integrating that equation gives the exponent
$1-m=(\kappa-1-\lambda^{*})/\kappa$. In the corresponding
coefficient, Eq.~(22) of the same reference contains $\kappa-1$ in
place of $\kappa-1-\lambda^{*}$. The two temperature expressions
therefore differ at order $\lambda^{*}\Pi_d$.

\subsection{Viscous magnetic accelerating flow}
Restoring the transverse field gives the NS energy equation
\begin{equation}
\begin{aligned}
 DT+\frac{T\theta}{3}-\frac{4(\eta_{\rm sh}/s)}{9}\theta^2
 +\frac{B^2}{12a_1T^3}
       \left[\theta-\frac{a\cosh(\Omega-\etas)}{\tau}\right]=0.
\end{aligned}
 \label{eq:thermalbalance}
\end{equation}
At $\etas=0$, reflection symmetry gives $D=\partial_\tau$, and the
prescribed slope fixes $\theta_p=(1+\lacc)/\tau$ at all times.
Equations~\eqref{eq:relaxE} and \eqref{eq:relaxPhi} then reduce to
\begin{equation}
\begin{aligned}
 \frac{dT_p}{d\tau}+\frac{(1+\lacc)T_p}{3\tau}
 &=\frac{(1+\lacc)\Phi_p+(a-1-\lacc)B^2}
             {12a_1\tau T_p^3},\\
 \tau_{\pi,p}\frac{d\Phi_p}{d\tau}+\Phi_p
 &=\frac{4\eta_{{\rm sh},p}(1+\lacc)}{3\tau}
   -\frac{4\tau_{\pi,p}(1+\lacc)\Phi_p}{3\tau}.
\end{aligned}
\label{eq:prescribedCentral}
\end{equation}
For the prescribed flow with $\Phi_p=0$, the energy equation at
midrapidity takes the form
\begin{equation*}
 \tau\frac{d\widetilde{\varepsilon}}{d\tau}
 +\frac{4}{3}(1+\lambda^{*})\widetilde{\varepsilon}
 =\frac{B_0^2}{\varepsilon_0}\left(\frac{\tau_0}{\tau}\right)^{2a}
   \bigl[a-(1+\lambda^{*})\bigr],
 \qquad
 \widetilde{\varepsilon}\equiv\frac{\varepsilon}{\varepsilon_0}.
\end{equation*}
For $c_s^2=1/3$ this equation agrees with Eq.~(12) of
Ref.~\cite{She2019} at $\eta_s=0$, whose magnetization parameter
$\sigma_0$ equals our $B_0^2/\varepsilon_0=\sigma/(3a_1)$. For this
flow, magnetic flux conservation requires $a=1+\lambda^{*}$. The
reference field with $a=1.3$ decays faster than the flux-conserving
field and therefore does positive work on the matter. In the limit
$\lambda^{*}\to0$, the equation yields the magnetic Bjorken solution
in Eq.~\eqref{eq:idealMagneticBaseline}. In the zero-field limit
$B_0\to0$, it gives the ideal accelerating solution
$\widetilde{T}_p=x^{-(1+\lambda^{*})/3}$. Including Navier--Stokes
shear stress lowers the longitudinal matter pressure to $p-\Phi_p$
for $\Phi_p>0$, further slowing the cooling.

\begin{figure}[tbp]
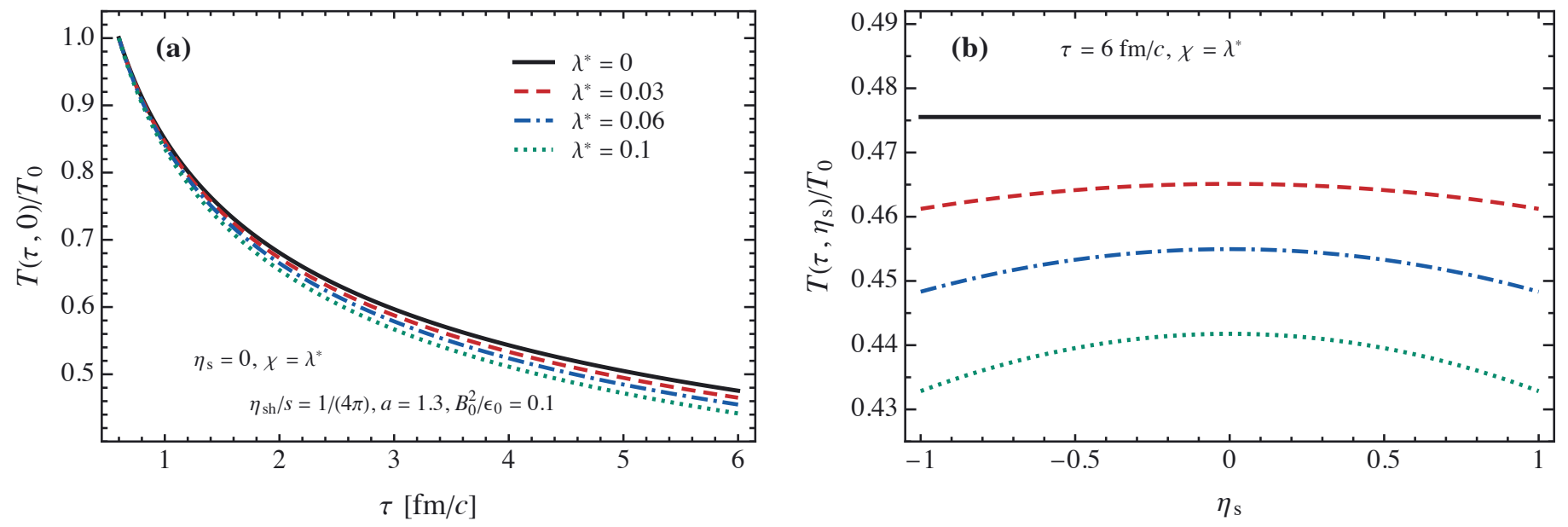

\centering\ReportFigureTwo
\caption{Prescribed accelerating flow at $\lambda^*=0,0.03,0.06,0.1$, with $\chi=\lambda^*$ in each initial state. Left: central normalized temperature. Right: temperature profiles at $\tau=6$ fm/$c$, with the same line styles as the left panel. Other parameters have their reference values.}
\label{fig:prescribed}
\end{figure}

In Fig.~\ref{fig:prescribed}, a larger prescribed slope increases the central expansion rate and lowers the central temperature. The profile at $6$~fm/$c$ also narrows along this scan, in which the initial thermal curvature increases with $\lacc$. At $a=1+\lacc$ the prescribed field obeys ideal induction for this velocity profile at every rapidity.

\subsection{Momentum residual of the prescribed flow}
\label{sec:prescribedResidual}
We obtain $T_p$ from the energy equation with $\Omega_p$ held fixed and then evaluate the longitudinal momentum equation along this solution. Near midrapidity, longitudinal momentum
conservation also requires the evolving temperature curvature to be
consistent with the prescribed rapidity slope. We define the curvature
as $\chi_p(\tau)\equiv-\tfrac{3}{2}[\partial_{\eta_s}^2\ln T_p]_0$,
with $\chi_p(\tau_0)=\chi$, where the subscript $0$ denotes evaluation
at $\eta_s=0$. To test this consistency condition, we evaluate the
left-hand side $\mathcal M_p$ of Eq.~\eqref{eq:relaxM} along the
prescribed-flow solution. By reflection symmetry, $\mathcal M_p$ is
odd in $\eta_s$ and vanishes at midrapidity. Its leading coefficient
is therefore $\mathcal R\equiv[\partial_{\eta_s}\mathcal M_p]_0$.

For the prescribed rapidity $\Omega_p=(1+\lambda^{*})\eta_s$, the
longitudinal projection of momentum conservation involves the total
longitudinal pressure $Q=a_1T_p^4+B^2/2-\Phi_p$ and the effective
inertia $\mathcal I=4a_1T_p^4+B^2-\Phi_p$. The residual takes the form
\begin{align}
 \tau\mathcal M_p
 &=\sinh(\lambda^{*}\eta_s)\,\tau\partial_\tau Q
   +\cosh(\lambda^{*}\eta_s)\,\partial_{\eta_s}Q \nonumber\\
 &\quad +(1+\lambda^{*})\mathcal I\sinh(\lambda^{*}\eta_s).
 \label{eq:prescribedMomentumResidual}
\end{align}
Differentiating with respect to $\eta_s$ and evaluating at midrapidity
gives
\begin{align}
 \tau\mathcal R
 &=\lambda^{*}[\tau\partial_\tau Q]_0
   +[\partial_{\eta_s}^2 Q]_0
   +\lambda^{*}(1+\lambda^{*})\mathcal I(\tau,0).
 \label{eq:residualExactStart}
\end{align}
The energy equation at midrapidity determines the time derivative of
the matter pressure,
\begin{equation}
 \tau\partial_\tau(a_1T_p^4)
 =-\frac{4}{3}(1+\lambda^{*})a_1T_p^4
  +\frac{(1+\lambda^{*})\Phi_p+(a-1-\lambda^{*})B^2}{3}.
 \label{eq:residualEnergyZero}
\end{equation}
Reflection symmetry gives $[\partial_{\eta_s}T_p]_0=0$. The definition
of $\chi_p$ then fixes the rapidity curvature of the matter pressure.
The prescribed field is uniform in $\eta_s$, and its time derivative
follows from its power-law decay. Thus,
\begin{align}
 [\partial_{\eta_s}^2(a_1T_p^4)]_0
 &=-\frac{8}{3}\chi_p a_1T_p^4,
 &\tau\partial_\tau(B^2/2)&=-aB^2.
 \label{eq:residualCurvature}
\end{align}
Here and below, quantities in the midrapidity equations are evaluated
at $\eta_s=0$. Substituting these relations into
Eq.~\eqref{eq:residualExactStart} and defining the shear-stress
curvature $C_\Phi\equiv[\partial_{\eta_s}^2\Phi_p]_0$, we obtain
\begin{align}
 \tau\mathcal R
 &=\frac{8}{3}a_1T_p^4
   \bigl[\lambda^{*}(1+\lambda^{*})-\chi_p\bigr] \nonumber\\
 &\quad+\lambda^{*}\biggl[
   \frac{2}{3}(1-a+\lambda^{*})B^2
   -\frac{2}{3}(1+\lambda^{*})\Phi_p
   -\tau\partial_\tau\Phi_p\biggr]-C_\Phi .
 \label{eq:prescribedResidual}
\end{align}
This identity is exact for the prescribed rapidity and magnetic
history. Requiring the term linear in $\eta_s$ in $\mathcal M_p$ to
vanish, $\mathcal R=0$, fixes the required curvature,
\begin{align}
 \chi_{\rm req}(\tau)
 &=\lambda^{*}(1+\lambda^{*})
   +\frac{3}{8a_1T_p^4}\Biggl\{
   \lambda^{*}\biggl[\frac{2}{3}(1-a+\lambda^{*})B^2 \nonumber\\
 &\qquad-\frac{2}{3}(1+\lambda^{*})\Phi_p
   -\tau\partial_\tau\Phi_p\biggr]-C_\Phi\Biggr\}.
 \label{eq:requiredCurvature}
\end{align}

In the ideal, zero-field limit, the required curvature is exactly
$\chi_{\rm req}=\lambda^{*}(1+\lambda^{*})$, so momentum balance
requires $\chi_p=\lambda^{*}$ to first order. The energy equation
governs the actual curvature along the prescribed flow,
\begin{equation}
 \tau\frac{d\chi_p}{d\tau}=-2\lambda^{*}\chi_p,
 \qquad
 \chi_p(\tau)=\chi\left(\frac{\tau_0}{\tau}\right)^{2\lambda^{*}}.
\end{equation}
Even if $\chi=\chi_{\rm req}$ initially, the evolving curvature
develops a mismatch of order $(\lambda^{*})^2\ln(\tau/\tau_0)$,
which generates a momentum residual through
Eq.~\eqref{eq:prescribedResidual}. Shear and magnetic stresses
introduce additional terms in the evolution of $\chi_p$ and modify
this power law.

\begin{figure}[tbp]
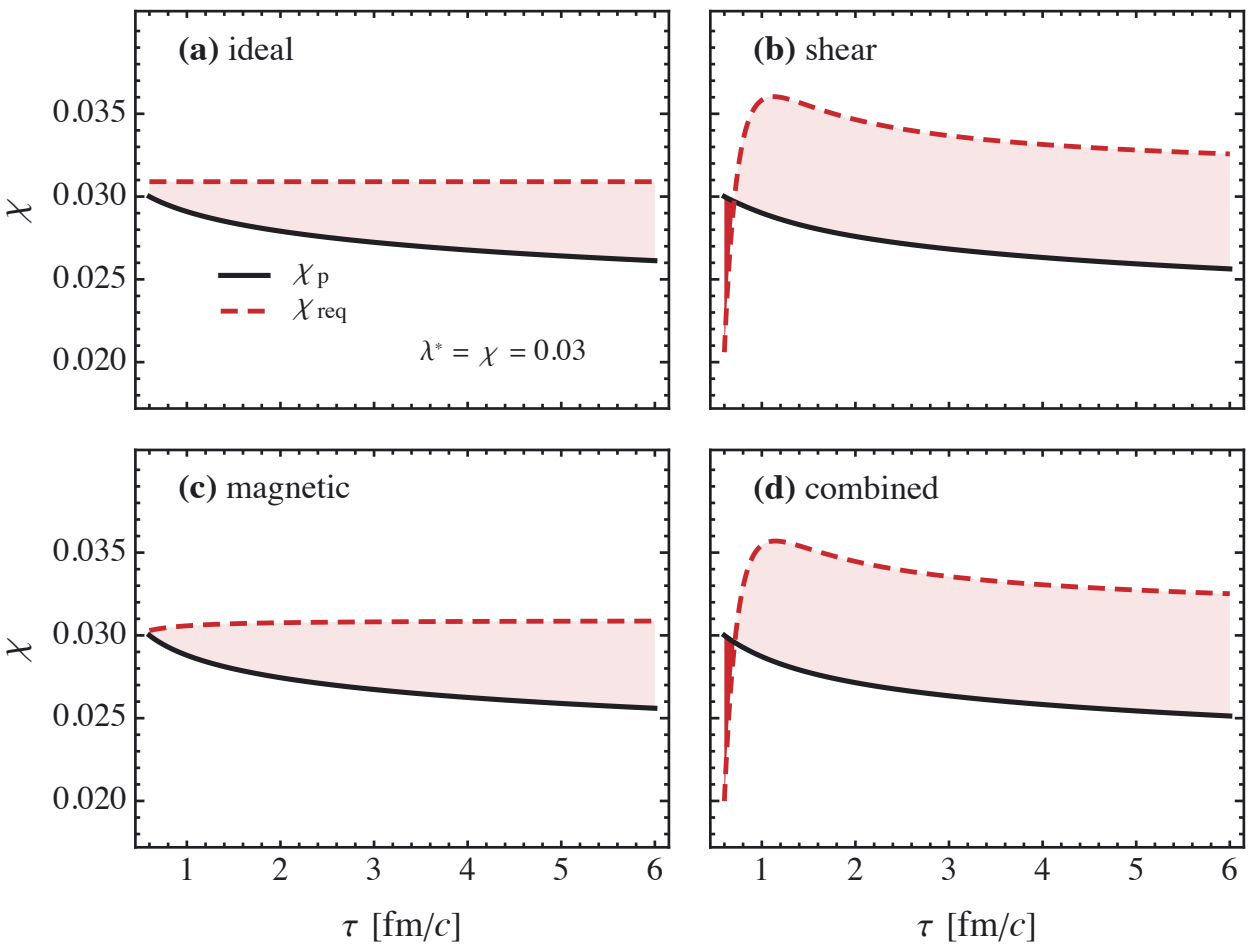

\centering\ReportFigureTen
\caption{Central temperature curvature for prescribed flow at $\lambda^*=\chi_p(\tau_0)=0.03$ with (a) ideal, (b) shear-only, (c) magnetic-only, and (d) combined stresses. Solid lines show the evolved curvature $\chi_p$ and dashed lines show $\chi_{\rm req}$ of Eq.~\eqref{eq:requiredCurvature}; their vertical gap equals $3\tau\mathcal R/(8a_1T_p^4)$.}
\label{fig:prescribedResidual}
\end{figure}

To obtain the evolution equation for $C_\Phi$, we differentiate
Eq.~\eqref{eq:relaxPhi} twice with respect to $\eta_s$ and evaluate
at midrapidity. We use $\tau_\pi=c_\pi(\eta_{\rm sh}/s)/T_p$,
$\theta_p=(1+\lambda^{*})\cosh(\lambda^{*}\eta_s)/\tau$, and
$[\partial_{\eta_s}^2\ln T_p]_0=-2\chi_p/3$. Retaining terms up to
first order in each of $\lambda^{*}$ and $\chi_p$, including the
mixed term proportional to $\lambda^{*}\chi_p$, gives
\begin{align}
 \tau_\pi\partial_\tau C_\Phi+C_\Phi
 &+\frac{2\lambda^{*}\tau_\pi}{\tau}C_\Phi
  +\frac{2\chi_p\tau_\pi}{3}\partial_\tau\Phi_p \nonumber\\
 &=-\frac{8\chi_p(1+\lambda^{*})}{3\tau}\eta_{\rm sh}
   -\frac{4(1+\lambda^{*})\tau_\pi}{3\tau}
    \left(C_\Phi+\frac{2\chi_p}{3}\Phi_p\right).
 \label{eq:shearCurvatureEvolution}
\end{align}
Figure~\ref{fig:prescribedResidual} compares $\chi_p$ with the required curvature of Eq.~\eqref{eq:requiredCurvature} for four choices of stress. Since $\tau\mathcal R=\frac83a_1T_p^4(\chi_{\rm req}-\chi_p)$ by Eq.~\eqref{eq:requiredCurvature}, the gap between the two curves measures the momentum residual. After the first fm/$c$, $\chi_p$ stays below $\chi_{\rm req}$ in all four cases, so the prescribed flow carries a persistent momentum residual. Finite shear and magnetic stress shift the curvature required for momentum balance because their gradients contribute alongside the thermal pressure gradient.

\section{Self-consistent accelerating flow}
\label{sec:selfConsistent}
The pressure profile found in Sec.~\ref{sec:prescribedResidual} generally fails to sustain a constant rapidity slope. We now let momentum conservation evolve the velocity from the same initial profile. The ideal limit shows how the pressure gradient changes the central expansion rate before shear or magnetic work affects the cooling.

\subsection{The pressure-gradient mechanism in the ideal limit}
\label{sec:idealReference}

We first consider an ideal conformal fluid in the absence of a magnetic
field. The energy and longitudinal momentum equations then reduce to
\begin{equation}
\begin{aligned}
 D\ln T+\frac{\theta}{3}&=0,\\
 D\Omega+n^\mu\partial_\mu\ln T&=0.
\end{aligned}
\label{eq:idealStarting}
\end{equation}
The longitudinal pressure gradient drives the evolution of the flow
rapidity through $D\Omega=-n^\mu\partial_\mu\ln T$. The resulting
flow determines the expansion rate $\theta$ and hence the comoving
cooling rate $D\ln T=-\theta/3$. We therefore solve for $T$ and
$\Omega$ together by expanding both fields about the Bjorken solution
while retaining the full proper-time dependence of the background.
The initial conditions in Eq.~\eqref{eq:initial} specify two
independent perturbations: the excess $\lambda^{*}$ of the initial
flow-rapidity slope above its Bjorken value and the initial
temperature curvature $\chi$. At linear order, the responses to
these perturbations add, and we write
\begin{equation}
\begin{aligned}
 \ln\frac{T(\tau,\etas)}{T(\tau,0)}&=-\frac{\chi\etas^2}{3},\\
 \Omega(\tau,\etas)&=\kappa_\Omega(\tau)\etas.
\end{aligned}
\label{eq:idealReferenceAnsatz}
\end{equation}
At this order the curvature $\chi$ stays constant, and the energy equation
uses $\theta=\kappa_\Omega/\tau$.  The two terms of the momentum equation
are
\[
D\Omega=\frac{\etas}{\tau}\left[\kappa_\Omega-1+\tau\frac{d\kappa_\Omega}{d\tau}\right],
\qquad
n^\mu\partial_\mu\ln T=-\frac{\etas}{3\tau}\,(\kappa_\Omega-1+2\chi),
\]
where the first term on the right of the second relation comes from the
cooling seen by a fluid element moving relative to the Milne frame, and the
second from the pressure gradient.  Equation~\eqref{eq:idealStarting}
therefore gives
\begin{equation}
\begin{aligned}
 \tau\frac{d}{d\tau}\ln\frac{T(\tau,0)}{T_0(\tau_0/\tau)^{1/3}}
 &=-\frac{\kappa_\Omega-1}{3},\\
 \tau\frac{d\kappa_\Omega}{d\tau}+\frac23(\kappa_\Omega-1)
 &=\frac23\chi.
\end{aligned}
\label{eq:idealReferenceODE}
\end{equation}
Multiplying the second equation by $(\tau/\tau_0)^{2/3}$ and using
$\kappa_\Omega(\tau_0)=1+\lacc$ yields
\begin{equation}
 \kappa_\Omega(\tau)=1+\chi+(\lacc-\chi)
                \left(\frac{\tau_0}{\tau}\right)^{2/3}.
\label{eq:idealReferenceSlope}
\end{equation}
At this order the slope relaxes from its initial value $1+\lacc$ to $1+\chi$, the value
supported by the pressure gradient, and the memory of the initial velocity
excess decays as $(\tau_0/\tau)^{2/3}$.  The slope therefore decreases for
$\lacc>\chi$ and increases for $\lacc<\chi$.  Integrating the first
equation from $T(\tau_0,0)=T_0$ gives
\begin{equation}
 \ln\frac{T(\tau,0)}{T_0(\tau_0/\tau)^{1/3}}
 =-\frac{\chi}{3}\ln\frac{\tau}{\tau_0}
 -\frac{\lacc-\chi}{2}\left[1-\left(\frac{\tau_0}{\tau}\right)^{2/3}\right].
\label{eq:idealReferenceTemperature}
\end{equation}
At late times the central cooling is controlled by $\chi$ rather than by
$\lacc$, whereas the prescribed flow cools as $(\tau_0/\tau)^{\lacc/3}$
relative to Bjorken flow.

\begin{figure}[tbp]
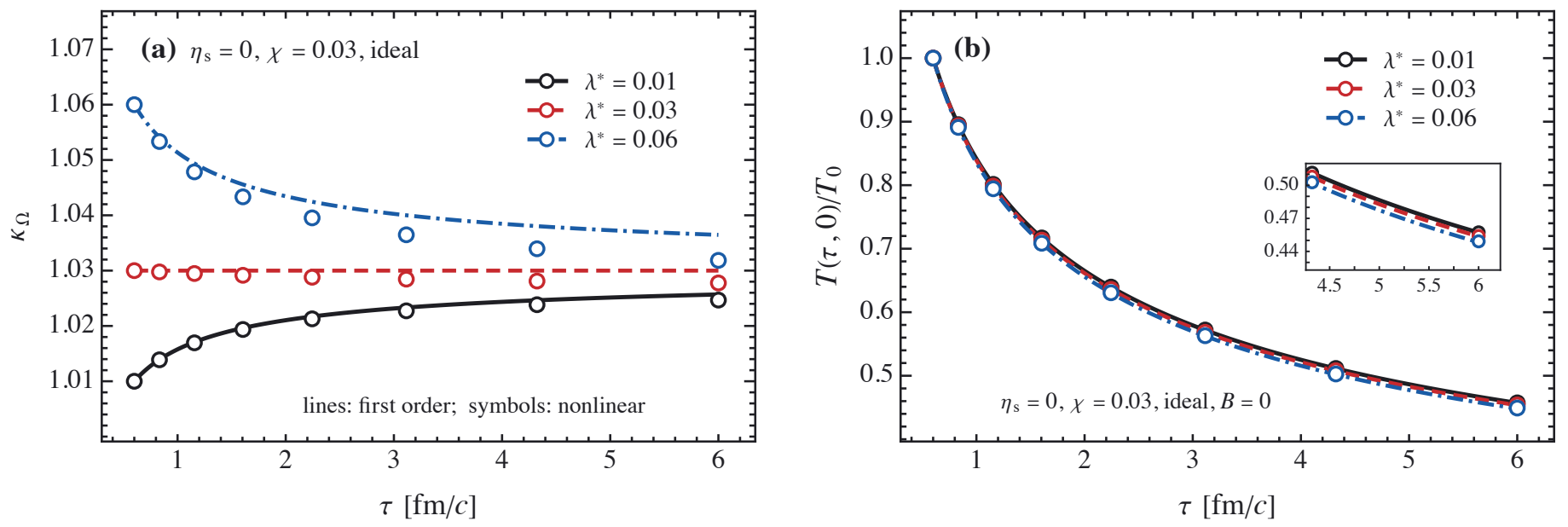

\centering\ReportFigureThree
\caption{Ideal self-consistent flow at fixed initial curvature $\chi=0.03$. Left: the central rapidity slope for $\lambda^*=0.01,0.03,0.06$. Right: central normalized temperatures. Lines are the first-order analytical results, Eqs.~\eqref{eq:idealReferenceSlope} and \eqref{eq:idealReferenceTemperature}; open symbols are the nonlinear ideal evolution at selected times.}
\label{fig:ideal}
\end{figure}

Figure~\ref{fig:ideal} shows that the slope grows for $\lambda^*<\chi$ and decreases for $\lambda^*>\chi$. At $\lambda^*=\chi$ it is constant through first order; the nonlinear solution retains the finite-amplitude correction. The analytic and numerical temperatures differ by terms beyond the retained order, even where their curves nearly coincide.

\subsection{Pressure gradients and the evolving velocity field}
For $\chi=\lacc$, the ideal rapidity slope remains constant to first order. We use this initial profile to examine how shear and magnetic stress change the slope. We first take instantaneous NS shear and later evolve $\Phi$ with its finite relaxation time.

At each order in the initial-gradient expansion, the Navier--Stokes
coefficient equations retain the full dependence on $\eta_{\rm sh}/s$
and $B_0^2$. Their integral form displays the pressure and inertia
terms explicitly and provides a reference for the shear-relaxation
calculation.

In this one-parameter expansion, the superscript $(0)$ denotes the
boost-invariant background at $\lambda^{*}=\chi=0$, with
$\eta_{\rm sh}/s$, $B_0^2$, and $a$ fixed at their chosen values.
The superscript $(1)$ denotes the derivative with respect to
$\lambda^{*}$ along $\chi=\lambda^{*}$, evaluated at $\lambda^{*}=0$.
Appendix~\ref{app:relaxationPrediction} treats $\chi$ and $\lambda^{*}$
independently and therefore uses a separate first-order coefficient
for each initial parameter. With the one-parameter convention used
here, we write
\begin{align}
 T&=\bgtemp+\lacc\tempresponse+
                         \mathcal O((\lacc)^2),\\
 \Omega&=\etas+\lacc\flowresponse+
                         \mathcal O((\lacc)^2),\\
 \eta_{\rm sh}^{(0)}&=4a_1(\eta_{\rm sh}/s)(\bgtemp)^3.
 \label{eq:responseNotation}
\end{align}
Thus $T^{(0)}$ is the temperature of the boost-invariant background, which
already contains shear viscosity and magnetic stress.  The coefficients
$T^{(1)}$ and $\Omega^{(1)}$ are the first-order temperature and rapidity
responses per unit $\lacc$, so the physical corrections are
$\lacc T^{(1)}$ and $\lacc\Omega^{(1)}$.

\paragraph*{Coupled temperature and momentum evolution.}

The zero-acceleration background obeys
\begin{equation}
 \tau\partial_\tau\bgtemp+\frac{\bgtemp}{3}
 -\frac{4(\eta_{\rm sh}/s)}{9\tau}
 +\frac{(1-a)B^2}{12a_1(\bgtemp)^3}=0.
 \label{eq:background}
\end{equation}
The background starts from $\bgtemp(\tau_0)=T_0$, and its shear coefficient
$\eta_{\rm sh}^{(0)}=4a_1(\eta_{\rm sh}/s)(\bgtemp)^3$ follows from
Eq.~\eqref{eq:eos}.  Differentiating both conservation equations once with
respect to $\lacc$ along $\chi=\lacc$, at fixed coordinates, $\tau_0$, $T_0$,
$\eta_{\rm sh}/s$, $B_0^2$, and $a$, gives
\begin{equation}
\begin{aligned}
 &\tau\partial_\tau\tempresponse
 +\left[\frac13+\frac{(a-1)B^2}{4a_1(\bgtemp)^4}\right]\tempresponse
 \\
 &\quad+\left[\frac{\bgtemp}{3}+\frac{B^2}{12a_1(\bgtemp)^3}
                    -\frac{8(\eta_{\rm sh}/s)}{9\tau}\right]
                \partial_{\etas}\flowresponse=0.
\end{aligned}
 \label{eq:responseE}
\end{equation}
\begin{equation}
\begin{aligned}
 &\tau\partial_\tau\!\left\{\tau^2
 \left[4a_1(\bgtemp)^4+B^2
       -\frac{16a_1(\eta_{\rm sh}/s)(\bgtemp)^3}{3\tau}\right]\flowresponse\right\}\\
 &\quad+4a_1\tau^2(\bgtemp)^3
       \left(1-\frac{4(\eta_{\rm sh}/s)}{\tau\bgtemp}\right)
                         \partial_{\etas}\tempresponse\\
 &\quad-\frac{16}{3}a_1(\eta_{\rm sh}/s)\tau(\bgtemp)^3
                         \partial_{\etas}^2\flowresponse=0.
\end{aligned}
 \label{eq:responseM}
\end{equation}
The momentum equation has been written in conservative form using
Eq.~\eqref{eq:background}.  The initial data of Eq.~\eqref{eq:initial}
with $\chi=\lacc$ give
\begin{equation}
 \tempresponse\big|_{\tau_0}=-\frac{T_0}{3}\etas^2,
 \qquad
 \flowresponse\big|_{\tau_0}=\etas .
\end{equation}
Equations~\eqref{eq:responseE} and \eqref{eq:responseM} preserve a
quadratic temperature correction and a linear rapidity correction, so that
\begin{align}
 \tempresponse(\tau,\etas)
 &=\tempresponse(\tau,0)
 +\frac{\etas^2}{2}\,\partial_{\etas}^2\tempresponse\big|_0,\\
 \flowresponse(\tau,\etas)
 &=\etas\,\partial_{\etas}\flowresponse\big|_0 ,
 \label{eq:responsePolynomial}
\end{align}
where $|_0$ denotes evaluation at $\etas=0$ at each $\tau$.  Higher powers
of $\etas$ first appear at second order in $\lacc$.  The problem thus
reduces to three time-dependent coefficients: the central temperature
correction, the temperature curvature, and the rapidity slope.  The
curvature supplies the pressure gradient that drives the slope, and the
slope feeds back into the temperature equation at the same order, which is
why both fields must be expanded together.

\subsubsection{Pressure forces and the temperature quadrature}
The transverse magnetic field contributes to the longitudinal inertia and changes the background temperature. Since $\eta_{\rm sh}\propto T^3$ at fixed $\eta_{\rm sh}/s$, this temperature change also modifies the shear stress. We determine the background first, then the thermal curvature, the rapidity slope, and finally the central temperature correction.
\label{sec:jointIntegration}

First, the background equation has the integrating-factor form
\begin{align}
 \bgtemp={}&\left(\frac{\tau_0}{\tau}\right)^{1/3}
 \left\{T_0+\int_{\tau_0}^{\tau}
 \left(\frac{\tau'}{\tau_0}\right)^{1/3}\right.\nonumber\\
 &\left.\quad\times
 \left[\frac{(a-1)B^2(\tau')}{12a_1[\bgtemp(\tau')]^3}+
       \frac{4(\eta_{\rm sh}/s)}{9\tau'}\right]
                  \frac{d\tau'}{\tau'}\right\}.
 \label{eq:backgroundIntegral}
\end{align}
Because the magnetic source depends on $\bgtemp$ itself,
Eq.~\eqref{eq:backgroundIntegral} is an implicit integral equation,
equivalent to a single first-order ordinary differential equation that we
integrate numerically.  It reduces to the closed forms of
Eqs.~\eqref{eq:baselineNS} and \eqref{eq:idealMagneticBaseline} when the
field or the viscosity vanishes.  Once $\bgtemp$ is known, the curvature,
the velocity slope, and the central temperature correction follow in turn
from explicit quadratures.
Taking two rapidity derivatives of Eq.~\eqref{eq:responseE} gives
\begin{equation}
 \left[\tau\partial_\tau+\frac13
       +\frac{(a-1)B^2}{4a_1(\bgtemp)^4}\right]
 \left.\partial_{\etas}^2\tempresponse\right|_0=0.
 \label{eq:curvatureODEjoint}
\end{equation}
The integrating factor of this linear equation, normalized to one at $\tau_0$, is
\begin{align}
 \mathcal I_T(\tau)={}&\left(\frac{\tau}{\tau_0}\right)^{1/3}
 \exp\left[\int_{\tau_0}^{\tau}
 \frac{(a-1)B^2(\tau')}{4a_1[\bgtemp(\tau')]^4}
                  \frac{d\tau'}{\tau'}\right].
 \label{eq:jointIntegratingFactor}
\end{align}
It records the accumulated dilution of the initial temperature curvature. The initial thermal curvature fixes
\begin{equation}
 \left.\partial_{\etas}^2\tempresponse\right|_0
       =-\frac{2T_0}{3\mathcal I_T(\tau)}.
 \label{eq:curvatureIntegraljoint}
\end{equation}
The longitudinal momentum equation then determines the rapidity slope.
Taking one rapidity derivative of Eq.~\eqref{eq:responseM} at the
center eliminates its last term in the linear polynomial sector.
The resulting equation and its integral are
\begin{equation}
\begin{aligned}
 &\tau\partial_\tau\Bigl[
 \tau^2\Bigl(4a_1(\bgtemp)^4+B^2 -\frac{16a_1(\eta_{\rm sh}/s)(\bgtemp)^3}{3\tau}\Bigr)
       \left.\partial_{\etas}\flowresponse\right|_0\Bigr] \\
 &\qquad=-4a_1\tau^2(\bgtemp)^3
       \left(1-\frac{4(\eta_{\rm sh}/s)}{\tau\bgtemp}\right)
       \left.\partial_{\etas}^2\tempresponse\right|_0.
\end{aligned}
\label{eq:slopeConservativeJoint}
\end{equation}
\begin{equation}
\begin{aligned}
 &\tau^2\left[4a_1(\bgtemp)^4+B^2
                   -\frac{16a_1(\eta_{\rm sh}/s)(\bgtemp)^3}{3\tau}\right]
       \left.\partial_{\etas}\flowresponse\right|_0\\
 &\quad=\tau_0^2\left[4a_1T_0^4+B_0^2
                 -\frac{16a_1(\eta_{\rm sh}/s) T_0^3}{3\tau_0}\right]\\
 &\qquad-\int_{\tau_0}^{\tau}\tau'^2
 4a_1[\bgtemp(\tau')]^3
 \left[1-\frac{4(\eta_{\rm sh}/s)}{\tau'\bgtemp(\tau')}\right]\\
 &\hspace{4em}\times
 \left.\partial_{\etas}^2\tempresponse\right|_0(\tau')
                    \frac{d\tau'}{\tau'}.
\end{aligned}
\label{eq:slopeIntegraljoint}
\end{equation}
The integration constant follows from the unit initial slope of
$\flowresponse$.  The bracket multiplying the slope is the effective
longitudinal inertia of the background, the total energy density plus the
total longitudinal pressure,
$\varepsilon+B^2/2+(p-\Phi^{(0)}+B^2/2)$, with
$\Phi^{(0)}=4\eta_{\rm sh}^{(0)}/(3\tau)$.  It stays positive as long as
$(\eta_{\rm sh}/s)/(\bgtemp\tau)<3/4$, which holds throughout our domain,
where this ratio is at most $0.04$.  Since the right-hand side involves
only $\bgtemp$ and the curvature found above, the slope follows directly by
dividing by this factor.  A negative temperature curvature makes the
integrand negative and therefore increases the slope.

Finally, the central part of Eq.~\eqref{eq:responseE} obeys
\begin{align}
 &\left[\tau\partial_\tau+\frac13
       +\frac{(a-1)B^2}{4a_1(\bgtemp)^4}\right]
                 \tempresponse(\tau,0)\nonumber\\
 &\qquad=-\left[\frac{\bgtemp}{3}+\frac{B^2}{12a_1(\bgtemp)^3}
                       -\frac{8(\eta_{\rm sh}/s)}{9\tau}\right]
                 \left.\partial_{\etas}\flowresponse\right|_0.
 \label{eq:centerODEjoint}
\end{align}
It has the same integrating factor $\mathcal I_T$ and zero central
initial temperature correction. Combining its integral with the curvature gives
\begin{equation}
\begin{aligned}
 \tempresponse(\tau,\etas)
 &=-\frac{1}{\mathcal I_T(\tau)}
 \Biggl\{\frac{T_0\etas^2}{3}
 +\int_{\tau_0}^{\tau}\mathcal I_T(\tau')
 \biggl[\frac{\bgtemp(\tau')}{3}
       +\frac{B^2(\tau')}{12a_1[\bgtemp(\tau')]^3}
 -\frac{8(\eta_{\rm sh}/s)}{9\tau'}\biggr]\\
 &\hspace{12em}\times
 \left.\partial_{\etas}\flowresponse\right|_0(\tau')
                         \frac{d\tau'}{\tau'}\Biggr\}.
\end{aligned}
 \label{eq:temperatureIntegraljoint}
\end{equation}
Thus background temperature, thermal curvature, rapidity slope, and
central temperature form a sequential initial-value problem.
The first-order temperature is simply
\begin{align}
 T(\tau,\etas)={}&
 \bgtemp(\tau)+\lacc\tempresponse(\tau,\etas)
 +\mathcal O((\lacc)^2).
 \label{eq:jointT}
\end{align}
Equations~\eqref{eq:backgroundIntegral},
\eqref{eq:curvatureIntegraljoint}, \eqref{eq:slopeIntegraljoint}, and
\eqref{eq:temperatureIntegraljoint} specify the joint solution by quadratures.

In the ideal, zero-field limit ($\eta_{\rm sh}=B_0=0$), the solutions
reduce to Eq.~\eqref{eq:idealReferenceSlope} along
$\chi=\lambda^{*}$. The flow-rapidity slope remains $1+\lambda^{*}$
to first order, and the temperature agrees with the conformal
weak-acceleration result of Ref.~\cite{Jiang2020}. In the
boost-invariant limit ($\lambda^{*}=\chi=0$), the temperature
satisfies the Navier--Stokes equation for Bjorken flow with a
prescribed magnetic field given in Ref.~\cite{Jiang2025}.

Away from these limits, the flow-rapidity slope generally evolves
with proper time. To isolate the effect of this evolution on the
temperature, we compare our zero-field solution with the viscous
prescribed-flow solution of Ref.~\cite{Jiang2020} and our inviscid
solution with the magnetic prescribed-flow solution of
Ref.~\cite{She2019}. In each comparison, we use the same equation
of state, magnetic-field history, and initial profiles for both flows.

\begin{figure}[tbp]
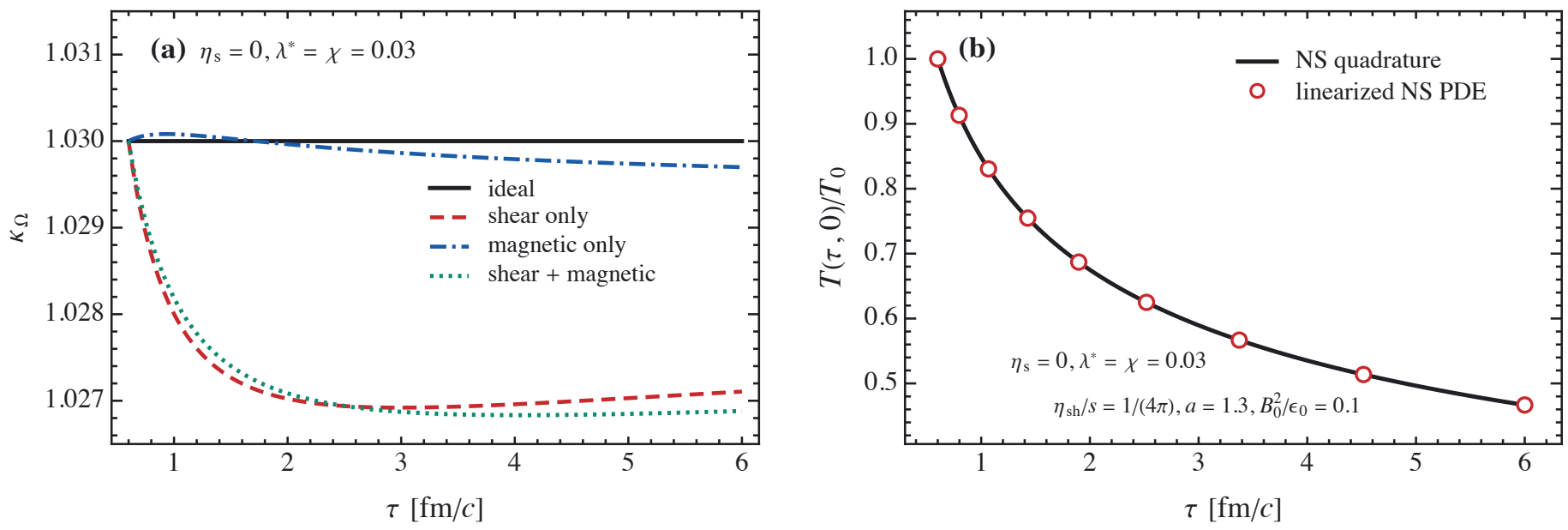

\centering\ReportFigureFour
\caption{First-order NS magnetic-viscous solutions along $\chi=\lambda^*=0.03$. Left: the central rapidity slope with neither stress, shear only, magnetic stress only, and both stresses. Right: normalized central temperature from the quadrature solution, with an additional numerical evolution of the linearized NS PDE shown by open symbols. Finite stress parameters have their reference values.}
\label{fig:rapidity}
\end{figure}

The lines in Fig.~\ref{fig:rapidity} give the NS quadrature solution. The open symbols in the temperature panel come from a finite-difference evolution of the linearized NS equations.

\subsubsection{Independent initial acceleration and thermal curvature}
The NS quadratures above take $\chi=\lacc$.  Because the first-order
equations are linear, they apply equally to independent initial
gradients.  We write the physical first-order corrections as $\delta T$
and $\delta\Omega$, with initial data $\delta T=-T_0\chi\etas^2/3$ and
$\delta\Omega=\lacc\etas$.  The background and integrating factor are
unchanged, and the homogeneous curvature equation gives
\begin{equation}
 \left.\partial_{\eta_s}^2\delta T\right|_0
       =-\frac{2T_0\chi}{3\mathcal I_T(\tau)}.
 \label{eq:independentNScurvature}
\end{equation}
For $\delta\kappa_\Omega=\left.\partial_{\eta_s}\delta\Omega\right|_0$, the momentum quadrature is
\begin{equation}
\begin{aligned}
 &\tau^2\left[4a_1(\bgtemp)^4+B^2
 -\frac{16a_1(\eta_{\rm sh}/s)(\bgtemp)^3}{3\tau}\right]\delta\kappa_\Omega(\tau)\\
 &=\tau_0^2\left[4a_1T_0^4+B_0^2
 -\frac{16a_1(\eta_{\rm sh}/s) T_0^3}{3\tau_0}\right]\lambda^*\\
 &\quad-\int_{\tau_0}^{\tau}4a_1t^2[\bgtemp(t)]^3
 \left[1-\frac{4(\eta_{\rm sh}/s)}{t\bgtemp(t)}\right]
 \left.\partial_{\eta_s}^2\delta T\right|_0(t)\frac{dt}{t}.
\end{aligned}
\label{eq:independentNSslope}
\end{equation}
The integration constant carries the initial velocity excess $\lacc$,
and the integral carries the initial thermal curvature $\chi$.  The central
energy equation then gives
\begin{equation}
\begin{aligned}
 \delta T(\tau,0)={}&-\frac{1}{\mathcal I_T(\tau)}
 \int_{\tau_0}^{\tau}\mathcal I_T(t) \times\left[\frac{\bgtemp(t)}{3}
 +\frac{B^2(t)}{12a_1[\bgtemp(t)]^3}
 -\frac{8(\eta_{\rm sh}/s)}{9t}\right] \times\delta\kappa_\Omega(t)\frac{dt}{t}, \\
 \delta T(\tau,\eta_s)={}&\delta T(\tau,0)
                  -\frac{T_0\chi\eta_s^2}{3\mathcal I_T(\tau)}.
\end{aligned}
\label{eq:independentNStemp}
\end{equation}
Setting $\chi=\lacc$ recovers the one-parameter solution of Sec.~\ref{sec:jointIntegration}. For vanishing magnetic field and shear viscosity, the background follows Bjorken cooling:
$\bgtemp=T_0(\tau_0/\tau)^{1/3}$, $\mathcal I_T=(\tau/\tau_0)^{1/3}$, and the
inertia factor is $4a_1(\bgtemp)^4\tau^2\propto\tau^{2/3}$.
Equation~\eqref{eq:independentNSslope} then gives
\begin{equation}
 \delta\kappa_\Omega=\lacc\left(\frac{\tau_0}{\tau}\right)^{2/3}
 +\chi\left[1-\left(\frac{\tau_0}{\tau}\right)^{2/3}\right],
\end{equation}
which is Eq.~\eqref{eq:idealReferenceSlope}.  The first term is the
initial velocity excess, whose longitudinal momentum
$\tau^2(\varepsilon+p)\,\delta\kappa_\Omega$ is conserved; because the
inertia grows as $\tau^{2/3}$, the slope excess decays as
$(\tau_0/\tau)^{2/3}$.  The second term is the velocity built up by the
pressure gradient of the curvature $\chi$, which approaches $\chi$ at late
times.  Inserting this slope into Eq.~\eqref{eq:independentNStemp} gives
$\delta T(\tau,0)/\bgtemp=-\tfrac13\int_{\tau_0}^{\tau}\delta\kappa_\Omega\,dt/t$,
the linearization of Eq.~\eqref{eq:idealReferenceTemperature}.

\subsection{Nonlinear evolution with shear relaxation}
A finite relaxation time allows the pressure anisotropy to lag behind the changing expansion rate. Starting from $\Phi_0=0$, we now evolve $T$, $\Omega$, and $\Phi$ together using Eqs.~\eqref{eq:relaxE}, \eqref{eq:relaxM}, and \eqref{eq:relaxPhi}. Appendix~\ref{sec:numericalMethod} describes the method of lines and its validation.

\begin{figure}[tbp]
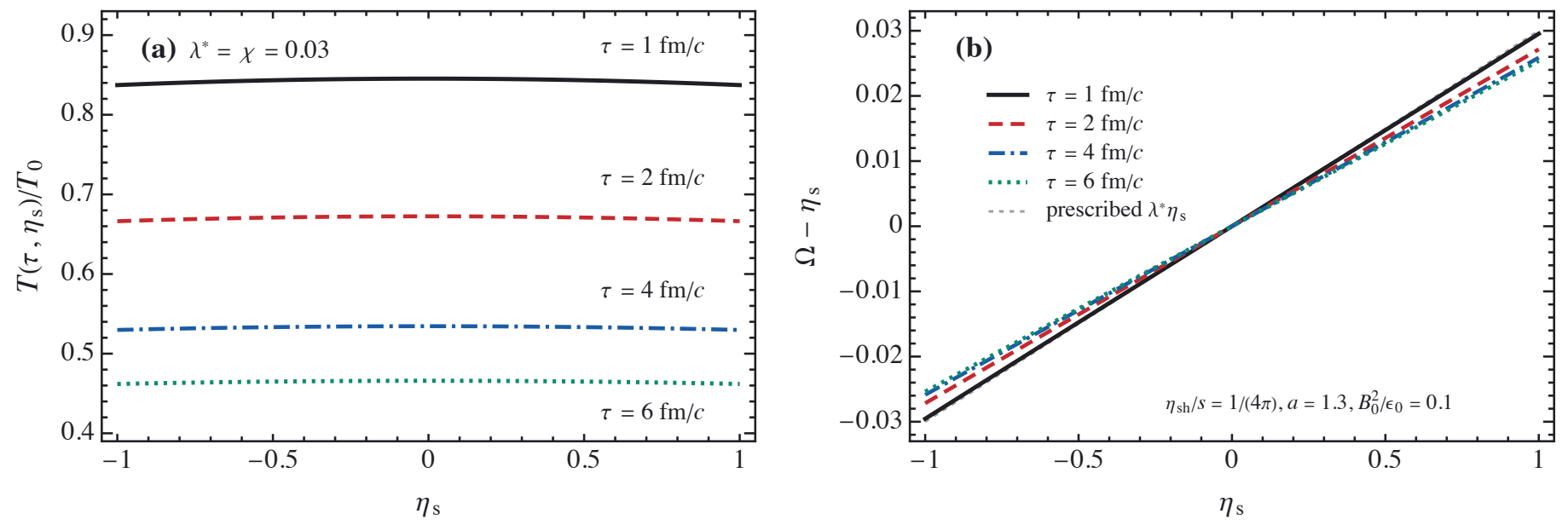

\centering\ReportFigureFive
\caption{Rapidity profiles from the nonlinear scalar shear-relaxation solution at $\lambda^*=\chi=0.03$. Left: $\widetilde T(\eta_s)$ at $\tau\simeq1$, $2$, $4$, and $6$ fm/$c$. Right: $\Omega-\eta_s$ at the same times. Parameters use their reference values.}
\label{fig:spacetime}
\end{figure}

In Fig.~\ref{fig:spacetime} the central temperature decreases between the four proper times shown. In the right panel, the tangent to $\Omega-\etas$ at midrapidity has slope $\kappa_\Omega-1$. The difference between the two central expansion rates is
$[\kappa_\Omega-(1+\lacc)]/\tau$.

Away from midrapidity, advection contributes to the temperature change at fixed $\etas$:
\begin{equation}
 \partial_\tau T
 =\frac{DT}{\cosh(\Omega-\etas)}
   -\frac{\tanh(\Omega-\etas)}{\tau}\partial_{\etas}T.
 \label{eq:eulerian}
\end{equation}
For outward motion relative to the Milne frame and a temperature decreasing toward positive rapidity, the advective term brings warmer matter outward. It vanishes at the reflection center. In the ideal weak-gradient solution the leading product contains both $\lacc\chi$ and $\chi^2$, since the initial thermal curvature itself generates a velocity perturbation.
\section{Prescribed versus self-consistent flow}
\label{sec:comparison}
A mismatch in longitudinal momentum balance modifies the expansion rate, and its thermal impact depends on how the resulting change in expansion accumulates. Comparing self-consistent and prescribed evolutions from identical initial temperature, rapidity, shear stress, and magnetic field profiles, we first evaluate the central temperature difference (Fig.~\ref{fig:centralTemperatureHistories}), then decompose it into expansion-work, shear, and magnetic sources (Fig.~\ref{fig:newMechanism}), and finally determine how the equal-temperature condition shifts with proper time, gradient amplitude, and initial shear (Fig.~\ref{fig:newPrediction}).
\subsection{Central temperature difference}
\label{sec:feedback}
We denote the central temperature difference and the two rapidity slopes by
\begin{equation}
\begin{aligned}
 \Delta T&=T_s-T_p,\\
 \kappa_{\Omega,s}&=\left.\partial_{\etas}\Omega_s\right|_0,\qquad
 \kappa_{\Omega,p}=1+\lacc ,
\end{aligned}
 \label{eq:feedbackDifference}
\end{equation}
where $s$ and $p$ denote the self-consistent and prescribed histories. For $\Delta T>0$, the self-consistent fluid is warmer at midrapidity. In the ideal limit, subtracting $\widetilde T_p=x^{-(1+\lacc)/3}$ from the first-order temperature in Eq.~\eqref{eq:idealReferenceTemperature} gives
\begin{equation}
 \frac{\Delta T}{T_0(\tau_0/\tau)^{1/3}}
 =\frac{\lacc-\chi}{3}\Biggl\{\ln\frac{\tau}{\tau_0}
 -\frac32\left[1-\left(\frac{\tau_0}{\tau}\right)^{2/3}\right]\Biggr\}
 +\mathcal O\bigl((\lacc)^2,\lacc\chi,\chi^2\bigr).
\label{eq:idealReferenceDifference}
\end{equation}
The central temperatures and their first time derivatives agree initially because the two flows start from the same expansion rate.
Writing $\xi=\ln(\tau/\tau_0)$, the small-$\xi$ expansion of Eq.~\eqref{eq:idealReferenceDifference} begins with $(\lacc-\chi)\xi^2/9$ at first order in the initial gradients. Thus a change in the slope precedes the separation of the central temperatures.

For $\tau>\tau_0$, the term in braces is positive, so $\Delta T$ has
the same sign as $\lambda^*-\chi$. When $\lambda^*>\chi$, the
flow-rapidity slope decreases from its prescribed value. The fluid
then does less longitudinal expansion work and remains warmer at
midrapidity than in the prescribed flow. When $\chi>\lambda^*$, the
initial temperature curvature produces a pressure gradient that
steepens the flow-rapidity profile, increasing the expansion work
and enhancing the cooling.

At late times, the first-order ideal solution gives a logarithmic
dependence for the relative temperature difference,
\begin{equation}
 \frac{\Delta T}{T_{\rm Bj}}\simeq
 \frac{\lambda^*-\chi}{3}
 \left[\ln\left(\frac{\tau}{\tau_0}\right)-\frac{3}{2}\right],
 \qquad \tau\gg\tau_0 .
 \label{eq:lateTimeDifference}
\end{equation}
Once the transient decays, the difference between the self-consistent
and prescribed rapidity slopes approaches $-(\lambda^*-\chi)$.
The corresponding difference in the midrapidity expansion rate
therefore scales as $1/\tau$. Integrating this contribution to the
cooling rate gives the logarithm in Eq.~\eqref{eq:lateTimeDifference}.

For $\lambda^*=\chi=0.03$, the ideal first-order contribution to
$\Delta T$ vanishes. The nonlinear ideal evolution gives
$\Delta T/T_0=3.64\times10^{-4}$ at $\tau=6$~fm/$c$, isolating the
finite-amplitude correction in this limit. At the same proper time,
$\Delta T/T_0$ is $8.28\times10^{-4}$ with shear alone,
$3.79\times10^{-4}$ with magnetic stress alone, and
$8.35\times10^{-4}$ with both stresses. At the reference parameters,
shear more than doubles the ideal temperature difference. Adding
magnetic stress increases $\Delta T/T_0$ by $0.15\times10^{-4}$ in
the inviscid case and by $0.07\times10^{-4}$ in the viscous case.
The magnetic contribution to the temperature difference thus depends
on the presence of shear.

\begin{figure}[tbp]
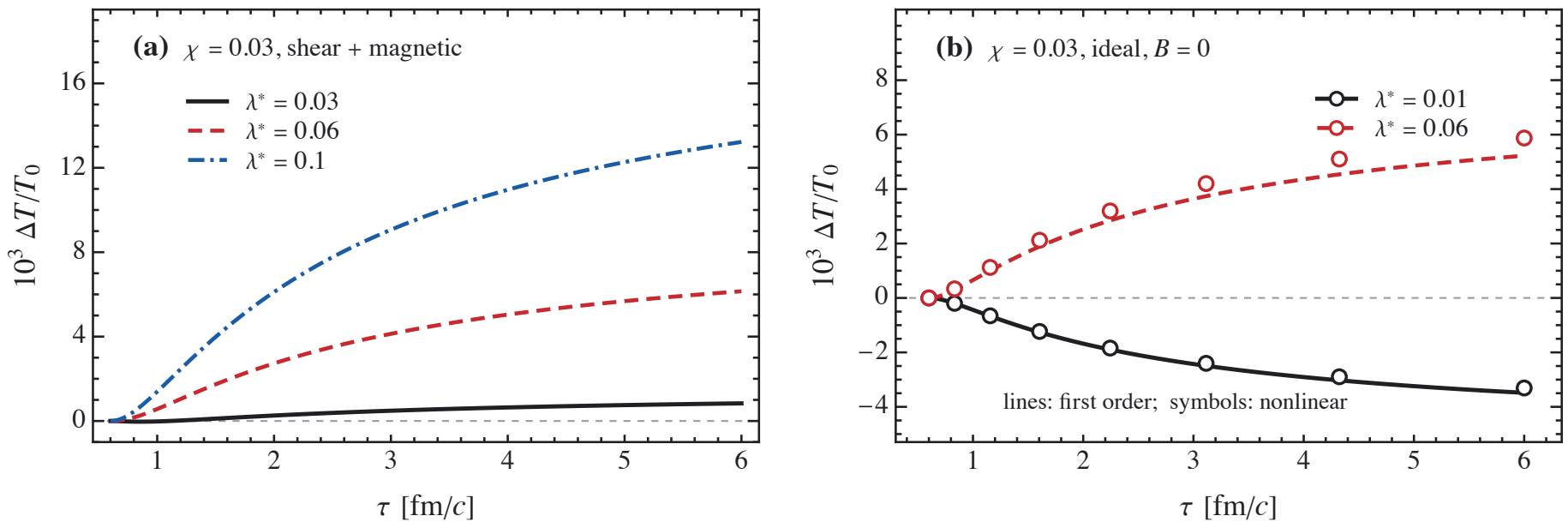

\centering\ReportFigureSix
\caption{Central temperature difference between self-consistent and prescribed flow. Left: reference magnetic and shear stresses, fixed $\chi=0.03$, and $\lambda^*=0.03,0.06,0.1$. Right: ideal zero-field differences at $\chi=0.03$, with $\lambda^*=0.01,0.06$. Lines in the right panel are Eq.~\eqref{eq:idealReferenceDifference}; open symbols are nonlinear calculations.}
\label{fig:centralTemperatureHistories}
\end{figure}

Figure~\ref{fig:centralTemperatureHistories} first shows the finite-stress comparison at fixed initial curvature. Increasing $\lambda^*$ increases the positive late difference from prescribed flow. The ideal comparisons in the right panel show both signs predicted by Eq.~\eqref{eq:idealReferenceDifference}; the departures of the symbols from the lines are finite-amplitude corrections. Both histories cool even as their temperature difference grows.

\subsection{Expansion work and the temperature balance}
The same initial state develops different central temperatures because the
two velocity slopes give different expansion rates
$\theta=\kappa_\Omega/\tau$, and hence different expansion work.

A fluid element starting at $\eta_{s,0}$ moves with the four-velocity, so
its trajectory obeys $d\etas/d\tau=u^{\etas}/u^\tau$:
\begin{equation}
 \frac{d\etas}{d\tau}
 =\frac{\tanh(\Omega-\etas)}{\tau},\qquad
 \etas(\tau_0)=\eta_{s,0}.
 \label{eq:materialPath}
\end{equation}
Since $D$ is the derivative with respect to the element's own proper time
and $u^\tau=\cosh(\Omega-\etas)$, the change along the path in Milne time
is $dT/d\tau=DT/u^\tau$.  The rest-frame volume of the element obeys
\begin{equation}
\begin{aligned}
 DV&=\theta V,\\
 \frac{d\ln(V/V_0)}{d\tau}
 &=\frac{\theta}{\cosh(\Omega-\etas)}.
\end{aligned}
 \label{eq:materialVolume}
\end{equation}
Here $V$ is the rest-frame volume of the same fluid element and $V_0$ its
initial value; a positive $\theta$ makes the element expand.  Combining
Eq.~\eqref{eq:materialVolume} with the energy equation relates the
temperature to this volume:
\begin{equation}
 D\!\left[T\left(\frac{V}{V_0}\right)^{1/3}\right] =\left(\frac{V}{V_0}\right)^{1/3}
       \left(DT+\frac{T\theta}{3}\right).
 \label{eq:temperatureVolume}
\end{equation}
The nonadiabatic source on the right follows from the temperature
equation~\eqref{eq:relaxE}.  When it vanishes, $TV^{1/3}$ is constant, as
in ideal Bjorken flow.  Departures from $T\propto V^{-1/3}$ therefore
measure the heating by shear and the energy exchanged with the prescribed
field.

At the reflection center the element remains at $\etas=0$, where
$u^\tau=1$, and Eq.~\eqref{eq:kinematics} gives
\begin{equation}
 \frac{d\ln(V/V_0)}{d\ln\tau}
 =\tau\theta(\tau,0)
 =\left.\partial_{\etas}\Omega\right|_0
 =\kappa_\Omega(\tau).
 \label{eq:centralVolumeLocal}
\end{equation}
The central slope thus sets the volume growth of the central element.  It
starts at $1+\lacc$, and momentum balance determines its later evolution.
For the prescribed magnetic history the local sources are those in Eq.~\eqref{eq:absoluteSources}. Magnetic work supplies energy when $a>\kappa_\Omega$ and extracts it when $a<\kappa_\Omega$. For an inviscid fluid, the temperature rises only if $B^2(a-\kappa_\Omega)>(\varepsilon+p)\kappa_\Omega$. The NS shear source is $4(\eta_{\rm sh}/s)\kappa_\Omega^2/(9\tau^2)$, whereas the relaxation source starts at zero for $\Phi_0=0$.
\subsubsection{Source decomposition of the temperature difference}
\label{sec:sourceDecomposition}
Integrating Eq.~\eqref{eq:centralVolumeLocal} gives
\begin{equation}
 \frac{V(\tau)}{V_0}=\exp\left[\int_{\tau_0}^{\tau}
            \frac{\kappa_{\Omega,s}(t)}{t}\,dt\right].
 \label{eq:centralVolume}
\end{equation}
For an ideal zero-field fluid, $TV^{1/3}$ is constant.  With shear and magnetic work, integrating the central temperature equation~\eqref{eq:absoluteSources} along one history gives Eq.~\eqref{eq:absoluteBudget}, and subtracting the prescribed history from the self-consistent one gives Eq.~\eqref{eq:rawHistoryBudget}:
\begin{equation}
 T(\tau)-T_0=-\int_{\tau_0}^{\tau}\frac{\kappa_\Omega(t)T(t)}{3t}\,dt +\int_{\tau_0}^{\tau}S_\eta(t)\,dt
       +\int_{\tau_0}^{\tau}S_B(t)\,dt.
\label{eq:absoluteBudget}
\end{equation}
\begin{equation}
\begin{aligned}
\Delta T={}&-\int_{\tau_0}^{\tau}
 \frac{\kappa_{\Omega,s}T_s-\kappa_{\Omega,p}T_p}{3t}\,dt\\
&+\int_{\tau_0}^{\tau}\frac{1}{12a_1t}
 \left(\frac{\Phi_s\kappa_{\Omega,s}}{T_s^3}-\frac{\Phi_p\kappa_{\Omega,p}}{T_p^3}\right)dt\\
&+\int_{\tau_0}^{\tau}\frac{B^2(t)}{12a_1t}
 \left(\frac{a-\kappa_{\Omega,s}}{T_s^3}-\frac{a-\kappa_{\Omega,p}}{T_p^3}\right)dt.
\end{aligned}
\label{eq:rawHistoryBudget}
\end{equation}
The three integrals in Eq.~\eqref{eq:rawHistoryBudget} represent the differences in expansion cooling, shear heating, and magnetic work between the two histories.  Each rate is evaluated along its respective history, so changes in either stress affect all three integrals through the modified flow and temperature.  The budget thus retains the dynamical coupling between the stresses.

The first integral of the raw budget combines the effects of the slope and temperature differences.  We separate these contributions by isolating the dilution of $\Delta T$ itself:
\begin{equation}
\begin{aligned}
 \kappa_{\Omega,s}T_s-\kappa_{\Omega,p}T_p
 &=\kappa_{\Omega,s}\Delta T+(\kappa_{\Omega,s}-\kappa_{\Omega,p})T_p,\\
 \frac{d\Delta T}{d\tau}
 +\frac{\kappa_{\Omega,s}}{3\tau}\Delta T
 &=S_{\rm exp}^\Delta+S_\eta^\Delta+S_B^\Delta.
\end{aligned}
\label{eq:restoredBudgetODE}
\end{equation}
Here the expansion source $S_{\rm exp}^\Delta=-(\kappa_{\Omega,s}-\kappa_{\Omega,p})T_p/(3\tau)$ represents the additional cooling due solely to the slope difference, while $S_\eta^\Delta$ and $S_B^\Delta$ are the differences in shear and magnetic rates.  Multiplying by $\exp[\int_{\tau_0}^{\tau}\kappa_{\Omega,s}(t)dt/(3t)]$ and integrating yields the weighted budget.  Both budgets are exact rearrangements of the same temperature equation; the weighted form additionally accounts for the dilution of each contribution by subsequent expansion:
\begin{equation}
\begin{aligned}
 \Delta T_j(\tau)&=\int_{\tau_0}^{\tau}
  \exp\!\left[-\int_t^\tau\frac{\kappa_{\Omega,s}(t')}{3t'}dt'\right]
  S_j^\Delta(t)dt,\\
 \Delta T&=\Delta T_{\rm exp}+\Delta T_\eta+\Delta T_B,
 \quad j\in\{\mathrm{exp},\eta,B\},
\end{aligned}
\label{eq:feedbackAttribution}
\end{equation}
with $S_j^\Delta$ denoting the three source differences defined above.

\begin{figure}[tbp]
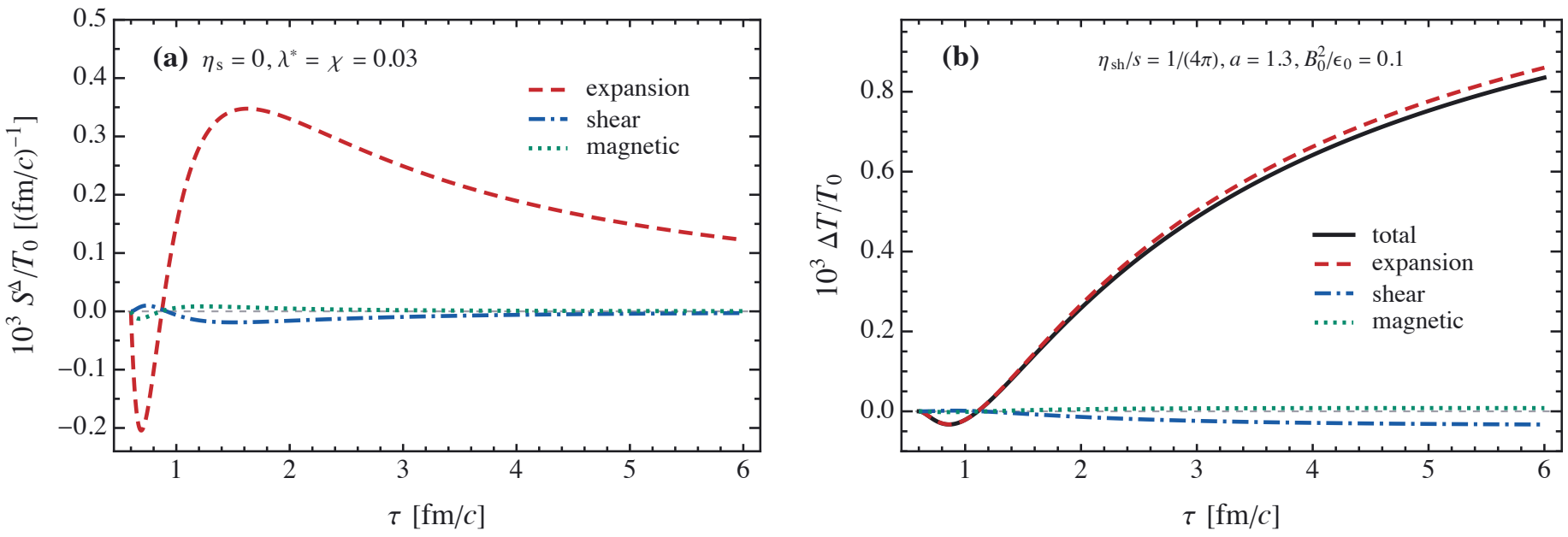

\centering\ReportFigureSeven
\caption{Temperature balance for the same pair of histories at $\lambda^*=\chi=0.03$. Left: the expansion, shear, and magnetic source differences after moving the common dilution term to the left. Right: their weighted integrals and the temperature difference from the nonlinear evolution. All quantities are normalized by $T_0$.}
\label{fig:newMechanism}
\end{figure}

Figure~\ref{fig:newMechanism} shows that expansion work dominates the temperature difference.  At $6$~fm/$c$, the expansion, shear, and magnetic contributions are $0.559$, $-0.0215$, and $0.00524$~MeV, respectively, giving $\Delta T\simeq0.543$~MeV.  The direct shear and magnetic contributions are small, but both stresses also affect the temperature indirectly by modifying the slope and hence the expansion term.  

The momentum residual and the central temperature difference probe distinct aspects of the prescribed-flow approximation: the residual measures the local force imbalance near midrapidity at a given proper time. Equal central temperatures at a later time require the weighted source differences in Eq.~\eqref{eq:feedbackAttribution} to integrate to zero over the expansion history. A persistent local momentum residual can therefore coexist with a small accumulated temperature error.
\subsection{Initial gradients and the moving cancellation condition}
The initial acceleration and the initial temperature curvature affect the cooling difference with opposite signs, and their competition determines when the two central temperatures coincide.  At first order, their contributions add linearly:
\begin{equation}
 T_{s,\mathrm{order}\,1}-T_{p,\mathrm{order}\,1}
={}\lacc\left.\frac{\partial(T_s-T_p)}{\partial\lacc}\right|_{\lacc=\chi=0} +\chi\left.\frac{\partial(T_s-T_p)}{\partial\chi}\right|_{\lacc=\chi=0}.
\label{eq:initialPrediction}
\end{equation}
At $\tau=6$ fm/$c$ and the reference parameters, the coefficient calculation of Appendix~\ref{app:relaxationPrediction} gives
\begin{equation}
 \Delta T^{(1)}
 = [109.36\lambda^*-98.45\chi]\ {\rm MeV}.
 \label{eq:latePrediction}
\end{equation}
To first order, the two central temperatures coincide at $\tau=6~\mathrm{fm}/c$ for $\chi/\lambda^{*}\simeq 1.11$, a terminal integral condition that balances the accumulated cooling over the full interval while leaving the two histories distinct at intermediate times.

\begin{figure}[tbp]
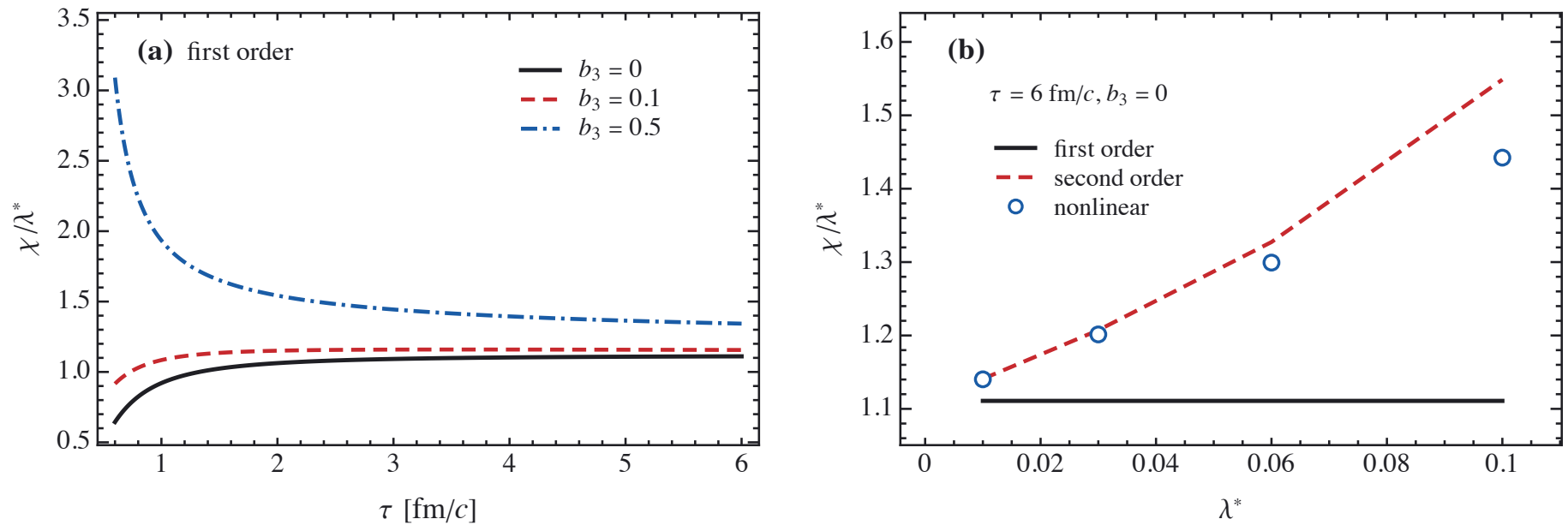

\centering\ReportFigureEight
\caption{Initial-gradient ratios giving zero central temperature difference. Left: first-order $\chi/\lambda^*$ as a function of comparison time for $b_3=0,0.1,0.5$. The common initial time, at which all differences vanish, is excluded. Right: terminal ratios at four acceleration amplitudes; lines join the first- and second-order predictions, and open symbols mark nonlinear roots. The right panel uses $b_3=0$.}
\label{fig:newPrediction}
\end{figure}

The onset condition concerns the quadratic term in the initial temperature difference. Both histories start with the same temperature, slope, and stress.
Consequently $\Delta T(\tau_0)=\Delta T'(\tau_0)=0$, and the initial
temperature difference begins with
$\tfrac12\Delta T''(\tau_0)(\tau-\tau_0)^2$. For zero initial shear,
\begin{equation}
\begin{aligned}
 \left.\tau_0\frac{d\kappa_{\Omega,s}}{d\tau}\right|_{\tau_0}
={}&\frac{1}{\frac43+B_0^2/(3a_1T_0^4)}\Bigg\{\frac89\chi +\lacc\bigg[\frac23a\frac{B_0^2}{3a_1T_0^4} \\
 &\hspace{2em}-\frac23\left(\frac43+\frac{B_0^2}{3a_1T_0^4}\right)(1+\lacc) +\frac{16(1+\lacc)}{9c_\pi}\bigg]\Bigg\}.
\end{aligned}
\label{eq:onsetSlope}
\end{equation}
\begin{equation}
 \left.\frac{d^2\Delta T}{d\tau^2}\right|_{\tau_0}
 =-\frac{T_0}{4\tau_0}
 \left(\frac43+\frac{B_0^2}{3a_1T_0^4}\right)
 \left.\frac{d\kappa_{\Omega,s}}{d\tau}\right|_{\tau_0}.
\label{eq:onsetTemperature}
\end{equation}
The leading temperature difference vanishes when the initial slope is stationary, which gives
\begin{equation}
 \chi=\lacc\Bigg[(1+\lacc)
 \left(1+\frac{B_0^2}{4a_1T_0^4}-\frac2{c_\pi}\right) -\frac{aB_0^2}{4a_1T_0^4}\Bigg].
\label{eq:onsetMatching}
\end{equation}
The term $-2/c_\pi$ arises from the buildup of shear starting from $\Phi_0=0$ and characterizes the finite-relaxation initial layer.  For the reference field and $b_3=0$, the onset ratio in the infinitesimal-gradient limit is approximately $0.64$, well below the terminal value $1.11$.  The cancellation condition therefore evolves with time.  Initial shear modifies this balance, as shown in Fig.~\ref{fig:newPrediction}.

\subsubsection{Initial shear and the onset condition}
For $\Phi_0=b_3p_0$ with constant $b_3$, the initial shear follows the thermal profile, so its curvature is determined by $\chi$.  The initial relaxation time is $\tau_{\pi,0}=c_\pi(\eta_{\rm sh}/s)/T_0$ in natural units.  At the center,
$\partial_{\eta_s}^2\varepsilon|_0=-8\varepsilon_0\chi/3$ and
$\partial_{\eta_s}^2\Phi|_0=(b_3/3)\partial_{\eta_s}^2\varepsilon|_0$.
Differentiating the momentum equation once with respect to rapidity and eliminating the central time derivatives of temperature and shear using their evolution equations yields the exact initialization identity
\begin{equation}
\begin{aligned}
 &\left(\frac43+\frac{B_0^2}{\varepsilon_0}-\frac{b_3}{3}\right)
       \left.\tau_0\frac{d\kappa_{\Omega,s}}{d\tau}\right|_{\tau_0}\\
 &=\frac89(1-b_3)\chi
  +\lambda^*\Bigg\{\frac{2aB_0^2}{3\varepsilon_0}
                     -\frac{b_3\tau_0}{3\tau_{\pi,0}}\\
 &\quad-\left[\frac89+\frac{2B_0^2}{3\varepsilon_0}
            +\frac{2b_3}{9}-\frac{16}{9c_\pi}\right](1+\lambda^*)\Bigg\}.
\end{aligned}
\label{eq:generalShearOnset}
\end{equation}
Because both histories initially have the same central temperature, slope, and shear, their first central temperature derivatives agree. Differentiating their energy equations once more gives
\begin{equation}
 \left.\frac{d^2\Delta T}{d\tau^2}\right|_{\tau_0}
 =-\frac{T_0}{4\tau_0}
 \left(\frac43+\frac{B_0^2}{\varepsilon_0}-\frac{b_3}{3}\right)
 \left.\frac{d\kappa_{\Omega,s}}{d\tau}\right|_{\tau_0}.
\label{eq:generalInitialTemperatureCurvature}
\end{equation}
Since the initial longitudinal inertia is positive, the leading temperature difference vanishes exactly when the initial slope is stationary.  For $b_3\ne1$ and nonzero $\lambda^*$, this condition requires
\begin{equation}
\begin{aligned}
 \frac{\chi}{\lambda^*}
  =\frac{1}{1-b_3}\Bigg\{&(1+\lambda^*)
   \left(1+\frac{3B_0^2}{4\varepsilon_0}
               +\frac{b_3}{4}-\frac2{c_\pi}\right)\\
  &-\frac{3aB_0^2}{4\varepsilon_0}
       +\frac{3b_3\tau_0}{8\tau_{\pi,0}}\Bigg\}.
\end{aligned}
\label{eq:generalShearCancellation}
\end{equation}
For zero initial shear, this reduces to Eq.~\eqref{eq:onsetMatching}.  Initial shear affects the balance in two ways: the factor $1-b_3$ reduces the initial curvature of the longitudinal pressure, while the term $\tau_0/\tau_{\pi,0}$ describes the relaxation of the shear present at $\tau_0$.  In the infinitesimal-gradient limit, the reference onset ratios are approximately $0.64$, $0.92$, and $3.09$ for $b_3=0$, $0.1$, and $0.5$, respectively.

At $b_3=1$, the initial curvature of the longitudinal matter pressure vanishes.  Equation~\eqref{eq:generalShearOnset} remains regular in this limit, while the expression obtained by dividing by $1-b_3$ becomes singular.

The left panel of Fig.~\ref{fig:newPrediction} shows how the first-order cancellation ratio varies with the comparison time and $b_3$, extending the initial-shear sensitivity identified in Ref.~\cite{Jiang2025} to cooling with an evolving flow velocity.  The right panel compares the terminal cancellation at finite amplitude with the initial-gradient prediction.  For the reference parameters, the second-order result is
\begin{equation}
 \Delta T^{(1+2)}={}\big[109.36\lambda^*-98.45\chi
                  +14.13(\lambda^*)^2 +183.56\lambda^*\chi+55.31\chi^2\big]\ {\rm MeV}.
\label{eq:quadraticCancellation}
\end{equation}
We obtain the quadratic coefficients by evolving the gradient equations to $6$ fm/$c$ at the reference transport and field parameters.  Near the linear balance, the quadratic terms are positive, so cancelling the temperature difference requires greater initial thermal curvature than predicted by the first-order ratio.

\begin{table}[tbp]
\caption{Terminal $\chi/\lambda^*$ giving equal central temperatures at $6$ fm/$c$. Roots and relative differences were computed before rounding.}
\label{tab:finiteCancellation}
\centering\small
\begin{tabular}{cccc}
\toprule
$\lambda^*$&First order&Second order&Nonlinear\\
\midrule
0.01&1.11&1.14&1.14\\
0.03&1.11&1.21&1.20\\
0.06&1.11&1.33&1.30\\
0.1&1.11&1.55&1.44\\
\bottomrule
\end{tabular}\end{table}

At $\lambda^*=0.03$, the nonlinear ratio is $1.20$, compared with the second-order value of $1.21$, giving a relative difference of $0.48\%$.  At $0.06$, the difference increases to $2.15\%$.  For larger amplitudes, up to $\lambda^*=0.1$, we take the nonlinear roots as the reference.

\subsection{Accuracy of the initial-gradient expansion}
\label{sec:accuracy}
The weak-gradient expansion predicts a difference of two temperatures. Near $\chi=\lacc$, their leading corrections largely cancel, so higher orders can matter even when each temperature lies close to its reference value. Table~\ref{tab:predictionSummary} measures the truncation error in $\Delta T$ against the nonlinear solution. Appendix~\ref{app:relaxationPrediction} gives the coefficient equations.

\begin{table}[tbp]
\caption{Maximum absolute prediction errors in $\Delta T$ on 121 logarithmic output times over $0.6$--$6$ fm/$c$, in MeV. All parameters are those of Eq.~\eqref{eq:referenceParameters}.}
\label{tab:predictionSummary}
\centering\small
\begin{tabular}{cccc}
\toprule
$\lambda^*$&$\chi$&First order&Second order\\
\midrule
0.03&0&0.0232&0.00147\\
0&0.03&0.0494&0.000430\\
0.03&0.03&0.216&0.0121\\
0.06&0&0.0901&0.0115\\
0&0.06&0.196&0.00345\\
0.06&0.06&0.818&0.0932\\
0.03&0.06&0.509&0.0330\\
0.06&0.03&0.387&0.0436\\
\bottomrule
\end{tabular}\end{table}

Near $\chi\approx\lacc$ the first-order difference nearly cancels, and the
first-order error becomes comparable to $\Delta T$ itself, for example
$0.216$~MeV at $\lacc=\chi=0.03$.  The initial-gradient solutions of the shear-relaxation model therefore
describe the separate responses to $\lacc$ and $\chi$ quantitatively, but
near the matched profile $\Delta T$ requires the second-order or nonlinear
calculation; second order reduces the error below $0.1$~MeV in all
displayed states.

\subsection{Prescribed-flow accuracy and stress dependence}
\label{sec:predictionValidation}
We quantify the accuracy of the prescribed-flow approximation for the central temperature using the maximum relative error
\begin{equation}
 100\max_{\tau_0\le\tau\le6\,{\rm fm}/c}
       \frac{|T_s(\tau,0)-T_p(\tau,0)|}{T_s(\tau,0)}.
\end{equation}

\begin{figure}[tbp]
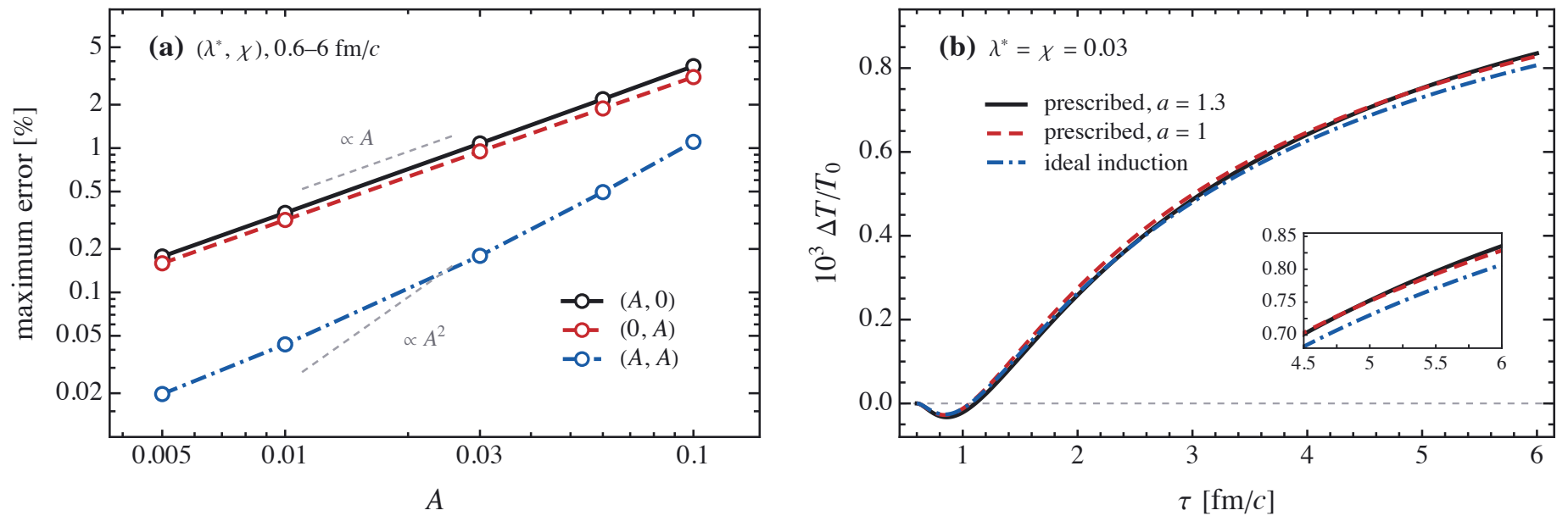

\centering\ReportFigureNine
\caption{Accuracy and magnetic-field dependence of prescribed flow. Left: maximum relative central-temperature error over $0.6$--$6$ fm/$c$ for $(\lambda^*,\chi)=(A,0),(0,A),(A,A)$. Right: normalized temperature differences at $\lambda^*=\chi=0.03$ for prescribed fields with $a=1.3$ and $a=1$, and an ideal-induction field evolved with the flow. The initial field and shear data agree between the paired histories.}
\label{fig:newRange}
\end{figure}

Figure~\ref{fig:newRange} shows that the error depends on both the relative size and the amplitude of the initial gradients.  At amplitude $A=0.03$, the velocity-only and curvature-only errors are $1.08\%$ and $0.95\%$, respectively, whereas $\chi=\lambda^*=0.03$ yields $0.18\%$.  At $A=0.01$, the matched error decreases to $0.044\%$.  At $A=0.1$, the corresponding three errors are $3.70\%$, $3.11\%$, and $1.11\%$.  Matching the ideal first-order profile therefore substantially improves the accuracy of the prescribed-flow approximation, while the time-dependent cancellation condition derived above further refines this balance in the presence of stresses.
\subsection{Stress and parameter dependence}
\begin{figure}[tbp]
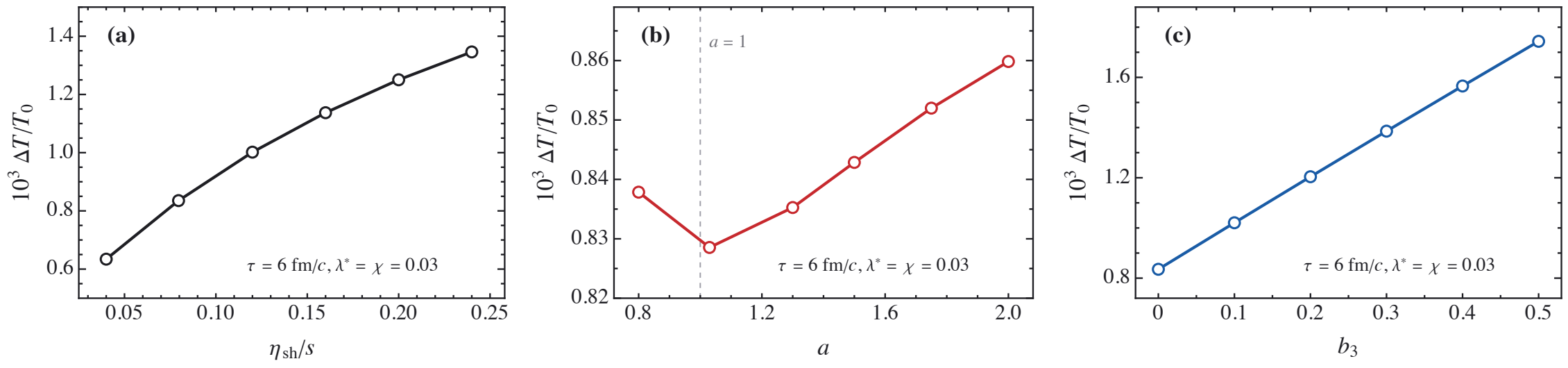

\centering\ReportFigureEleven
\caption{Terminal central temperature difference $\Delta T=T_s-T_p$ at $\tau=6$ fm/$c$, with $\lambda^*=\chi=0.03$. Panels (a), (b), and (c) vary the shear-viscosity ratio $\eta_{\rm sh}/s$, the magnetic decay exponent $a$, and the initial shear parameter $b_3$, respectively. The other parameters are held at their reference values. Points show the calculated values; lines guide the eye.}
\label{fig:parameterDependence}
\end{figure}

The terminal temperature difference depends on the shear-viscosity
ratio, magnetic decay history, and initial shear. Figure~\ref{fig:parameterDependence}
compares six values of each parameter at fixed $\lambda^*=\chi=0.03$.
Both viscosity and initial shear change the longitudinal pressure and
the velocity history. Changing $a$ alters the energy supplied by the
prescribed magnetic stress, the pressure-gradient force, and the
duration of the magnetic contribution to inertia. In
Fig.~\ref{fig:parameterDependence}(b), the terminal temperature
difference falls and then rises as $a$ increases, with a minimum near
$a=1$, the flux-conserving exponent for Bjorken flow. Varying $a$
changes this difference much less than it changes the absolute
temperature, as quantified below.

\subsubsection{Magnetic decay and the three contributions to cooling}
Table~\ref{tab:decayEffects} compares the magnetized evolutions with their unmagnetized counterparts launched from identical initial states, separating the field's impact on absolute cooling from its effect on the paired difference $T_{s}-T_{p}$. At $a=1.30$, the prescribed magnetic stress raises the final self-consistent central temperature by $1.48$~MeV and increases $T_s-T_p$ by $4.59$~keV. Most of the direct magnetic work is common to both evolutions and cancels in the difference. The smaller net increment combines the remaining magnetic work with field-induced changes in the pressure force, rapidity slope, and expansion work.
The decay rate changes the cooling through three coupled mechanisms.
First, the center receives the direct work
$B^2(a-\kappa_\Omega)/\tau$ in Eq.~\eqref{eq:powerExchangeCenter}.
Second, in the NS limit the power-law history contributes $-(a-1)B^2[\tfrac23+\Phi/(\varepsilon+p)](\kappa_\Omega-1)$ to the coefficient terms governing the slope evolution relative to flux-conserving decay. Third, $B^2$ changes the longitudinal inertia throughout the expansion. On the power-law background the relative magnetic stress and thermal curvature obey
\begin{align}
 \frac{d}{d\ln\tau}\ln\frac{B^2}{\varepsilon+p}
 &=\frac43-2a-\frac43\,\frac{\Phi+(a-1)B^2}{\varepsilon+p},\nonumber\\
 \frac{d}{d\ln\tau}\ln\big(-[\partial_{\etas}^2\ln T]_0\big)
 &=-\frac{\Phi+4(a-1)B^2}{3(\varepsilon+p)}.
 \label{eq:decayRatioEvolution}
\end{align}
Positive magnetic work softens the thermal curvature and hence its
pressure force. This change competes with the weaker explicit restoring
term at $a>1$. A faster field decay also shifts the inertial contribution
toward earlier times.

At $a=2/3$, even the ideal background has
$B^2/(\varepsilon+p)=(3B_0^2/4\varepsilon_0)/[1-(B_0^2/3\varepsilon_0)\ln x]$.
The logarithm in the denominator makes the magnetic stress grow relative to the matter enthalpy. For $a>1$, a shorter interval of appreciable magnetic
inertia can coexist with a larger accumulated energy transfer.

\begin{table}[tbp]
\centering\small
\caption{Magnetic contributions at $\tau=6$ fm/$c$ for
$\lambda^*=\chi=0.03$, $B_0^2/\varepsilon_0=0.10$, and the reference
shear-relaxation parameters. Each entry is the change relative to the unmagnetized evolution from
identical initial temperature, velocity, curvature, and shear.
Temperatures are in keV.}
\label{tab:decayEffects}
\begin{tabular}{lrrr}
\toprule
Field history & Change in $T_s$ & Change in $T_s-T_p$ & Change in $10^4\kappa_\Omega$\\
\midrule
$2/3$ & -6398.52 & 22.27 & 5.77 \\
$1.00$ & -261.52 & 0.21 & 0.98 \\
$1.30$ & 1482.76 & 4.59 & -1.71 \\
$2.00$ & 2659.97 & 20.56 & -4.57 \\
Ideal induction & -13.78 & -13.78 & 0.72 \\
\bottomrule
\end{tabular}
\end{table}

Using the full linear propagator, we assign the remaining field-dependent
energy and curvature terms to exchange, the explicit restoring term to
restoration, and the $B^2$ in
$\varepsilon+p+B^2-\Phi$ to inertia.
At $a=1.30$, these contributions to $T_s-T_p$ are $43.03$, $-31.03$,
and $-4.17$ keV, summing to $7.83$ keV at first order, compared with
$4.59$ keV from the unexpanded evolution.
The corresponding absolute-temperature contributions are $1.52$ MeV,
$-31.03$ keV, and $-4.17$ keV.
For shear relaxation, the actual background shear-to-enthalpy ratio
replaces its NS value, and mixed terms are assigned to recover the NS expression
in its instantaneous limit; this assignment affects the individual
contributions but leaves their total unchanged.

Magnetic inertia and shear affect the slope on different time
histories. Equality of accumulated inverse inertia is defined by
\begin{equation}
 \int_{\tau_0}^{\tau_f}
 \left[\frac{1}{1+B^2/(\varepsilon+p)-\Phi/(\varepsilon+p)}-1\right]
 \frac{d\tau}{\tau}=0,
 \label{eq:equalInertiaHistory}
\end{equation}
with the ratios evaluated on the corresponding uniform background.
For the reference viscosity and $a=1.30$, this condition gives
$B_0^2/\varepsilon_0\simeq0.11$, close to the reference $0.10$.
Equal decay of the homogeneous slope perturbation instead gives
$B_0^2/\varepsilon_0\simeq0.14$, because the restoring coefficient
also changes. Equal central temperatures require the integrated thermal balance in Eq.~\eqref{eq:feedbackAttribution}.

\subsubsection{Joint magnetic and shear contributions}
Magnetic stress and shear modify both the temperature and the flow responsible for expansion work, so their thermal effects are coupled.  We isolate their interaction using four self-consistent histories with identical initial temperature, curvature, rapidity, and zero shear.  The subscripts $B$ and $\eta$ indicate the inclusion of the magnetic field and shear viscosity, respectively.  The cross contribution is
\begin{equation}
 [T_{B,\eta}-T_{0,0}]-[T_{B,0}-T_{0,0}] -[T_{0,\eta}-T_{0,0}] =T_{B,\eta}-T_{B,0}-T_{0,\eta}+T_{0,0}.
\label{eq:restoredCrossContribution}
\end{equation}
All increments are measured from the same ideal reference.  The cross contribution quantifies how the two stresses modify each other, whereas $T_s-T_p$ compares two velocity histories within one stress prescription.  Applied to $\Delta T/T_0$ at $6$~fm/$c$, the values of Sec.~\ref{sec:feedback} give $(8.35-8.28-3.79+3.64)\times10^{-4}=-0.08\times10^{-4}$: the magnetic increment is $0.15\times10^{-4}$ without shear and $0.07\times10^{-4}$ with shear.  The velocity slopes of the four histories are compared in Sec.~\ref{sec:observables}.

\subsection{Replacing the prescribed field by ideal induction}

Three-dimensional ideal magnetohydrodynamic simulations evolve the magnetic field together with the plasma and resolve the spatial dependence of the field and flow~\cite{InghiramiMHD,BhacQgp}. The comparison below keeps the initial field and matter profiles fixed and changes the field evolution law. It isolates the role of longitudinal magnetic work from that of magnetic pressure and inertia.

The right panel of Fig.~\ref{fig:newRange} compares two prescribed fields with a field evolved by ideal induction. Along each flow, $D(B^2/2)=-B^2\theta$ cancels the direct magnetic work in the energy equation. For a relaxing shear stress the matter balances become
\begin{align}
 DT+\frac{T\theta}{3}&=\frac{\Phi\theta}{12a_1T^3},&
 \nabla_\mu(su^\mu)&=\frac{\Phi\theta}{T}.
\end{align}
For instantaneous shear, $\Phi=4\eta_{\rm sh}\theta/3$ gives
\begin{align}
 DT+\frac{T\theta}{3}-\frac{4(\eta_{\rm sh}/s)}{9}\theta^2&=0,\\
 \nabla_\mu(su^\mu)&=\frac{4\eta_{\rm sh}}{3T}\theta^2.
 \label{eq:inductionMatterBalance}
\end{align}
Consider first an ideal fluid with a homogeneous initial transverse
field. At first order in the independent initial gradients, the central
slope obeys
\begin{equation}
 (\varepsilon+p+B^2)\frac{d\kappa_\Omega}{d\ln\tau}
 +\frac23(\varepsilon+p)(\kappa_\Omega-1-\chi)=0,
 \label{eq:inductionSlopeLinear}
\end{equation}
with coefficients evaluated on the Bjorken background. The magnetic stress relative to the enthalpy decays as
\[
 B^2/(\varepsilon+p)=(3B_0^2/4\varepsilon_0)\,x^{-2/3}.
\]
The pressure gradient sustains the slope $1+\chi$, while the magnetic
contribution $B^2$ to the inertia delays the loss of the initial
velocity excess. Integration gives
\begin{equation}
 \kappa_\Omega
 =1+\chi+(\lambda^*-\chi)\,
   \frac{4\varepsilon_0+3B_0^2}{4\varepsilon_0x^{2/3}+3B_0^2}.
 \label{eq:inductionSlopeClosed}
\end{equation}
The initial contribution $\lambda^*-\chi$ decreases by a factor $e$ at
$\tau=\tau_0[e+3(e-1)B_0^2/(4\varepsilon_0)]^{3/2}$, which is
$4.48\,\tau_0$ without a field and $4.80\,\tau_0$ for
$B_0^2/\varepsilon_0=0.1$.

Within this ideal-induction comparison, the given flow has
$T_p=T_0x^{-(1+\lambda^*)/3}$. The temperature ratio is therefore
\begin{align}
 \ln\frac{T_s}{T_p}
 ={}&\frac{\lambda^*-\chi}{2}
 \left\{\frac23\ln x\right.\nonumber\\
 &\left.\quad-\frac{4\varepsilon_0+3B_0^2}{3B_0^2}
 \ln\frac{x^{2/3}(4\varepsilon_0+3B_0^2)}
         {4\varepsilon_0x^{2/3}+3B_0^2}\right\}
 +O\big((\lambda^*)^2,\lambda^*\chi,\chi^2\big).
 \label{eq:inductionGivenTemperature}
\end{align}
Its zero-field limit gives Eq.~\eqref{eq:idealReferenceDifference}.
The reference in Eq.~\eqref{eq:inductionGivenTemperature} is the prescribed
accelerating flow, so the accumulated slope difference directly
determines the difference in expansion cooling.

For instantaneous NS shear the same projection gives
\begin{align}
 &(\varepsilon+p+B^2-\Phi)\frac{d\kappa_\Omega}{d\ln\tau}
 +\left[\frac23(\varepsilon+p)+\frac43\Phi
        -\frac{\Phi^2}{\varepsilon+p}\right](\kappa_\Omega-1)\nonumber\\
 &\qquad=\frac23\,(\varepsilon+p-3\Phi)\,
        \chi\,\frac{T_0x^{-1/3}}{\bgtemp},
 \label{eq:inductionNSSlope}
\end{align}
with the background values $\varepsilon+p=4a_1(\bgtemp)^4$,
$\Phi=4\eta_{\rm sh}/(3\tau)$, and $B^2=B_0^2x^{-2}$, where $\bgtemp$
is the NS Bjorken temperature in Eq.~\eqref{eq:baselineNS}. The
combination $\varepsilon+p+B^2-\Phi$ is the longitudinal inertia: the
field increases it and shear reduces it. The factor
$\chi T_0x^{-1/3}/\bgtemp$ is the thermal curvature at time $\tau$,
and the spatial dependence of $\eta_{\rm sh}\propto T^3$ changes the
pressure-gradient source. The nonlinear comparisons below retain the
shear-relaxation equation and the initial condition $\Phi_0=0$.
Under ideal induction, $D\ln B=-\theta$ cancels the direct magnetic work before the gradient expansion. Along $\chi=\lambda^*$, expanding the induction equation gives the field correction to the NS quadratures.  For a homogeneous initial transverse field,
\begin{align}
 B^2(\tau)={}&B_0^2\left(\frac{\tau_0}{\tau}\right)^2
 \bigg[1-2\lacc\int_{\tau_0}^{\tau}\frac{d\tau'}{\tau'}
 \left.\partial_{\etas}\flowresponse(\tau',\etas)\right|_0\bigg]
 \nonumber\\
 &+\mathcal O((\lacc)^2).
 \label{eq:inductionControl}
\end{align}
This result follows by substituting $D=\partial_\tau+\lambda^*\Omega^{(1)}\partial_{\eta_s}/\tau+\cdots$ and $\tau\theta=1+\lambda^*\partial_{\eta_s}\Omega^{(1)}+\cdots$ into $D\ln B^2=-2\theta$ and integrating from the common initial field.  The first-order correction is homogeneous because the first-order rapidity slope is homogeneous.  At second order, the field develops a constant and a quadratic-rapidity correction, which evolve together with the flow:
\begin{equation}
 B^2(\tau,\etas)
 =B_0^2(\tau_0/\tau)^2+\lacc(B^2)^{(1)}(\tau) +(\lacc)^2(B^2)^{(2)}(\tau,\etas)
       +\mathcal O((\lacc)^3),
\label{eq:inductionSecond}
\end{equation}
where $(B^2)^{(2)}$ contains a constant and an $\eta_s^2$ term, both initialized consistently with the homogeneous field.

Magnetic pressure and inertia continue to enter the momentum equation, allowing the field to influence the temperature indirectly through the flow.  The prescribed $a=1$ history coincides with ideal induction in Bjorken flow, while longitudinal acceleration causes the two field histories to diverge.  Each paired induction run starts from the same field, evolves it with its own velocity, and employs the same scalar shear-relaxation law, thereby isolating the effect of flux conservation.

Flux conservation also changes the curvature required for initial and terminal temperature cancellation. With zero initial shear, ideal induction gives
\begin{equation}
 \frac{\chi}{\lambda^*}=(1+\lambda^*)\left(1-\frac{2}{c_\pi}\right)
 \label{eq:inductionInitialCancellation}
\end{equation}
for a stationary initial slope, $[d\kappa_\Omega/d\tau]_{\tau_0}=0$. The homogeneous initial magnetic field
cancels from this ratio, while it remains in the inertia governing the
rate of slope evolution. Table~\ref{tab:closureCancellation} gives the
first-order and finite-amplitude onset conditions and the terminal roots.
The onset entries use this stationary-slope condition; the terminal
entries concern $T_s-T_p$ at $6$ fm/$c$.
\begin{table}[tbp]
\centering\small
\caption{Initial-gradient ratios $\chi/\lambda^*$ for onset and terminal
temperature cancellation, at the reference shear and field parameters.
The finite-amplitude onset follows the full initialization equations.}
\label{tab:closureCancellation}
\begin{tabular}{lccccc}
\toprule
History & $\lambda^*$ & \shortstack{Onset\\linear} & \shortstack{Onset\\finite}
 & \shortstack{Terminal\\linear} & \shortstack{Terminal\\nonlinear}\\
\midrule
$a=1.30$ & 0.03 & 0.6442 & 0.6664 & 1.1108 & 1.2014 \\
$a=1.30$ & 0.06 & 0.6442 & 0.6887 & 1.1108 & 1.2993 \\
Ideal induction & 0.03 & 0.6667 & 0.6867 & 1.1070 & 1.1973 \\
Ideal induction & 0.06 & 0.6667 & 0.7067 & 1.1070 & 1.2948 \\
\bottomrule
\end{tabular}
\end{table}

\subsection{Implications for observables}
\label{sec:observables}
Measured charged-meson rapidity distributions and harmonic-flow decorrelations constrain different aspects of the longitudinal structure~\cite{BrahmsRapidity,CmsDecorrelation,AtlasDecorrelation}.
Models generate longitudinal fluctuations through fluctuating entropy deposition from the participants~\cite{BozekTorque,BroniowskiRapidity} or through the rapidity evolution of the initial Glasma fields~\cite{SchenkeSchlichting}. Event-by-event calculations also distinguish effects from the initial fluctuations and fluctuations generated during hydrodynamic evolution~\cite{HydroRapidityFluct}.

Two flows with nearly equal central temperatures can distribute entropy differently in fluid rapidity. We first identify the matter entropy carried by a fluid element and then transform from spacetime rapidity to fluid rapidity on a constant-$\tau$ surface.
Using $d\varepsilon=T\,ds$ and $\varepsilon+p=Ts$, we obtain the local
matter entropy balance from the energy equation. In the Navier--Stokes
limit, the prescribed magnetic field gives
\begin{equation}
 \nabla_\mu(su^\mu)
 =\frac{4\eta_{\rm sh}}{3T}\theta^2
 +\frac{B^2}{T}
  \left[\frac{a\cosh(\Omega-\eta_s)}{\tau}-\theta\right].
 \label{eq:entropyBalance}
\end{equation}
The first term is the nonnegative entropy production due to shear
viscosity~\cite{Romatschke2010}. The second term is the rate of
magnetic work per unit proper volume divided by $T$. Its general
form, $-B^2(D\ln B+\theta)/T$, vanishes under ideal induction,
for which $D\ln B=-\theta$. A prescribed field can either supply
energy to the matter or extract it. When energy extraction exceeds
shear heating, the entropy of a comoving fluid element decreases.

Using $s\propto T^3$ and
$D(sV)=V\nabla_\mu(su^\mu)$, we obtain
\begin{equation}
 \frac{DT}{T}=-\frac{\theta}{3}
              +\frac{D(sV)}{3sV}.
 \label{eq:entropyCooling}
\end{equation}
The first term is the cooling at fixed entropy per element.  An increase
in $sV$ offsets this cooling and a decrease enhances it, so the element
cools whenever its fractional entropy growth is slower than its fractional
volume growth, $D\ln(sV)<D\ln V=\theta$.  The temperature thus reflects the
competition between dilution by expansion, shear dissipation, and energy
exchange with the prescribed field; the source integrals of
Sec.~\ref{sec:sourceDecomposition} quantify these contributions.

With the relaxation equation, the same identity holds with $\Phi\theta$ in
place of $4\eta_{\rm sh}\theta^2/3$.  Here $sV$ is the equilibrium entropy
of the element.  In second-order theories the full entropy current also
contains a term quadratic in $\Phi$. We use the equilibrium matter entropy density for the rapidity estimate below.

On a constant-proper-time hypersurface, the flux of the equilibrium matter entropy current gives
\begin{equation}
\frac{dS}{d\etas}\propto s\tau\cosh(\Omega-\etas).
\end{equation}
At the center, $\Omega=\etas=0$ and $dS/d\etas\propto s\tau$.  Transforming to fluid rapidity introduces the inverse slope:
\begin{align}
\left.\frac{dS}{d\Omega}\right|_0
&\propto\frac{s\tau}{\kappa_\Omega},&
\kappa_\Omega&=\left.\partial_{\etas}\Omega\right|_0,\\
\frac{(dS/d\Omega)_s}{(dS/d\Omega)_p}
&=\left(\frac{T_s}{T_p}\right)^3
\frac{\kappa_{\Omega,p}}{\kappa_{\Omega,s}}.
\label{eq:idealEntropyDifference}
\end{align}
A smaller slope therefore concentrates the same entropy within a narrower interval of fluid rapidity.  For the ideal weak-gradient solution, with $q=(\tau_0/\tau)^{2/3}$, the temperature and slope differences combine to give
\begin{align}
\frac{\Delta(dS/d\Omega)}{(dS/d\Omega)_p}
&\simeq3\frac{T_s-T_p}{T_{\rm Bj}}
-\frac{\kappa_{\Omega,s}-\kappa_{\Omega,p}}{\kappa_{\Omega,p}}\nonumber\\
&=(\lambda^*-\chi)\left[\ln\frac{\tau}{\tau_0}-\frac{1-q}{2}\right].
\end{align}
The temperature and slope contributions have the same sign, making the entropy estimate more sensitive than the temperature alone.  The change of variables requires a nonzero central slope.  A particle spectrum requires the phase-space distribution and its flux through a particlization hypersurface, as in the Cooper--Frye prescription~\cite{CooperFrye}. Viscous particlization additionally specifies how dissipative stresses enter that distribution~\cite{McNelisHeinz2021}.

\begin{figure}[tbp]
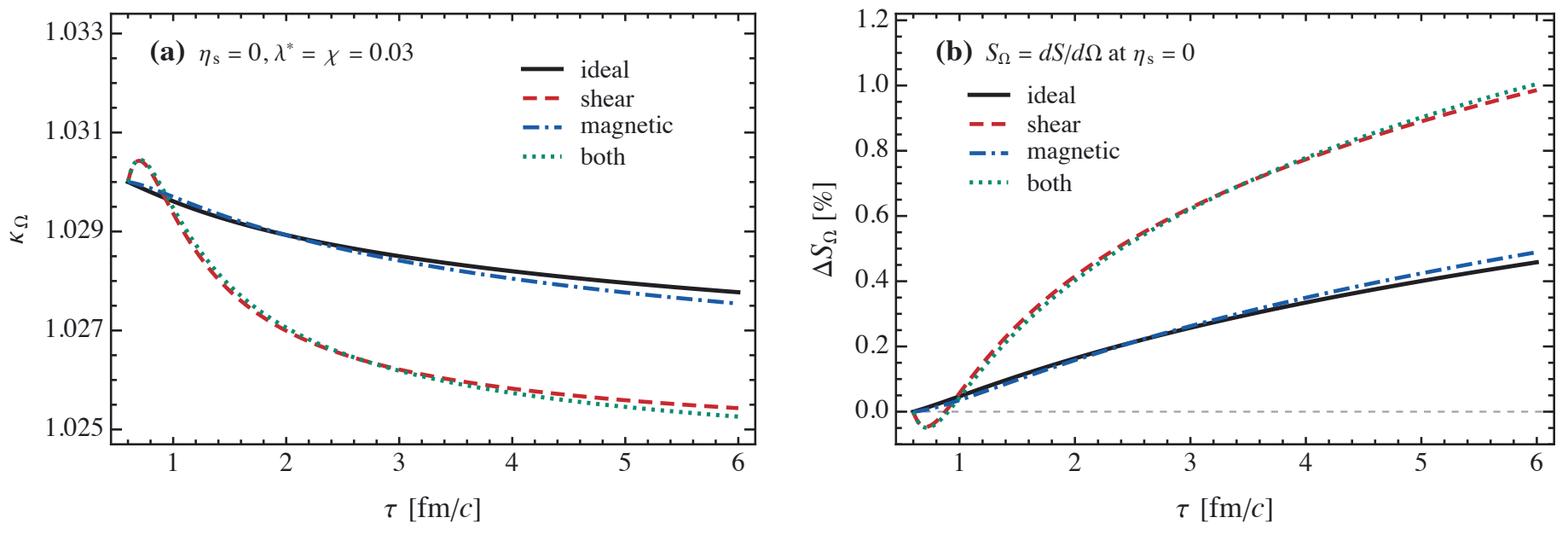

\centering\ReportFigureTwelve
\caption{Evolution at $\lambda^*=\chi=0.03$ for ideal, shear-only, magnetic-only, and combined stresses. (a) Central rapidity slope $\kappa_\Omega=\partial_{\etas}\Omega|_0$ of the self-consistent flow. (b) Relative change of the central entropy per unit fluid rapidity, $\Delta S_\Omega=100[(dS/d\Omega)_s/(dS/d\Omega)_p-1]$, from Eq.~\eqref{eq:idealEntropyDifference}.}
\label{fig:entropySlope}
\end{figure}

For orientation, Landau's width estimate $\sigma_\eta^2=\ln(\sqrt{s_{NN}}/2m_N)$~\cite{Steinberg2005} gives $\sigma_\eta\approx2.2$ at $200$~GeV and $2.7$--$2.8$ at $2.76$--$5.02$~TeV. If this width is assigned to a Gaussian initial energy density, $T\propto\exp[-\etas^2/(8\sigma_\eta^2)]$ gives $\chi=3/(8\sigma_\eta^2)\approx0.05$--$0.08$. This estimate motivates varying the curvature independently of the initial rapidity slope.
Table~1 of Ref.~\cite{Jiang2018Fits} reports $0.035\pm0.003$ and $0.032\pm0.002$ for Pb+Pb at $2.76$ and $5.02$ TeV. For Au+Au at $200$ GeV it reports $0.062\pm0.002$.
The acceleration parameters fitted in Ref.~\cite{Jiang2018Fits} correspond to a time-independent slope.  In the present model, we interpret them as initial slopes.  In the ideal case, the subsequent slope evolves toward $1+\chi$, with further modifications arising from shear and magnetic stress.  Figure~\ref{fig:entropySlope} shows the central rapidity slope and entropy per unit fluid rapidity for the four stress histories.  At $6$~fm/$c$, $\Delta S_\Omega$ is $0.46\%$ for the ideal fluid and $1.0\%$ with both stresses.  For $\lambda^*=\chi$, the first-order ideal contribution vanishes, making the ideal result a finite-amplitude effect.  Shear approximately doubles this contribution, while the reference power-law field adds $0.02$ percentage points. The ideal-induction value is $0.96\%$, reflecting the different velocity history.

\begin{table}[tbp]
\caption{Illustrative evolution using the acceleration parameters in Table~1 of Ref.~\cite{Jiang2018Fits} as initial values $\lambda^*$. The Xe+Xe value follows from the energy dependence in Eq.~(21) of that reference. At $\tau=6$ fm/$c$, we show the central slope excess $100(\kappa_\Omega-1)$ and the relative entropy estimate $100[(dS/d\Omega)_s/(dS/d\Omega)_p-1]$ for two initial curvatures. Other parameters are those of Eq.~\eqref{eq:referenceParameters}.}
\label{tab:fitInitialGradients}
\centering\small
\begin{tabular}{lccccc}
\toprule
System, $\sqrt{s_{NN}}$ [GeV] & $\lambda^*$ & \multicolumn{2}{c}{$\chi=\lambda^*$} & \multicolumn{2}{c}{$\chi=0$} \\
\cmidrule(lr){3-4}\cmidrule(lr){5-6}
 & & $100(\kappa_\Omega-1)$ & $\Delta S_\Omega$ [\%] & $100(\kappa_\Omega-1)$ & $\Delta S_\Omega$ [\%] \\
\midrule
Au+Au, 130 & 0.076 & 5.69 & 4.03 & 1.36 & 15.54 \\
Au+Au, 200 & 0.062 & 4.81 & 2.93 & 1.15 & 12.44 \\
Cu+Cu, 200 & 0.060 & 4.67 & 2.79 & 1.11 & 12.01 \\
Pb+Pb, 2760 & 0.035 & 2.91 & 1.25 & 0.69 & 6.78 \\
Pb+Pb, 5020 & 0.032 & 2.68 & 1.10 & 0.63 & 6.17 \\
Xe+Xe, 5440 & 0.030 & 2.53 & 1.00 & 0.60 & 5.77 \\
\bottomrule
\end{tabular}
\end{table}

At a common initial acceleration, Table~\ref{tab:fitInitialGradients} gives a larger late entropy difference for $\chi=0$ than for $\chi=\lacc$. For example, the Pb+Pb entry at $5.02$~TeV gives $6.17\%$ and $1.10\%$, respectively. All systems use the same reference thermodynamic parameters. The same Pb+Pb entry with $\chi=0$ also separates the two factors in Eq.~\eqref{eq:idealEntropyDifference}. The tabulated slopes give $\kappa_{\Omega,p}/\kappa_{\Omega,s}\simeq1.032/1.0063\simeq1.026$. Dividing the tabulated entropy ratio $1.0617$ by this factor gives $(T_s/T_p)^3\simeq1.035$, or $T_s/T_p\simeq1.012$. These estimates use the rounded entries at the same proper time and initial curvature. The slope mapping and the thermal density therefore both contribute to the entropy enhancement in this same initial state.

\section{Conclusions}
\label{sec:discussionNew}
Near midrapidity, a prescribed rapidity slope requires a specific temperature curvature $\chi_{\rm req}$ to satisfy longitudinal momentum conservation. In an ideal fluid with weak gradients, the self-consistent slope $\kappa_\Omega$ moves from $1+\lambda^*$ to $1+\chi$; for $\chi=\lambda^*$, this reproduces the constant first-order slope of Ref.~\cite{Jiang2020}. At finite gradient amplitudes, however, $\chi_p$ and $\chi_{\rm req}$ diverge even without viscosity or magnetic fields. In the inviscid prescribed-flow limit, we obtain the magnetic accelerating solution of Ref.~\cite{She2019}. In the boost-invariant Navier--Stokes limit, the system reduces to the viscous magnetic cooling equation of Ref.~\cite{Jiang2025}.

When the two flows are evolved from identical initial data, a persistent local momentum residual can occur alongside a small central temperature difference, since central cooling depends on integrated expansion work. At first order in the ideal limit, $\Delta T/T_{\mathrm{Bj}}$ grows logarithmically at late times in proportion to $\lambda^{*}-\chi$. With the reference shear and prescribed magnetic stresses, the maximum central temperature error is $0.18\%$ for $\chi=\lambda^{*}=0.03$ and $1.08\%$ for $\lambda^{*}=0.03$ with $\chi=0$. At $\tau=6~\mathrm{fm}/c$, equal central temperatures are reached at $\chi/\lambda^{*}\simeq 1.11$ to first order and around $1.20$ in the nonlinear run at $\lambda^{*}=0.03$. A nonzero initial shear stress shifts this ratio further. For $a=1.30$, the prescribed field raises the final self-consistent central temperature by $1.48~\mathrm{MeV}$ but increases $T_s-T_p$ by only $4.59~\mathrm{keV}$, as most of the direct magnetic work cancels between the two histories. In the ideal-induction closure, direct magnetic work drops out of the energy equation, and magnetic pressure and inertia act only through longitudinal momentum balance.

At midrapidity, the entropy per unit fluid rapidity scales as $dS/d\Omega \propto s\tau/\kappa_{\Omega}$ and thus depends on the velocity slope directly. Taking the fitted acceleration parameters as initial slopes with a flat initial temperature ($\chi=0$) gives a $6\text{--}16\%$ change in $dS/d\Omega$ at $\tau=6$~fm/$c$.

\paragraph{Outlook.} Dissipative spin hydrodynamics describes the relaxation of spin density~\cite{HattoriSpin}, with analytic solutions available for Bjorken expansion~\cite{WangBjorkenSpin}. The self-consistent temperature, velocity, and magnetic-field evolutions obtained here provide longitudinal backgrounds for extending this comparison to finite gradients using relativistic spin magnetohydrodynamics~\cite{BhadurySpinMHD,FangSpinMHD}. Starting from the same small spin perturbation in the prescribed and self-consistent flows would allow direct evaluation of how their expansion rates, cooling histories, and field closures affect spin transport.

\acknowledgments We used AI-assisted tools (ChatGPT, Claude, and Gemini) to revise English grammar and improve the logical flow of the manuscript, and to assist in reviewing the Mathematica code used for the calculations. The authors take responsibility for the final manuscript and its results. \par\appendix

\section{Numerical method and validation}
\label{sec:numericalMethod}
The nonlinear calculation evolves $(T/T_0)^4$, $\Omega-\eta_s$, and $\Phi/\varepsilon_0$ on $|\eta_s|\le3$, using fourth-order spatial differences and RK4 time stepping in $\xi=\ln(\tau/\tau_0)$.  Five-point one-sided derivatives close the outer boundary, and physical comparisons are restricted to $|\eta_s|\le1$.  The reference grid spacing is $0.05$, and the maximum logarithmic time step is $0.002$.  Under ideal induction, $B^2/\varepsilon_0$ is evolved as a fourth field.  The coefficient predictions employ polynomial arithmetic through second order in the initial gradients, with the energy-to-temperature reconstruction retained at that order.

\subsection{Method of lines and parity conditions}
After discretization in rapidity, the nonlinear evolution reduces to a coupled system of ordinary differential equations in $\xi=\ln(\tau/\tau_0)$, with $\partial_\tau=\tau^{-1}\partial_\xi$.  The state vector contains the matter energy, flow rapidity, and shear stress, with $B^2$ included as an additional variable in the ideal-induction calculation.  The temperature is reconstructed as $T=(\varepsilon/3a_1)^{1/4}$, and the prescribed magnetic history is evaluated at each integration stage.

The spatial formulas below are the uniform-grid case of finite-difference weights obtained by polynomial interpolation~\cite{FornbergStencils}.

At interior grid points, the first rapidity derivative is approximated by the fourth-order centered stencil
\begin{equation}
(D_1 f)_i=\frac{f_{i-2}-8f_{i-1}+8f_{i+1}-f_{i+2}}{12h},
\end{equation}
and the second derivative uses
\begin{equation}
(D_2 f)_i=\frac{-f_{i-2}+16f_{i-1}-30f_i+16f_{i+1}-f_{i+2}}{12h^2}.
\end{equation}
The outer boundary uses one-sided five-point formulas.  Reflection symmetry makes the temperature and shear even and $\Omega-\eta_s$ odd, which enforces $\partial_{\eta_s}T=\partial_{\eta_s}\Phi=0$ and $\Omega-\eta_s=0$ at the center; the grid contains the center, so no interpolation is needed there.

At each Runge--Kutta stage, we evaluate the expansion scalar and the shear source from the same intermediate fields.  The prescribed-flow run uses the same energy and shear equations, with $\Omega_p=(1+\lambda^*)\eta_s$ entering the kinematic coefficients, so the two runs differ only in the momentum constraint.

\subsection{Conservative form and independent checks}
For the conservation check, the Milne components are expressed in terms of the temperature, magnetic field, and longitudinal shear:
\begin{align}
 T^{\tau\tau}&=(4a_1T^4+B^2-\Phi)\cosh^2(\Omega-\etas) -a_1T^4-B^2/2+\Phi, \\
 \tau T^{\tau\etas}&=(4a_1T^4+B^2-\Phi) \times\cosh(\Omega-\etas)\sinh(\Omega-\etas), \\
 \tau^2T^{\etas\etas}&=(4a_1T^4+B^2-\Phi)\sinh^2(\Omega-\etas) +a_1T^4+B^2/2-\Phi.
\end{align}
These components represent the energy density, longitudinal momentum density, and longitudinal stress in Milne coordinates.  Covariant conservation then gives
\begin{equation}
\begin{aligned}
 &\partial_\tau(\tau T^{\tau\tau})+\partial_{\etas}(\tau T^{\tau\etas})
       +\tau^2T^{\etas\etas}=0,\\
 &\partial_\tau(\tau^3T^{\tau\etas})
       +\partial_{\etas}(\tau^3T^{\etas\etas})=0.
\end{aligned}
\label{eq:conservativeCheck}
\end{equation}
To test a nonvanishing momentum budget, we integrate over $0\le\etas\le1$:
\begin{align}
 \frac{d}{d\tau}\int_0^1\tau T^{\tau\tau}\,d\etas
 &=-\tau[T^{\tau\etas}]_0^1 -\int_0^1\tau^2T^{\etas\etas}\,d\etas, \\
 \frac{d}{d\tau}\int_0^1\tau^3T^{\tau\etas}\,d\etas
 &=-\tau^3[T^{\etas\etas}]_0^1.
\end{align}
The brackets denote the difference between the outer and central fluxes.  We use the half interval because reflection symmetry makes the total momentum vanish over the symmetric interval, and we retain the geometric energy source and all boundary fluxes.  These relations define the conservation budgets independently of the evolution stencil; Table~\ref{tab:numericalChecks} reports the corresponding residuals at the reference parameters.

\subsection{Convergence diagnostics and coefficient comparison}
The evolution uses fourth-order spatial differences. We evaluate the conservation residuals with centered second-order differences. Halving $h$ and $\Delta\xi$ together reduces both residuals by approximately a factor of four (Table~\ref{tab:numericalChecks}).

The rapidity domain extends beyond the interval used for physical comparisons, keeping the reported central region separated from the outer boundary.  Runs are compared over the same physical interval and at identical sampling times.  Varying the boundary position quantifies sensitivity to the finite domain, varying the time step measures temporal convergence, and the spatial refinement sequence in Table~\ref{tab:numericalChecks} measures spatial convergence.

We integrate the gradient equations and the nonlinear PDE separately. Their difference in $\Delta T$ gives the truncation error reported in Table~\ref{tab:predictionSummary}.

Table~\ref{tab:numericalChecks} gives the energy and momentum residuals at each resolution for the reference state. The residuals are normalized by the local $4a_1T^4+B^2$, and temporal endpoints are excluded from the centered residual norms.

For the prescribed magnetic field, the squared local characteristic
speed of longitudinal disturbances in the fluid rest frame follows from
the principal part of Eqs.~\eqref{eq:relaxE}, \eqref{eq:relaxM}, and
\eqref{eq:relaxPhi}:
\begin{equation}
 v_{\rm char}^2=\frac{1}{3}
 +\frac{4\eta_{\rm sh}/(3\tau_\pi)-4\Phi/3}
        {\varepsilon+p+B^2-\Phi}.
\end{equation}
In the reference matched run, $T>0.300$~GeV, the longitudinal inertia
$\varepsilon+p+B^2-\Phi$ stays positive, and
$0.502<v_{\rm char}^2<0.541$. The longitudinal characteristics are
therefore real and subluminal throughout the sampled evolution, satisfying
longitudinal causality for this run.

Under ideal induction,
Eq.~\eqref{eq:inductionMatterBalance} gives the longitudinal principal
part in a local fluid rest frame. With $z$ the local longitudinal
coordinate and $v_z$ the longitudinal velocity, the terms containing
derivatives of the perturbations are
\begin{align*}
 \partial_t\delta\varepsilon
   +(\varepsilon+p-\Phi)\partial_z\delta v_z &=0,\\
 (\varepsilon+p+B^2-\Phi)\partial_t\delta v_z
   +\partial_z(\delta\varepsilon/3+\delta B^2/2-\delta\Phi)&=0,\\
 \partial_t\delta\Phi
   -\left(\frac{4\eta_{\rm sh}}{3\tau_\pi}-\frac43\Phi\right)
      \partial_z\delta v_z &=0,\\
 \partial_t\delta B^2+2B^2\partial_z\delta v_z &=0.
\end{align*}
Here $\delta B^2$ denotes the perturbation of $B^2$. The first, third,
and fourth equations express the stress perturbations in terms of
$\delta v_z$ for a mode of nonzero speed. Substitution in the
longitudinal momentum equation gives
\begin{equation*}
 \left.v_{\rm char}^2\right|_{\rm ind}
 =\frac13+
 \frac{4\eta_{\rm sh}/(3\tau_\pi)-4\Phi/3+2B^2/3}
      {\varepsilon+p+B^2-\Phi}.
\end{equation*}
The remaining two characteristics have zero speed in this frame.
For the self-consistent reference induction run ($\lambda^*=\chi=0.03$)
over $|\eta_s|\le1$ and $0.6\le\tau\le6$~fm/$c$, the sampled
inertia has a minimum of $0.0596\,\varepsilon_0$ and the sampled
squared speeds lie between $0.5341$ and $0.5877$.

Halving the evolution time step alone changes the terminal temperature difference by $3.3\times10^{-12}$~MeV, while extending the boundary to $|\eta_s|=3.5$ changes it by $2.4\times10^{-11}$~MeV.  The numerical quadrature of the thermal budget closes to $3.8\times10^{-6}$~MeV using 481 output times.  These numerical errors are much smaller than the truncation errors of the initial-gradient expansion reported in Table~\ref{tab:predictionSummary} and represent a separate source of uncertainty.

Several analytic identities of the main text are also checked
numerically.  Along the same prescribed-flow trajectories, with the
nonlinear shear curvature $C_\Phi$ in both expressions,
Eqs.~\eqref{eq:residualExactStart} and \eqref{eq:prescribedResidual} give
values of $\tau\mathcal R$ that differ by at most $3.4\times10^{-11}$ in
units of $\tfrac83a_1T_p^4\lambda^*$ for all four stress cases.  The
finite-amplitude initialization identity, Eq.~\eqref{eq:generalShearOnset},
agrees with the unexpanded initialization equations within
$1.7\times10^{-12}$ in the dimensionless slope derivative at
$\lambda^*=\chi=0.03$ for $b_3=0$, $0.1$, and $0.5$.  The NS quadratures
agree with an independent finite-difference evolution of the linearized NS
equations within $1.7\times10^{-8}$ in the normalized central temperature,
and the source decomposition of Eq.~\eqref{eq:feedbackAttribution}
reproduces the nonlinear temperature difference within
$3.8\times10^{-6}$~MeV.

\begin{table}[tbp]
\caption{Resolution and independent conservation checks at the reference matched state. $N_\xi$ is the number of diagnostic output times; $\Delta\xi$ is the maximum evolution step. Residuals are maxima over $|\eta_s|\le1$ and interior output times.}
\label{tab:numericalChecks}
\centering\small
\begin{tabular}{ccccc}
\toprule
$\Delta\eta_s$&$\Delta\xi$&$N_\xi$&Energy&Momentum\\
\midrule
0.10&0.0040&241&$7.00\times10^{-5}$&$5.54\times10^{-6}$\\
0.050&0.0020&481&$1.76\times10^{-5}$&$1.41\times10^{-6}$\\
0.025&0.0010&961&$4.42\times10^{-6}$&$3.55\times10^{-7}$\\
\bottomrule
\end{tabular}\end{table}

For the reference parameters, Table~\ref{tab:terminalRefinement} reports
the joint refinement of the two terminal quantities used in Sec.~\ref{sec:comparison}.
We locate the nonlinear root by a secant iteration initialized at
$\chi/\lambda^*=1.1$ and $1.3$, stopping with
$|\Delta T(6\,\mathrm{fm}/c)|<10^{-8}$~MeV.

\begin{table}[tbp]
\centering\small
\caption{Joint refinement of the central magnetic increment in the
  paired temperature difference and of the nonlinear central
  equal-temperature ratio at $\tau=6$~fm/$c$. The increment uses
  $\chi=\lambda^*=0.03$ and is measured relative to the same pair with
  $B_0=0$. The root uses $\lambda^*=0.03$ and the reference prescribed
  field. $\Delta\xi$ is the actual uniform step.}
\label{tab:terminalRefinement}
\begin{tabular}{cccc}
\toprule
$\Delta\eta_s$ & $\Delta\xi$ &
$(T_s-T_p)_{B}-(T_s-T_p)_{B=0}$ [keV] &
$\chi/\lambda^*$ at $T_s=T_p$\\
\midrule
0.100 & 0.00383764 & 4.589678 & 1.20144757 \\
0.050 & 0.00191882 & 4.589693 & 1.20144736 \\
0.025 & 0.00095941 & 4.589694 & 1.20144735 \\
\bottomrule
\end{tabular}
\end{table}

\subsection{Initial-gradient expansion with shear relaxation}
\label{app:relaxationPrediction}

We expand the energy, momentum, and shear equations through second order in the initial gradients, with finite transport coefficients.  To track the common amplitude, we replace $(\lacc,\chi)$ by $(\rho\lacc,\rho\chi)$, retain the required powers of $\rho$, and set $\rho=1$ at the end.  The fields are expanded as
\begin{equation}
\begin{aligned}
 T&=T^{(0)}+\rho T^{(1)}+\rho^2T^{(2)}+O(\rho^3),\\
 \Omega&=\etas+\rho\Omega^{(1)}+\rho^2\Omega^{(2)}+O(\rho^3),\\
 \Phi&=\Phi^{(0)}+\rho\Phi^{(1)}+\rho^2\Phi^{(2)}+O(\rho^3).
\end{aligned}
\label{eq:relaxationPredictionExpansion}
\end{equation}
The background is independent of rapidity.  At first order, the temperature and shear are even polynomials through $\etas^2$, while the rapidity is linear.  At second order, the temperature and shear contain even powers through $\etas^4$, and the rapidity contains odd powers through $\etas^3$.  For the specified initial data, these powers form a closed polynomial sector at each retained order.

The background is the magnetic--viscous Bjorken solution of Sec.~\ref{sec:baselines}, expressed in the expansion notation.  Its temperature and shear stress satisfy
\begin{equation}
\begin{aligned}
 &\frac{dT^{(0)}}{d\tau}+\frac{T^{(0)}}{3\tau}
 -\frac{\Phi^{(0)}+(a-1)B^2}{12a_1\tau[T^{(0)}]^3}=0,\\
 &\frac{d\Phi^{(0)}}{d\tau}
 +\frac{T^{(0)}\Phi^{(0)}}{c_\pi(\eta_{\rm sh}/s)}
 +\frac{4\Phi^{(0)}}{3\tau}
 -\frac{16a_1[T^{(0)}]^4}{3c_\pi\tau}=0.
\end{aligned}
\label{eq:relaxationPredictionBackground}
\end{equation}
At first order, the equations couple changes in the temperature profile to the pressure-driven flow and the associated additional expansion work.  The energy equation reads
\begin{equation}
\begin{aligned}
 &\partial_\tau T^{(1)}+\frac{T^{(1)}}{3\tau}
  +\frac{T^{(0)}\partial_{\etas}\Omega^{(1)}}{3\tau}\\
 &-\frac{\Phi^{(1)}+\Phi^{(0)}\partial_{\etas}\Omega^{(1)}
                 -3\Phi^{(0)}T^{(1)}/T^{(0)}}
             {12a_1\tau[T^{(0)}]^3}\\
 &+\frac{B^2}{12a_1\tau[T^{(0)}]^3}
  \left[\partial_{\etas}\Omega^{(1)}
             +3(a-1)\frac{T^{(1)}}{T^{(0)}}\right]=0.
\end{aligned}
\label{eq:relaxationPredictionEnergy1}
\end{equation}
The first-order momentum equation is
\begin{equation}
\begin{aligned}
 &[4a_1(T^{(0)})^4+B^2-\Phi^{(0)}]\partial_\tau\Omega^{(1)} \\
 &+\Bigg[\frac{4a_1(T^{(0)})^4+(1-a)B^2-\Phi^{(0)}}{\tau} +4a_1(T^{(0)})^3\frac{dT^{(0)}}{d\tau}
                     -\frac{d\Phi^{(0)}}{d\tau}\Bigg]\Omega^{(1)} \\
 &+\frac{4a_1(T^{(0)})^3\partial_{\etas}T^{(1)}
                       -\partial_{\etas}\Phi^{(1)}}{\tau}=0.
\end{aligned}
\label{eq:relaxationPredictionMomentum1}
\end{equation}
The first-order shear equation is
\begin{equation}
\begin{aligned}
 &\partial_\tau\Phi^{(1)}
 +\frac{T^{(0)}\Phi^{(1)}+\Phi^{(0)}T^{(1)}}{c_\pi(\eta_{\rm sh}/s)}\\
 &+\frac{4\Phi^{(1)}}{3\tau}
 -\frac{64a_1(T^{(0)})^3T^{(1)}}{3c_\pi\tau}\\
 &+\left[\frac{4\Phi^{(0)}}{3}
       -\frac{16a_1(T^{(0)})^4}{3c_\pi}\right]
       \frac{\partial_{\etas}\Omega^{(1)}}{\tau}=0.
\end{aligned}
\label{eq:relaxationPredictionShear1}
\end{equation}
Equating coefficients of the independent rapidity powers yields five coupled first-order ODEs.  For the reference initialization $b_3=0$, the initial conditions are
\begin{equation}
\begin{aligned}
 T^{(0)}(\tau_0)&=T_0,& \Phi^{(0)}(\tau_0)&=0,\\
 T^{(1)}(\tau_0,\etas)&=-T_0\chi\etas^2/3,&
 \Omega^{(1)}(\tau_0,\etas)&=\lacc\etas,\\
 \Phi^{(1)}(\tau_0,\etas)&=0.&
\end{aligned}
\end{equation}

\subsubsection{Quadratic sources and initial values}
At second order, substituting Eq.~\eqref{eq:relaxationPredictionExpansion} into Eqs.~\eqref{eq:relaxE}--\eqref{eq:relaxPhi} and collecting the coefficients of $\rho^2\etas^j$ gives the quadratic equations. In particular,
\begin{equation}
\begin{aligned}
 \cosh(\Omega-\etas)&=1+\tfrac12\rho^2(\Omega^{(1)})^2+O(\rho^3), \\
 \sinh(\Omega-\etas)&=\rho\Omega^{(1)}+\rho^2\Omega^{(2)}+O(\rho^3), \\
 T^{-3}&=[T^{(0)}]^{-3}\Bigg[1-3\rho\frac{T^{(1)}}{T^{(0)}} +\rho^2\left(6\frac{[T^{(1)}]^2}{[T^{(0)}]^2}
                  -3\frac{T^{(2)}}{T^{(0)}}\right)\Bigg]+O(\rho^3).
\end{aligned}
\label{eq:relaxationPredictionProducts}
\end{equation}
These quadratic products generate the finite-amplitude shift of the cancellation condition reported in Table~\ref{tab:finiteCancellation}.  The eight second-order coefficient equations are driven by the previously computed lower-order fields.  For the polynomial temperature initial condition and zero initial shear, all second-order initial coefficients vanish.  When expressed in terms of $3a_1T^4$, the quadratic energy coefficients are nonzero, and reconstruction of the temperature requires expanding the fourth root to the same order.  This yields the equivalent quadratic energy coefficients used in the calculation.  The self-consistent prediction therefore involves $2+5+8=15$ ODEs.  For the prescribed-flow prediction, we set $\Omega^{(1)}=\lacc\etas$ and $\Omega^{(2)}=0$, omit the momentum projection, and solve the remaining $2+4+6=12$ ODEs with the same initial temperature and shear.  The temperatures are subtracted after consistent reconstruction.  Under ideal induction, $B^2$ is also expanded.  Its background, uniform first-order term, and second-order constant and $\etas^2$ terms introduce four additional equations, yielding nineteen self-consistent and sixteen prescribed ODEs.  The relation $DB^2=-2B^2\theta$ then replaces the prescribed field history in the energy and momentum equations.  The hierarchy retains the pressure--motion coupling that determines the central cooling.

Table~\ref{tab:predictionSummary} gives the first- and second-order errors in $\Delta T$ relative to the nonlinear PDE solution.  Second order reduces the absolute error in every displayed state, while the nonlinear solution serves as the reference at larger amplitudes.  The same procedure applies to an equal-curvature Gaussian profile or a different relaxation time, using the corresponding initial coefficients and shear law.

\subsection{Higher-gradient structure and temperature reconstruction}
\label{sec:higherOmega}
Reflection symmetry makes the temperature and shear even and the rapidity odd in $\etas$, so locally
\begin{equation}
 \Omega(\tau,\etas)=\left.\partial_{\etas}\Omega\right|_0\etas
 +\frac16\left.\partial_{\etas}^3\Omega\right|_0\etas^3
 +O(\etas^5).
 \label{eq:rapidityTaylor}
\end{equation}
The cubic coefficient vanishes initially but is generated by nonlinear products of the temperature gradient and rapidity correction.  At second order in the initial gradients, the temperature and shear contain even powers through $\eta_s^4$, while the rapidity contains odd powers through $\eta_s^3$.  Third order extends these expansions to $\eta_s^6$ and $\eta_s^5$, respectively.  Powers of $\eta_s$ characterize the local rapidity profile, while the superscript denotes the total degree in the two initial gradients.

We extend Eq.~\eqref{eq:relaxationPredictionExpansion} by adding $\rho^3T^{(3)}$, $\rho^3\Omega^{(3)}$, and $\rho^3\Phi^{(3)}$.  For the polynomial initial temperature and linear initial rapidity, the second- and third-order temperature and rapidity coefficients vanish initially.  The energy density, however, has nonzero higher-order initial coefficients because of its nonlinear dependence on temperature:
\begin{equation}
\begin{aligned}
 \varepsilon^{(0)}={}&3a_1(\bgtemp)^4,\\
 \varepsilon^{(1)}={}&12a_1(\bgtemp)^3\tempresponse,\\
 \varepsilon^{(2)}={}&12a_1(\bgtemp)^3T^{(2)}
          +18a_1(\bgtemp)^2(\tempresponse)^2,\\
 \varepsilon^{(3)}={}&12a_1(\bgtemp)^3T^{(3)}
          +36a_1(\bgtemp)^2\tempresponse T^{(2)}\\
 &+12a_1\bgtemp(\tempresponse)^3.
\end{aligned}
\label{eq:temperatureCoefficientMap}
\end{equation}
The superscripts follow the Taylor convention in the common amplitude: at $\rho=0$, $T^{(2)}$ equals half the second derivative, and $T^{(3)}$ equals one sixth of the third derivative.  Products of the first-order temperature profile generate the higher-order energy coefficients.  The fourth-root reconstruction and the expansions of $T^3$, $T^{-3}$, and the hyperbolic velocity factors are truncated at the same degree.  The inverse reconstruction is
\begin{equation}
\begin{aligned}
 T^{(2)}&=\frac{\varepsilon^{(2)}}{12a_1(\bgtemp)^3}
                  -\frac{3[T^{(1)}]^2}{2\bgtemp},\\
 T^{(3)}&=\frac{\varepsilon^{(3)}}{12a_1(\bgtemp)^3}
     -\frac{3T^{(1)}T^{(2)}}{\bgtemp}
     -\frac{[T^{(1)}]^3}{(\bgtemp)^2}.
\end{aligned}
\label{eq:inverseTemperatureMap}
\end{equation}
In addition to the quadratic terms above, the cubic coefficient equations contain
\begin{equation}
\begin{aligned}
 \cosh(\Omega-\eta_s)&=1+\frac{\rho^2}{2}[\Omega^{(1)}]^2
                +\rho^3\Omega^{(1)}\Omega^{(2)}+O(\rho^4), \\
 \sinh(\Omega-\eta_s)&=\rho\Omega^{(1)}+\rho^2\Omega^{(2)} +\rho^3\left(\Omega^{(3)}
              +\frac{[\Omega^{(1)}]^3}{6}\right)+O(\rho^4).
\end{aligned}
\label{eq:thirdHyperbolicProducts}
\end{equation}
In particular, the inverse-temperature source is
\begin{equation}
\begin{aligned}
 \frac{(\bgtemp)^3}{T^3}={}&1-3\rho\frac{T^{(1)}}{\bgtemp} +\rho^2\left(6\frac{[T^{(1)}]^2}{(\bgtemp)^2}
                    -3\frac{T^{(2)}}{\bgtemp}\right) \\
 &+\rho^3\Biggl[-3\frac{T^{(3)}}{\bgtemp}
       +12\frac{T^{(1)}T^{(2)}}{(\bgtemp)^2} -10\frac{[T^{(1)}]^3}{(\bgtemp)^3}\Biggr]
 +O(\rho^4).
\end{aligned}
\label{eq:thirdInverseTemperature}
\end{equation}
Substituting these products into the energy, momentum, and shear equations and equating the coefficients of each $\rho^3\eta_s^j$ term determines the cubic sources.

For NS shear, the parity sector contains six energy-density and three rapidity coefficients through quadratic order.  Cubic order adds four energy-density and three rapidity coefficients, yielding nine- and sixteen-equation NS hierarchies.  Because the instantaneous NS stress depends on the velocity gradient, the momentum projection contains higher time derivatives of lower-order rapidity coefficients.  We eliminate these derivatives by perturbative order reduction, differentiating the previously solved lower-order equations.  The expansion orders count powers of the initial gradients, while the shear constitutive law remains first-order NS.

The third-order numerical test retains the scalar shear-relaxation law.  The third-order sector adds four temperature, four shear, and three rapidity coefficients to the quadratic hierarchy, yielding $2+5+8+11=26$ equations for the self-consistent flow and $2+4+6+8=20$ for the prescribed flow.  The lower-order solutions provide the sources, and the third-order coefficients enter linearly.  At $\rho=1$,
\begin{equation}
 T_{\mathrm{order}\,2}=T^{(0)}+T^{(1)}+T^{(2)},\qquad
 T_{\mathrm{order}\,3}=T_{\mathrm{order}\,2}+T^{(3)}.
\end{equation}

\begin{table}[tbp]
\caption{Maximum absolute error in the central cooling difference on 121 logarithmic output times, in MeV, at the reference relaxation parameters and $\chi=\lambda^*$.}
\label{tab:thirdOrderRestore}
\centering\small
\begin{tabular}{ccc}
\toprule
$\lambda^*$&Second order&Third order\\
\midrule
0.03&0.0121&0.000430\\
0.06&0.0932&0.00671\\
\bottomrule
\end{tabular}\end{table}

Third order reduces the error in $\Delta T$ by a factor of approximately $28$ at $\lambda^*=0.03$ and $14$ at $0.06$ (Table~\ref{tab:thirdOrderRestore}). The lower-order coefficients obtained from the quadratic and cubic integrations agree within $2.3\times10^{-13}$~MeV in the central cooling difference. Doubling the amplitude increases the third-order truncation error by a factor of approximately $15.7$, in agreement with the $2^4$ scaling of a leading fourth-order remainder.

\bibliographystyle{JHEP}
\bibliography{main8_references}

\providecommand{\href}[2]{#2}\begingroup\raggedright\begin{thebibliography}{10}

\bibitem{Kharzeev2008}
D.~E. Kharzeev, L.~D. McLerran, and H.~J. Warringa, {\it The effects of topological charge change in heavy ion collisions: "event by event {P} and {CP} violation"},  {\em Nucl. Phys. A} {\bf 803} (2008) 227--253, [\href{http://arxiv.org/abs/0711.0950}{{\tt arXiv:0711.0950}}].

\bibitem{DengHuang2012}
W.-T. Deng and X.-G. Huang, {\it Event-by-event generation of electromagnetic fields in heavy-ion collisions},  {\em Phys. Rev. C} {\bf 85} (2012) 044907, [\href{http://arxiv.org/abs/1201.5108}{{\tt arXiv:1201.5108}}].

\bibitem{Jing2026}
J.~Jing, D.~She, and Z.-F. Jiang, {\it Magnetization and magnetic field-induced correction: Implications for {QGP} thermal photon production in magnetohydrodynamic},  {\em Phys. Rev. D} {\bf 114} (2026) 056002, [\href{http://arxiv.org/abs/2608.13364}{{\tt arXiv:2608.13364}}].

\bibitem{Bjorken1983}
J.~D. Bjorken, {\it Highly relativistic nucleus-nucleus collisions: The central rapidity region},  {\em Phys. Rev. D} {\bf 27} (1983) 140--151. \href{https://sites.ohio.edu/frantz/phys728/bjorken.pdf}{Public full text}.

\bibitem{Roy2015}
V.~Roy, S.~Pu, L.~Rezzolla, and D.~H. Rischke, {\it Analytic {Bjorken} flow in one-dimensional relativistic magnetohydrodynamics},  {\em Phys. Lett. B} {\bf 750} (2015) 45--52, [\href{http://arxiv.org/abs/1506.06620}{{\tt arXiv:1506.06620}}].

\bibitem{Pu2016}
S.~Pu, V.~Roy, L.~Rezzolla, and D.~H. Rischke, {\it Bjorken flow in one-dimensional relativistic magnetohydrodynamics with magnetization},  {\em Phys. Rev. D} {\bf 93} (2016) 074022, [\href{http://arxiv.org/abs/1602.04953}{{\tt arXiv:1602.04953}}].

\bibitem{Jiang2025}
Z.-F. Jiang, S.-Y. Liu, T.-Y. Hu, H.-J. Zheng, and D.~She, {\it 1+1 dimensional relativistic viscous non-resistive magnetohydrodynamics with longitudinal boost invariance},  {\em Chin. Phys. C} {\bf 49} (2025) 114104, [\href{http://arxiv.org/abs/2411.11398}{{\tt arXiv:2411.11398}}].

\bibitem{Csorgo2008}
T.~Cs{\"o}rg{\H{o}}, M.~I. Nagy, and M.~Csan{\'a}d, {\it New family of simple solutions of relativistic perfect fluid hydrodynamics},  {\em Phys. Lett. B} {\bf 663} (2008) 306--311, [\href{http://arxiv.org/abs/nucl-th/0605070}{{\tt nucl-th/0605070}}].

\bibitem{Nagy2008}
M.~I. Nagy, T.~Cs{\"o}rg{\H{o}}, and M.~Csan{\'a}d, {\it Detailed description of accelerating, simple solutions of relativistic perfect fluid hydrodynamics},  {\em Phys. Rev. C} {\bf 77} (2008) 024908, [\href{http://arxiv.org/abs/0709.3677}{{\tt arXiv:0709.3677}}].

\bibitem{CKCJ2018}
T.~Cs{\"o}rg{\H{o}}, G.~Kasza, M.~Csan{\'a}d, and Z.-F. Jiang, {\it New exact solutions of relativistic hydrodynamics for longitudinally expanding fireballs},  {\em Universe} {\bf 4} (2018) 69, [\href{http://arxiv.org/abs/1805.01427}{{\tt arXiv:1805.01427}}].

\bibitem{She2019}
D.~She, Z.-F. Jiang, D.~Hou, and C.~B. Yang, {\it 1+1 dimensional relativistic magnetohydrodynamics with longitudinal acceleration},  {\em Phys. Rev. D} {\bf 100} (2019) 116014, [\href{http://arxiv.org/abs/1907.01250}{{\tt arXiv:1907.01250}}].

\bibitem{Jiang2020}
Z.-F. Jiang, D.~She, C.~B. Yang, and D.~Hou, {\it Perturbation solutions of relativistic viscous hydrodynamics for longitudinally expanding fireballs},  {\em Chin. Phys. C} {\bf 44} (2020) 084107, [\href{http://arxiv.org/abs/2001.09416}{{\tt arXiv:2001.09416}}].

\bibitem{SongHeinz2008}
H.~Song and U.~Heinz, {\it {Causal viscous hydrodynamics in 2 + 1 dimensions for relativistic heavy-ion collisions}},  {\em Phys. Rev. C} {\bf 77} (2008), no.~6 064901, [\href{http://arxiv.org/abs/0712.3715}{{\tt arXiv:0712.3715}}].

\bibitem{ShenVISHNU}
C.~Shen, Z.~Qiu, H.~Song, J.~Bernhard, S.~Bass, and U.~Heinz, {\it {The iEBE-VISHNU code package for relativistic heavy-ion collisions}},  {\em Comput. Phys. Commun.} {\bf 199} (2016) 61--85, [\href{http://arxiv.org/abs/1409.8164}{{\tt arXiv:1409.8164}}].

\bibitem{MusicIdeal}
B.~Schenke, S.~Jeon, and C.~Gale, {\it {(3+1)D hydrodynamic simulation of relativistic heavy-ion collisions}},  {\em Phys. Rev. C} {\bf 82} (2010), no.~1 014903, [\href{http://arxiv.org/abs/1004.1408}{{\tt arXiv:1004.1408}}].

\bibitem{MusicViscous}
B.~Schenke, S.~Jeon, and C.~Gale, {\it {Higher flow harmonics from (3+1)D event-by-event viscous hydrodynamics}},  {\em Phys. Rev. C} {\bf 85} (2012), no.~2 024901, [\href{http://arxiv.org/abs/1109.6289}{{\tt arXiv:1109.6289}}].

\bibitem{KarpenkoVHLLE}
I.~Karpenko, P.~Huovinen, and M.~Bleicher, {\it {A 3 + 1 dimensional viscous hydrodynamic code for relativistic heavy ion collisions}},  {\em Comput. Phys. Commun.} {\bf 185} (2014), no.~11 3016--3027, [\href{http://arxiv.org/abs/1312.4160}{{\tt arXiv:1312.4160}}].

\bibitem{DelZannaECHO}
L.~Del~Zanna, V.~Chandra, G.~Inghirami, V.~Rolando, A.~Beraudo, A.~De~Pace, G.~Pagliara, A.~Drago, and F.~Becattini, {\it {Relativistic viscous hydrodynamics for heavy-ion collisions with ECHO-QGP}},  {\em Eur. Phys. J. C} {\bf 73} (2013), no.~8 2524, [\href{http://arxiv.org/abs/1305.7052}{{\tt arXiv:1305.7052}}].

\bibitem{BozekViscous}
P.~Bo{\.z}ek, {\it {Flow and interferometry in (3 + 1)-dimensional viscous hydrodynamics}},  {\em Phys. Rev. C} {\bf 85} (2012), no.~3 034901, [\href{http://arxiv.org/abs/1110.6742}{{\tt arXiv:1110.6742}}].

\bibitem{PangCLVisc}
L.-G. Pang, H.~Petersen, and X.-N. Wang, {\it {Pseudorapidity distribution and decorrelation of anisotropic flow within the open-computing-language implementation CLVisc hydrodynamics}},  {\em Phys. Rev. C} {\bf 97} (2018), no.~6 064918, [\href{http://arxiv.org/abs/1802.04449}{{\tt arXiv:1802.04449}}].

\bibitem{SongEta}
H.~Song, S.~A. Bass, U.~Heinz, T.~Hirano, and C.~Shen, {\it {200 A GeV Au + Au Collisions Serve a Nearly Perfect Quark-Gluon Liquid}},  {\em Phys. Rev. Lett.} {\bf 106} (2011), no.~19 192301, [\href{http://arxiv.org/abs/1011.2783}{{\tt arXiv:1011.2783}}].

\bibitem{BernhardBayes}
J.~E. Bernhard, J.~S. Moreland, S.~A. Bass, J.~Liu, and U.~Heinz, {\it {Applying Bayesian parameter estimation to relativistic heavy-ion collisions: Simultaneous characterization of the initial state and quark-gluon plasma medium}},  {\em Phys. Rev. C} {\bf 94} (2016), no.~2 024907, [\href{http://arxiv.org/abs/1605.03954}{{\tt arXiv:1605.03954}}].

\bibitem{JetscapeBayes}
D.~Everett et~al., {\it {Multisystem Bayesian constraints on the transport coefficients of QCD matter}},  {\em Phys. Rev. C} {\bf 103} (2021), no.~5 054904, [\href{http://arxiv.org/abs/2011.01430}{{\tt arXiv:2011.01430}}].

\bibitem{FlorkowskiRapidity}
W.~Florkowski, R.~Ryblewski, M.~Strickland, and L.~Tinti, {\it {Non-boost-invariant dissipative hydrodynamics}},  {\em Phys. Rev. C} {\bf 94} (2016), no.~6 064903, [\href{http://arxiv.org/abs/1609.06293}{{\tt arXiv:1609.06293}}].

\bibitem{Jiang2018Fits}
Z.-F. Jiang, C.~B. Yang, C.~Ding, and X.-Y. Wu, {\it Pseudo-rapidity distribution from a perturbative solution of viscous hydrodynamics for heavy ion collisions at rhic and lhc},  {\em Chin. Phys. C} {\bf 42} (2018) 123103, [\href{http://arxiv.org/abs/1808.10287}{{\tt arXiv:1808.10287}}].

\bibitem{Hatta2014}
Y.~Hatta, J.~Noronha, and B.-W. Xiao, {\it Exact analytical solutions of second-order conformal hydrodynamics},  {\em Phys. Rev. D} {\bf 89} (2014) 051702, [\href{http://arxiv.org/abs/1401.6248}{{\tt arXiv:1401.6248}}].

\bibitem{Shi2022}
S.~Shi, S.~Jeon, and C.~Gale, {\it Family of new exact solutions for longitudinally expanding ideal fluids},  {\em Phys. Rev. C} {\bf 105} (2022) L021902, [\href{http://arxiv.org/abs/2201.06670}{{\tt arXiv:2201.06670}}].

\bibitem{Chen2024}
S.~Chen and S.~Shi, {\it Exact solution of boltzmann equation in a longitudinal expanding system},  {\em Phys. Rev. C} {\bf 109} (2024) L051901, [\href{http://arxiv.org/abs/2311.09575}{{\tt arXiv:2311.09575}}].

\bibitem{Chen2025a}
S.~Chen and S.~Shi, {\it Anisotropic hydrodynamics with a boost-noninvariant expansion},  {\em Phys. Rev. D} {\bf 111} (2025) 014001, [\href{http://arxiv.org/abs/2409.19897}{{\tt arXiv:2409.19897}}].

\bibitem{Chen2025b}
S.~Chen and S.~Shi, {\it Attractor for 1+1d viscous hydrodynamics with general rapidity distribution},  {\em Phys. Rev. C} {\bf 111} (2025) L021902, [\href{http://arxiv.org/abs/2407.15209}{{\tt arXiv:2407.15209}}].

\bibitem{Kurgyis2017}
B.~Kurgyis and M.~Csan{\'a}d, {\it Perturbative accelerating solutions of relativistic hydrodynamics},  {\em Universe} {\bf 3} (2017), no.~4 84, [\href{http://arxiv.org/abs/1711.05446}{{\tt arXiv:1711.05446}}].

\bibitem{Moghaddam2020}
M.~Haddadi~Moghaddam, W.~M. Alberico, D.~She, A.~F. Kord, and B.~Azadegan, {\it Accelerating longitudinal expansion of resistive relativistic-magneto-hydrodynamics in heavy ion collisions},  {\em Phys. Rev. D} {\bf 102} (2020) 014017, [\href{http://arxiv.org/abs/2002.09752}{{\tt arXiv:2002.09752}}].

\bibitem{Kord2022}
A.~F. Kord, A.~Ghaani, and M.~Haddadi~Moghaddam, {\it Analytical solution of magneto-hydrodynamics with acceleration effects of bjorken expansion in heavy-ion collisions},  {\em Eur. Phys. J. Plus} {\bf 137} (2022) 53, [\href{http://arxiv.org/abs/1911.05455}{{\tt arXiv:1911.05455}}].

\bibitem{Nakamura2023}
K.~Nakamura, T.~Miyoshi, C.~Nonaka, and H.~R. Takahashi, {\it Relativistic resistive magneto-hydrodynamics code for high-energy heavy-ion collisions},  {\em Eur. Phys. J. C} {\bf 83} (2023) 229, [\href{http://arxiv.org/abs/2211.02310}{{\tt arXiv:2211.02310}}].

\bibitem{HotQCDCrossover}
A.~Bazavov et~al., {\it The chiral and deconfinement aspects of the {QCD} transition},  {\em Phys. Rev. D} {\bf 85} (2012) 054503, [\href{http://arxiv.org/abs/1111.1710}{{\tt arXiv:1111.1710}}].

\bibitem{HotQCD2014}
{\bf HotQCD} Collaboration, A.~Bazavov et~al., {\it The equation of state in (2+1)-flavor qcd},  {\em Phys. Rev. D} {\bf 90} (2014) 094503, [\href{http://arxiv.org/abs/1407.6387}{{\tt arXiv:1407.6387}}].

\bibitem{SkokovField}
V.~V. SKOKOV, A.~Y. ILLARIONOV, and V.~D. TONEEV, {\it {Estimate of the magnetic field strength in heavy-ion collisions}},  {\em Int. J. Mod. Phys. A} {\bf 24} (2009), no.~31 5925--5932, [\href{http://arxiv.org/abs/0907.1396}{{\tt arXiv:0907.1396}}].

\bibitem{TuchinField}
K.~Tuchin, {\it {Time and space dependence of the electromagnetic field in relativistic heavy-ion collisions}},  {\em Phys. Rev. C} {\bf 88} (2013), no.~2 024911, [\href{http://arxiv.org/abs/1305.5806}{{\tt arXiv:1305.5806}}].

\bibitem{MclerranSkokov}
L.~McLerran and V.~Skokov, {\it {Comments about the electromagnetic field in heavy-ion collisions}},  {\em Nucl. Phys. A} {\bf 929} (2014) 184--190, [\href{http://arxiv.org/abs/1305.0774}{{\tt arXiv:1305.0774}}].

\bibitem{AartsConductivity}
G.~Aarts, C.~Allton, A.~Amato, P.~Giudice, S.~Hands, and J.-I. Skullerud, {\it {Electrical conductivity and charge diffusion in thermal QCD from the lattice}},  {\em JHEP} {\bf 2015} (2015), no.~2 186, [\href{http://arxiv.org/abs/1412.6411}{{\tt arXiv:1412.6411}}].

\bibitem{PuYangField}
S.~Pu and D.-L. Yang, {\it {Transverse flow induced by inhomogeneous magnetic fields in the Bjorken expansion}},  {\em Phys. Rev. D} {\bf 93} (2016), no.~5 054042, [\href{http://arxiv.org/abs/1602.04954}{{\tt arXiv:1602.04954}}].

\bibitem{HernandezKovtun}
J.~Hernandez and P.~Kovtun, {\it {Relativistic magnetohydrodynamics}},  {\em JHEP} {\bf 2017} (2017), no.~5 1, [\href{http://arxiv.org/abs/1703.08757}{{\tt arXiv:1703.08757}}].

\bibitem{IsraelStewart}
W.~Israel and J.~Stewart, {\it {Transient relativistic thermodynamics and kinetic theory}},  {\em Ann. Phys.} {\bf 118} (1979), no.~2 341--372. \href{https://doi.org/10.1016/0003-4916(79)90130-1}{Publisher record}.

\bibitem{Romatschke2010}
P.~Romatschke, {\it New developments in relativistic viscous hydrodynamics},  {\em Int. J. Mod. Phys. E} {\bf 19} (2010) 1, [\href{http://arxiv.org/abs/0902.3663}{{\tt arXiv:0902.3663}}].

\bibitem{Denicol2012}
G.~S. Denicol, H.~Niemi, E.~Moln{\'a}r, and D.~H. Rischke, {\it Derivation of transient relativistic fluid dynamics from the boltzmann equation},  {\em Phys. Rev. D} {\bf 85} (2012) 114047, [\href{http://arxiv.org/abs/1202.4551}{{\tt arXiv:1202.4551}}].

\bibitem{BetzMoments}
B.~Betz, D.~Henkel, and D.~H. Rischke, {\it From kinetic theory to dissipative fluid dynamics},  {\em Prog. Part. Nucl. Phys.} {\bf 62} (2009) 556--558, [\href{http://arxiv.org/abs/0812.1440}{{\tt arXiv:0812.1440}}].

\bibitem{BaierBRSSS}
R.~Baier, P.~Romatschke, D.~T. Son, A.~O. Starinets, and M.~A. Stephanov, {\it {Relativistic viscous hydrodynamics, conformal invariance, and holography}},  {\em JHEP} {\bf 2008} (2008), no.~04 100--100, [\href{http://arxiv.org/abs/0712.2451}{{\tt arXiv:0712.2451}}].

\bibitem{JaiswalRTA}
A.~Jaiswal, {\it {Relativistic dissipative hydrodynamics from kinetic theory with relaxation-time approximation}},  {\em Phys. Rev. C} {\bf 87} (2013), no.~5 051901, [\href{http://arxiv.org/abs/1302.6311}{{\tt arXiv:1302.6311}}].

\bibitem{ShenSchenkeInitial}
C.~Shen and B.~Schenke, {\it {Dynamical initial-state model for relativistic heavy-ion collisions}},  {\em Phys. Rev. C} {\bf 97} (2018), no.~2 024907, [\href{http://arxiv.org/abs/1710.00881}{{\tt arXiv:1710.00881}}].

\bibitem{ShenSchenkeDynamic}
C.~Shen and B.~Schenke, {\it {Dynamical initialization and hydrodynamic modeling of relativistic heavy-ion collisions}},  {\em Nucl. Phys. A} {\bf 982} (2019) 411--414, [\href{http://arxiv.org/abs/1807.05141}{{\tt arXiv:1807.05141}}].

\bibitem{ShenAlzhrani}
C.~Shen and S.~Alzhrani, {\it {Collision-geometry-based 3D initial condition for relativistic heavy-ion collisions}},  {\em Phys. Rev. C} {\bf 102} (2020), no.~1 014909, [\href{http://arxiv.org/abs/2003.05852}{{\tt arXiv:2003.05852}}].

\bibitem{BozekTilt}
P.~Bo{\.z}ek and I.~Wyskiel, {\it {Directed flow in ultrarelativistic heavy-ion collisions}},  {\em Phys. Rev. C} {\bf 81} (2010), no.~5 054902, [\href{http://arxiv.org/abs/1002.4999}{{\tt arXiv:1002.4999}}].

\bibitem{PangInitialFlow}
L.~Pang, Q.~Wang, and X.-N. Wang, {\it {Effects of initial flow velocity fluctuation in event-by-event (3+1)D hydrodynamics}},  {\em Phys. Rev. C} {\bf 86} (2012), no.~2 024911, [\href{http://arxiv.org/abs/1205.5019}{{\tt arXiv:1205.5019}}].

\bibitem{Kovtun2005}
P.~Kovtun, D.~T. Son, and A.~O. Starinets, {\it Viscosity in strongly interacting quantum field theories from black hole physics},  {\em Phys. Rev. Lett.} {\bf 94} (2005) 111601, [\href{http://arxiv.org/abs/hep-th/0405231}{{\tt hep-th/0405231}}].

\bibitem{Denicol2018MHD}
G.~S. Denicol, X.-G. Huang, E.~Moln{\'a}r, G.~M. Monteiro, H.~Niemi, J.~Noronha, D.~H. Rischke, and Q.~Wang, {\it Non-resistive dissipative magnetohydrodynamics from the {B}oltzmann equation in the 14-moment approximation},  {\em Phys. Rev. D} {\bf 98} (2018) 076009, [\href{http://arxiv.org/abs/1804.05210}{{\tt arXiv:1804.05210}}].

\bibitem{Loizides2015}
C.~Loizides, J.~Nagle, and P.~Steinberg, {\it Improved version of the {PHOBOS} {G}lauber {M}onte {C}arlo},  {\em SoftwareX} {\bf 1--2} (2015) 13--18, [\href{http://arxiv.org/abs/1408.2549}{{\tt arXiv:1408.2549}}].

\bibitem{Muronga2004}
A.~Muronga, {\it Causal theories of dissipative relativistic fluid dynamics for nuclear collisions},  {\em Phys. Rev. C} {\bf 69} (2004) 034903, [\href{http://arxiv.org/abs/nucl-th/0309055}{{\tt nucl-th/0309055}}].

\bibitem{FlorkowskiExact}
W.~Florkowski, R.~Ryblewski, and M.~Strickland, {\it {Testing viscous and anisotropic hydrodynamics in an exactly solvable case}},  {\em Phys. Rev. C} {\bf 88} (2013), no.~2 024903, [\href{http://arxiv.org/abs/1305.7234}{{\tt arXiv:1305.7234}}].

\bibitem{FlorkowskiAniso}
W.~Florkowski, R.~Ryblewski, and M.~Strickland, {\it {Anisotropic hydrodynamics for rapidly expanding systems}},  {\em Nucl. Phys. A} {\bf 916} (2013) 249--259, [\href{http://arxiv.org/abs/1304.0665}{{\tt arXiv:1304.0665}}].

\bibitem{HellerSpalinski}
M.~P. Heller and M.~Spalinski, {\it {Hydrodynamics Beyond the Gradient Expansion: Resurgence and Resummation}},  {\em Phys. Rev. Lett.} {\bf 115} (2015), no.~7 072501, [\href{http://arxiv.org/abs/1503.07514}{{\tt arXiv:1503.07514}}].

\bibitem{GubserSymmetry}
S.~S. Gubser, {\it {Symmetry constraints on generalizations of Bjorken flow}},  {\em Phys. Rev. D} {\bf 82} (2010), no.~8 085027, [\href{http://arxiv.org/abs/1006.0006}{{\tt arXiv:1006.0006}}].

\bibitem{DenicolGubser}
G.~Denicol, U.~Heinz, M.~Martinez, J.~Noronha, and M.~Strickland, {\it {Studying the validity of relativistic hydrodynamics with a new exact solution of the Boltzmann equation}},  {\em Phys. Rev. D} {\bf 90} (2014), no.~12 125026, [\href{http://arxiv.org/abs/1408.7048}{{\tt arXiv:1408.7048}}].

\bibitem{InghiramiMHD}
G.~Inghirami, L.~Del~Zanna, A.~Beraudo, M.~H. Moghaddam, F.~Becattini, and M.~Bleicher, {\it {Numerical magneto-hydrodynamics for relativistic nuclear collisions}},  {\em Eur. Phys. J. C} {\bf 76} (2016), no.~12 659, [\href{http://arxiv.org/abs/1609.03042}{{\tt arXiv:1609.03042}}].

\bibitem{BhacQgp}
M.~Mayer, A.~Dash, G.~Inghirami, H.~Elfner, L.~Rezzolla, and D.~H. Rischke, {\it {Three-dimensional magnetohydrodynamics simulations of relativistic heavy-ion collisions with a black-hole accretion code adapted for quark-gluon plasma (BHAC-QGP). I. Methods and tests}},  {\em Phys. Rev. C} {\bf 111} (2025), no.~4 044907, [\href{http://arxiv.org/abs/2403.08668}{{\tt arXiv:2403.08668}}].

\bibitem{BrahmsRapidity}
I.~G. Bearden et~al., {\it {Charged Meson Rapidity Distributions in Central Au+Au Collisions at $\sqrt{s_{NN}}=200$ GeV}},  {\em Phys. Rev. Lett.} {\bf 94} (2005), no.~16 162301, [\href{http://arxiv.org/abs/nucl-ex/0403050}{{\tt nucl-ex/0403050}}].

\bibitem{CmsDecorrelation}
{CMS Collaboration}, {\it {Evidence for transverse momentum and pseudorapidity dependent event plane fluctuations in PbPb and pPb collisions}},  {\em Phys. Rev. C} {\bf 92} (2015) 034911, [\href{http://arxiv.org/abs/1503.01692}{{\tt arXiv:1503.01692}}].

\bibitem{AtlasDecorrelation}
{ATLAS Collaboration}, {\it {Measurement of longitudinal flow de-correlations in Pb+Pb collisions at $\sqrt{s_\mathrm{NN}} = 2.76$ and 5.02 TeV with the ATLAS detector}},  {\em Eur. Phys. J. C} {\bf 78} (2018) 142, [\href{http://arxiv.org/abs/1709.02301}{{\tt arXiv:1709.02301}}].

\bibitem{BozekTorque}
P.~Bo{\.z}ek and W.~Broniowski, {\it {The torque effect and fluctuations of entropy deposition in rapidity in ultra-relativistic nuclear collisions}},  {\em Phys. Lett. B} {\bf 752} (2016) 206--211, [\href{http://arxiv.org/abs/1506.02817}{{\tt arXiv:1506.02817}}].

\bibitem{BroniowskiRapidity}
W.~Broniowski and P.~Bo{\.z}ek, {\it {Simple model for rapidity fluctuations in the initial state of ultrarelativistic heavy-ion collisions}},  {\em Phys. Rev. C} {\bf 93} (2016), no.~6 064910, [\href{http://arxiv.org/abs/1512.01945}{{\tt arXiv:1512.01945}}].

\bibitem{SchenkeSchlichting}
B.~Schenke and S.~Schlichting, {\it {3-D Glasma initial state from small-x evolution}},  {\em Nucl. Phys. A} {\bf 967} (2017) 285--288, [\href{http://arxiv.org/abs/1704.03018}{{\tt arXiv:1704.03018}}].

\bibitem{HydroRapidityFluct}
A.~Sakai, K.~Murase, and T.~Hirano, {\it {Effects of hydrodynamic and initial longitudinal fluctuations on rapidity decorrelation of collective flow}},  {\em Phys. Lett. B} {\bf 829} (2022) 137053, [\href{http://arxiv.org/abs/2111.08963}{{\tt arXiv:2111.08963}}].

\bibitem{CooperFrye}
F.~Cooper and G.~Frye, {\it {Single-particle distribution in the hydrodynamic and statistical thermodynamic models of multiparticle production}},  {\em Phys. Rev. D} {\bf 10} (1974), no.~1 186--189. \href{https://doi.org/10.1103/physrevd.10.186}{Publisher record}.

\bibitem{McNelisHeinz2021}
M.~McNelis and U.~Heinz, {\it Modified equilibrium distributions for {Cooper--Frye} particlization},  {\em Phys. Rev. C} {\bf 103} (2021) 064903, [\href{http://arxiv.org/abs/2103.03401}{{\tt arXiv:2103.03401}}].

\bibitem{Steinberg2005}
P.~Steinberg, {\it Landau hydrodynamics and {RHIC} phenomena},  {\em Acta Phys. Hung. A} {\bf 24} (2005) 51--57, [\href{http://arxiv.org/abs/nucl-ex/0405022}{{\tt nucl-ex/0405022}}].

\bibitem{HattoriSpin}
K.~Hattori, M.~Hongo, X.-G. Huang, M.~Matsuo, and H.~Taya, {\it {Fate of spin polarization in a relativistic fluid: An entropy-current analysis}},  {\em Phys. Lett. B} {\bf 795} (2019) 100--106, [\href{http://arxiv.org/abs/1901.06615}{{\tt arXiv:1901.06615}}].

\bibitem{WangBjorkenSpin}
D.-L. Wang, S.~Fang, and S.~Pu, {\it {Analytic solutions of relativistic dissipative spin hydrodynamics with Bjorken expansion}},  {\em Phys. Rev. D} {\bf 104} (2021), no.~11 114043, [\href{http://arxiv.org/abs/2107.11726}{{\tt arXiv:2107.11726}}].

\bibitem{BhadurySpinMHD}
S.~Bhadury, W.~Florkowski, A.~Jaiswal, A.~Kumar, and R.~Ryblewski, {\it {Relativistic Spin Magnetohydrodynamics}},  {\em Phys. Rev. Lett.} {\bf 129} (2022), no.~19 192301, [\href{http://arxiv.org/abs/2204.01357}{{\tt arXiv:2204.01357}}].

\bibitem{FangSpinMHD}
Z.~Fang, K.~Hattori, and J.~Hu, {\it {Anisotropic linear waves and breakdown of the momentum expansion in spin magnetohydrodynamics}},  {\em Phys. Rev. D} {\bf 112} (2025), no.~7 076016, [\href{http://arxiv.org/abs/2409.07096}{{\tt arXiv:2409.07096}}].

\bibitem{FornbergStencils}
B.~Fornberg, {\it {Generation of finite difference formulas on arbitrarily spaced grids}},  {\em Math. Comp.} {\bf 51} (1988), no.~184 699--706. \href{https://people.sc.fsu.edu/~jburkardt/m_src/differ_arbitrary/fornberg.pdf}{Public full text}.

\end{thebibliography}\endgroup
\end{document}